\documentclass[iop]{emulateapj}
\usepackage{xcolor}
\usepackage[colorinlistoftodos]{todonotes}
\usepackage{amssymb}
\usepackage{apjfonts}
\usepackage{graphicx}
\usepackage{soul}
\usepackage{CJK}

\newcommand{\teff}{T$_{\rm eff}$}
\newcommand{\logg}{$\log{g}$}
\newcommand{\feh}{[Fe/H]}
\newcommand{\afe}{[$\alpha$/Fe]}
\newcommand{\Vstar}{v$_{\rm stellar}$}
\newcommand{\Vrad}{v$_{\rm rad}$}
\newcommand{\Vmicro}{v$_{\rm micro}$}
\newcommand{\Av}{A$_{\rm V}$}

\newcommand{\MS}{\texttt{MINESweeper}}
\newcommand{\Msun}{\,M_{\odot}}

\newcommand{\kms}{km s$^{-1}$}

\shorttitle{Spectrophotometric Modeling of Stars}
\shortauthors{Cargile et al.}

\begin{document}

\begin{CJK*}{UTF8}{gbsn}

\title{\MS: Spectrophotometric Modeling of Stars in the {\it Gaia} Era}

\author{Phillip A. Cargile\altaffilmark{1,*}, Charlie Conroy\altaffilmark{1}, Benjamin D. Johnson\altaffilmark{1}, Yuan-Sen Ting (丁源森)\altaffilmark{2,3,4,5,6}, Ana Bonaca\altaffilmark{1}, Aaron Dotter\altaffilmark{1},  Joshua S. Speagle\altaffilmark{1}}
\altaffiltext{1}{Center for Astrophysics $\vert$ Harvard \& Smithsonian, Cambridge, MA, 02138, USA} 
\altaffiltext{2}{Institute for Advanced Study, Princeton, NJ 08540, USA} 
\altaffiltext{3}{Department of Astrophysical Sciences, Princeton University, Princeton, NJ 08544, USA} 
\altaffiltext{4}{Observatories  of  the  Carnegie  Institution  of  Washington,  813  Santa Barbara Street, Pasadena, CA 91101, USA} 
\altaffiltext{5}{Research School of Astronomy \& Astrophysics, Australian National University, Cotter Rd., Weston, ACT 2611, Australia}
\altaffiltext{6}{Hubble Fellow} 
\altaffiltext{*}{Correspondence to: \email{pcargile@cfa.harvard.edu}}

\begin{abstract}

We present \MS, a tool to measure stellar parameters by jointly fitting observed spectra and broadband photometry to model isochrones and spectral libraries.  This approach enables the measurement of spectrophotometric distances, in addition to stellar parameters such as \teff, \logg, \feh, \afe, and radial velocity.  \MS\, employs a Bayesian framework and can easily incorporate a variety of priors, including {\it Gaia} parallaxes.  Mock data are fit in order to demonstrate how the precision of derived parameters depends on evolutionary phase and SNR.  We then fit a selection of data in order to validate the model outputs.  Fits to a variety of  benchmark stars including Procyon, Arcturus, and the Sun result in derived stellar parameters that are in good agreement with the literature.  We then fit combined spectra and photometry of stars in the open and globular clusters M92, M13, M3, M107, M71, and M67.  Derived distances, \feh, \afe, and \logg$-$\teff\, relations are in overall good agreement with literature values, although there are trends between metallicity and \logg\, within clusters that point to systematic uncertainties at the $\approx0.1$ dex level.  Finally, we fit a large sample of stars from the H3 Spectroscopic Survey in which high quality {\it Gaia} parallaxes are also available.  These stars are fit {\it without} the {\it Gaia} parallaxes so that the geometric parallaxes can serve as an independent test of the spectrophotometric distances.  Comparison between the two reveals good agreement within their formal uncertainties after accounting for the {\it Gaia} zero point uncertainties.

\end{abstract}

\section{Introduction}
\label{sec.intro}

The field of Galactic astronomy is undergoing a dramatic change driven by multiple large-scale ground-based photometric and spectroscopic surveys, and the space-based {\it Gaia} mission \citep{Gaia2016}.  Together, these observational efforts promise to provide new, much sharper maps of the stellar components of the Galaxy.  In turn, these maps should deliver new insight into the formation and assembly history of the Galaxy, its dynamical state, and possibly the nature of dark matter.

In order to make progress, the raw observations must be converted into physical quantities.  These include stellar properties such as \teff, \logg, \feh, \afe, mass, and age, and 3D positions and velocities.  In previous work the measurement of these quantities has been performed in separate steps.  For example, different methods for analysis of stellar spectra can deliver, with varying degrees of reliability, estimates of \teff, \logg, radial velocities, and abundances \citep[e.g.,][]{Smiljanic2014, Jofre2019}. In order to estimate masses, ages, and distances, one often combines these derived stellar parameters with model stellar isochrones \citep[e.g.,][]{Pont2004, Jorgensen2005, Takeda2007, Juric2008, Breddels2010, Burnett2010, Binney2014, Xue2014, Wang2016, Mints2017, Mints2018, Anders2019}.  In essence, this approach translates stellar properties that can be measured independent of distance and age such as \teff, \logg, and \feh\, into a predicted age and absolute luminosity.  Comparison to the observed flux enables a measurement of distance (an estimate of the extinction must also be available).

For relatively bright and nearby sources, {\it Gaia} DR2 now provides very precise parallaxes.  For fainter and more distant sources the {\it Gaia} parallax becomes quite uncertain and hence distances are also uncertain, and depend on various subtleties such as the method of parallax inversion and assumed priors \citep[e.g.,][]{Bailer-Jones2018, Luri18}.  Therefore, even in the {\it Gaia} era, there is a need for complementary distance constraints provided by stellar isochrones \citep[e.g.,][]{Sanders2018, Mints2018} or empirically-calibrated data-driven models \citep[e.g.,][]{Leistedt2017, Anderson2018}.  Furthermore, even with precise parallaxes, model isochrones are still necessary for estimating masses and ages.

A limitation of existing isochrone-based methods is the reliance on stellar parameters that have been derived elsewhere (e.g., from a separate spectroscopic pipeline).  This presents several challenges.  It is difficult to accurately propagate the uncertainties in the stellar parameters to the isochrone-based parameters (unless the full covariance matrices are preserved).  Moreover, there is no guarantee that the derived stellar parameters will coincide with any isochrone.  In fact, owing to various uncertainties associated with deriving stellar parameters, there are many cases in which unphysical \teff$-$\logg\, relations along the main sequence have been derived from fitting spectroscopic data \citep[e.g.,][]{Valenti2005, Adibekyan2012, Holtzman2015}. Consequently, deriving isochrone-based properties in such cases would incur large systematic uncertainties.

\citet{Schonrich2014} proposed a unified framework in which isochrone-based stellar parameters are derived directly from the available data, including spectra, photometry, and parallaxes, as well as prior information on other quantities where available.  They argue that this approach is optimal in that the all available data are brought to bear on the problem, resulting in the most robust available parameter constraints.  In this paper we present \MS, a new tool that follows the same general philosophy of \citet{Schonrich2014}.  Specifically, the program incorporates a variety of observational constraints, including {\it Gaia} parallaxes, optical spectra, and broadband photometry, in order to estimate stellar parameters and distances.  

\section{Model}\label{sec.minesweeper}

\MS\, is a program to jointly model observed stellar spectra and photometry using stellar isochrones.  Brief descriptions of early versions of this code can be found in several previous studies where we have determined stellar parameters \citep[e.g.,][]{Dotter2017, Rodriguez2017}. In this section we provide a detailed overview of the basic ingredients and our approach to the computation of the likelihood and priors.  

\subsection{Generative Model for Stellar Spectra \& Photometry} \label{sec.minesweeper.model}

Our model consists of two main parts, (1) a stellar evolution model that predicts the physical parameters (\teff, \logg, \feh, etc.) of a star given its mass, evolutionary state, and initial composition, and (2) a stellar spectral model that takes these parameters and, along with a model for interstellar reddening, predicts the observable spectrum and photometry from this star.  These two components are described in the sections below.

\subsubsection{\texttt{MIST} Stellar Evolution Models}

We use the \texttt{MIST} stellar evolution models, full details of which can be found in \citet[][]{Choi2016} and Dotter et al. (in prep). \texttt{MIST} uses the stellar evolution code \texttt{MESA} \citep{Paxton2011} to compute evolutionary sequences in 1D for stars in the mass range $0.1-300\Msun$.  Stars are evolved from pre-main sequence contraction through either the end of carbon burning or until the white dwarf cooling stage.  A variety of free parameters, including those controlling mass-loss rates and convection, are tuned to match observational constraints.  \texttt{MIST} models distinguish between the initial composition of the star and its current surface composition.  These two can differ due to diffusion and mixing near the surface layers \citep[e.g.,][]{Dotter2017}.  We refer to the initial composition with a $i$ subscript, e.g., \feh$_i$.

In this work we use \texttt{MIST} v2.0 rotating models (Dotter et al. in prep).  A key improvement in v2.0 compared to previous models is the inclusion of non-solar abundance ratios (i.e., variable \afe$_i$).  In addition, the treatment of the surface boundary conditions are significantly improved, which results in better agreement with the red giant branch temperature scale.

We incorporate the \texttt{MIST} v2.0 models into \MS\, using an N-D linear interpolator in order to predict the effective temperature, surface gravity, surface composition, and bolometric luminosity of stars. Adopting the recommendation given in \citet[][]{Dotter2016}, we have chosen to interpolate the \texttt{MIST} models on equivalent evolutionary points (EEP), initial mass, and initial composition.  EEP is monotonically related to the age, but it is defined in such a way so that evolutionary phases in which there are rapid changes in either surface or interior stellar properties are adequately captured in the isochrone tables.  

The program samples in initial composition (\feh$_i$ and \afe$_i$), initial mass, $M_i$, and evolutionary stage (EEP).  The specification of these parameters makes a unique prediction for the spectrum and photometry, which we describe next.

\subsubsection{Model Spectra \& Photometry}

A given EEP, initial mass, and composition fully specifies the stellar surface properties, including \teff, \logg, \feh, and \afe.  Model spectra and photometry are assigned to these parameters using our custom grid of stellar spectral models. The synthetic spectra are calculated using the 1D LTE plane-parallel atmosphere and radiative transfer codes \texttt{ATLAS12} and \texttt{SYNTHE} maintained by R. Kurucz \citep[][]{Kurucz1970,Kurucz1981,Kurucz1993}. The line list used in the radiative transfer calculation was provided by R. Kurucz, and has been empirically tuned to the observed, ultra-high resolution spectra of the Sun and Arcturus.  We have adopted a constant microturbulence of \Vmicro$=1$ \kms.  We have computed models with variable $\alpha$-abundance by scaling the $\alpha$ elements (O, Ne, Mg, Si, S, Ar, Ca and Ti) in lock-step.   The grid is computed with $\Delta$\logg$\,=0.5$ over the range $[-1,5]$, $\Delta$\feh$\,=0.25$ over the range [$-4,+0.5$], $\Delta$\afe$\,=0.2$ over the range [$-0.2,+0.6$], and an irregular (approximately log-spaced) \teff\, grid with 40 points between 3500 K and 15,000K.The synthetic spectra in our custom grid we computed at a resolving power of R$=$1,000,000, and were subsequently smoothed to a resolution of R$=$42,000 and re-binned to critical sampling before building the generative model. We note, that the \MS code is capable of including different wavelength ranges and resolutions, investigating the effect of these changes on inferred stellar parameters is left to future work.

Photometry is computed from each synthetic spectrum for a large set of photometric systems. Details of the specific filter curves and zeropoints can be found in \citet[][]{Choi2016}. Additionally, we have updated the filter curves and zeropoints for {\it Gaia} G, BP and RP with the latest DR2 photometric calibration \citep{Evans2018,MaizApellaniz2018}.  We also include extinction in our model, using the extinction curve of \citet[][]{Cardelli1989}, characterized by \Av\ with $R_{\rm V}=3.1$.

\subsubsection{Neural Networks for Rapid Prediction of Stellar Spectra and Photometry}

In order to determine the spectrum and photometry for any set of interpolated \texttt{MIST} stellar labels, we need to be able to quickly evaluate the spectral models at arbitrary \teff, \logg, and \feh. As explained in \citet[][]{Ting2019}, the use of artificial neural networks to predict synthetic spectra (and in our case also photometry) has advantages over other interpolation methods like quadratic functions or Gaussian Processes. Namely, neural networks are very flexible and can easily handle non-linear behavior of spectra, and can be evaluated extremely rapidly which is important in Monte Carlo-like sampling techniques. 

Here, we take a similar approach to \texttt{The Payne} as described in \citet[][]{Ting2019}, to which we direct the interested reader for more details. The main difference is that instead of emulating the flux variation of individual pixels, here we emulate the model spectrum as a whole. Emulating the whole spectrum allows one to extract correlations between adjacent pixels which generally leads to better emulations. This change is also reflected in the latest version of The Payne\footnote{\href{https://github.com/tingyuansen/The_Payne}{https://github.com/tingyuansen/The\_Payne}}. We train a multi-layered feed-forward artificial neural network on the grid of normalized synthetic spectra. We adopt a training set of 18000 model spectra and hold out 2000 spectra as the validation set. The network we use in \MS\ has four layers (two hidden layers) with 300 neurons per layer and rectified linear unit (ReLU) activation functions. We employ an adaptive cross-validation procedure to optimize the weights and biases (i.e., the coefficients for the individual ReLU activation functions) of the neural network, where we investigate the training at regular epochs, and through standard validation techniques (e.g., holdout method, leave-one-out method, etc.). We adopt a batch size of 512, and train for $\approx10^6$ epoch. In a similar way, we train a multi-layered feed-forward artificial neural network on synthetic photometry derived from our same synthetic spectral grid. A three layer (one hidden layer) network with 128 neurons per layer is trained on 10000 model photometric datasets for all of the bands of interest (e.g., Pan-STARRS ugriz, Gaia DR2, 2MASS, etc). We also use a multi-epoch training schema and cross-validation techniques to validate the accuracy of these networks. 

Our final validation of the \MS\ neural network results in a predicted spectral and photometric fluxes that agreed with a ``held-out" test model data set to a median residual of $\le0.01-0.1$\% over all wavelength pixels and photometric bands, and over \teff $=$ 3500 -- 15000 K, \logg $=-1$ to 5, \feh $=-4$ to $+0.5$, \afe $=-0.2$ to $+0.6$, and \Av $=0-5$. The code is written in Python using the machine learning package \texttt{PyTorch} \citep{Paszke2017} and takes $\sim$24 hours to train the full network on modern CPUs with GPU acceleration.

Once the neural network is trained, it is used within \MS\, to provide spectra and photometry at arbitrary locations in the grid.

\begin{deluxetable*}{llll}[t!]
\tablewidth{0.95\linewidth}
\tabletypesize{\scriptsize}
\tablecaption{Summary of parameters and priors used in \MS \label{tab.priors}}
\tablehead{
\colhead{Parameter} & \multicolumn{3}{c}{Prior Functions} \\
\colhead{}          & \colhead{Mocks} & \colhead{Clusters} & \colhead{H3}
}
\startdata
\cutinhead{Sampled Parameters}
Mass  [M$_{\odot}$]  & Kroupa IMF     & Kroupa IMF     & Kroupa IMF    \\
EEP                  & U($200$,$808$) & U($200$,$808$) & U($200$,$808$)\\
\feh$_{i}$           & U($-4.0,+0.5$) & U($-4.0,+0.5$) & U($-4.0,+0.5$)\\
\afe$_{i}$           & U($-0.2,+0.6$) & U($-0.2,+0.6$) & U($-0.2,+0.6$)\\
distance [kpc]       & U($1,50$)      & U($0,50$)      & Galactic Model\tablenotemark{a} \\
\Vrad\tablenotemark{b} [\kms] & U($v_{\rm rad,init}-25$,\,$v_{\rm rad,init}+25$) & U($v_{\rm rad,init}-25$,\,$v_{\rm rad,init}+25$) & U($v_{\rm rad,init}-25$,\,$v_{\rm rad,init}+25$) \\
\Vstar [\kms]  & U($0,15$)$\times$S(\Vstar)\tablenotemark{c} & U($0,15$)$\times$S(\Vstar) & U($0,15$)$\times$S(\Vstar)\\
T$_{n}$ & \nodata\tablenotemark{d} & N(1,[$0.5,0.1,0.05,0.05$]) & N(1,[$0.5,0.1,0.05,0.05$]) \\
\Av\ & N(0.1,0.015) & U(0,1.5) & N(\Av$_{\rm ,SFD}$,0.15$\times$\Av$_{\rm ,SFD})$\\
\cutinhead{Derived Parameters}
Age [Gyr]  & U($4,14$) & M67: U($0.5,8$), others: U($6,14$) & Galactic Model\tablenotemark{a} \\
$\pi$ [mas]   & N($\pi$,$\sigma_{\pi}$) & N($\pi_{\rm Gaia DR2}$,$\sigma_{\pi_{\rm Gaia DR2}}$) & N($\pi_{\rm Gaia DR2}$,$\sigma_{\pi_{\rm Gaia DR2}}$) \\ 
\enddata
\tablecomments{U($x$,$y$) indicates a uniform prior from $x$ to $y$, while N($\mu$,$\sigma$): normal prior.}
\tablenotetext{a}{See text for a description of our Galactic model.}
\tablenotetext{b}{$v_{\rm rad,init}$ is an initial guess on the radial velocity of the star, typically taken from a preliminary optimization (see text for details).}
\tablenotetext{c}{S(\Vstar) is a Sigmoid prior on the stellar broadening with an exponential drop-off at 7 \kms for giants and 10 \kms for dwarfs and subgiants.}
\tablenotetext{d}{Mock spectra are synthesized as continuum-normalized models, therefore, no Chebyshev polynomial is included when fitting them.}
\end{deluxetable*}

\subsubsection{Continuum normalization}

Accurate flux calibration of observed spectra is very challenging.  Since we are simultaneously able to fit the broadband photometry in \MS, which captures the overall shape of the stellar spectral energy distribution, we can forego the information content contained in the broadband shape of the spectrum.  In particular, we work with continuum-normalized model spectra, and therefore must also continuum-normalize the observed spectra.  In the example cases presented below we have performed an initial preparation of the data by using twilight observations in order to approximately continuum-normalize the observations (this procedure takes into account wavelength-dependent throughput variations). In addition, during the fitting process we multiply the observed spectrum by a 4th order Chebyshev polynomial whose coefficients, T$_n$, are free parameters.

\subsection{Fitting Spectrophotometric Data}\label{sec.minesweeper.fitting}

In \MS\ we use a Bayesian framework to infer the physical properties of a star given a set of input data.  Using Bayes's theorem, we sample the posterior distribution by evaluating the likelihood of observing the stellar spectrum and/or photometric data given a set of model parameters (known as the likelihood probability distribution, or just likelihood), the probability of our model (the prior), and normalizing these probabilities over all likelihood space (the evidence). Calculating the evidence is necessary when performing model comparison, as well as formally weighting individual posterior distributions in hierarchical modeling. In the following sections, we describe the calculation of the combined spectroscopic and photometric likelihood probability, a general description of typical priors, as well as the method we use to estimate the posterior probability and infer stellar properties.

\begin{figure*}[t!]
    \centering
    \includegraphics[width=1.0\linewidth]{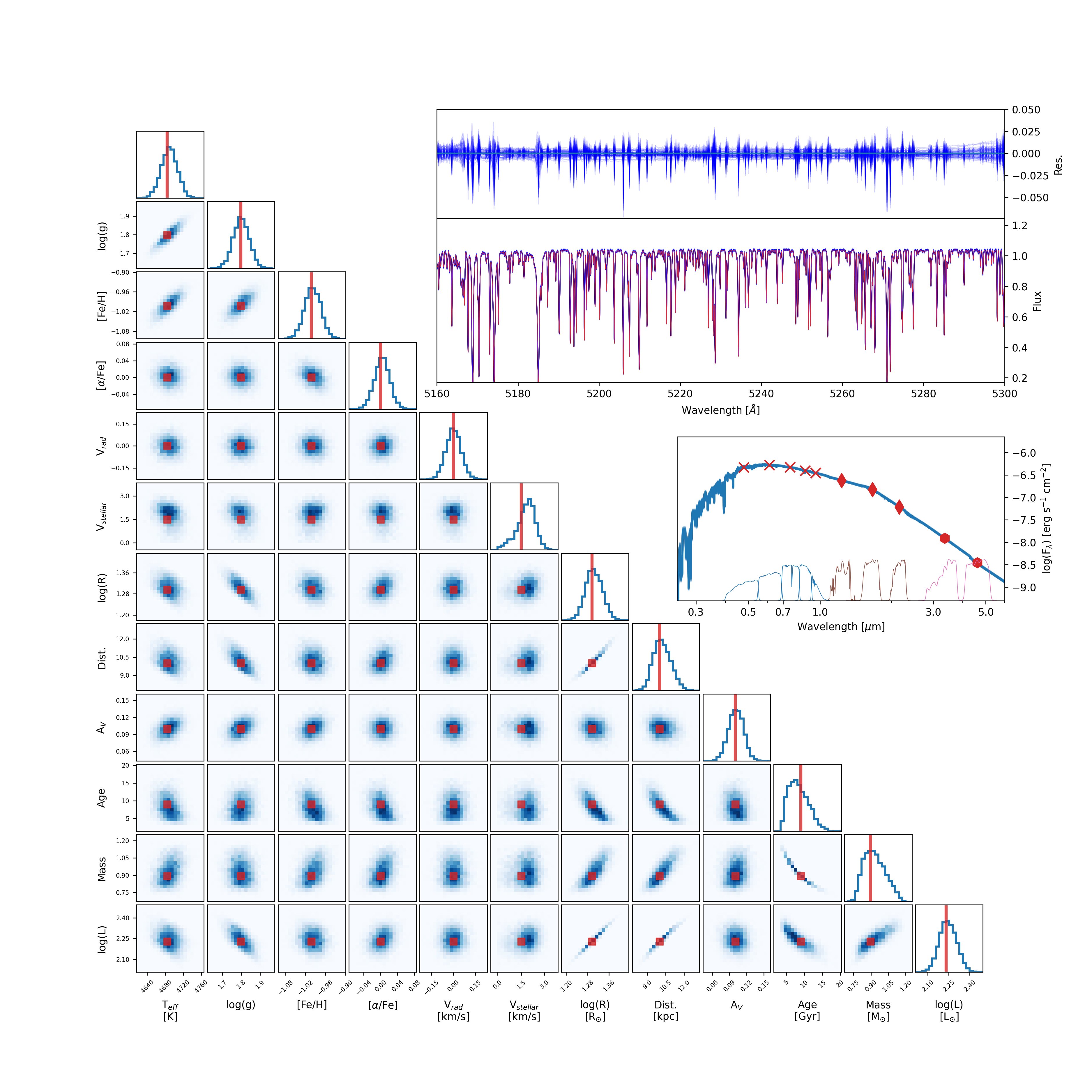}
    \caption{
    Example posterior distributions for a \MS\ fit to a solar metallicity  red giant branch star with spectrum SNR$_{\rm spec}=10$ pix$^{-1}$ and parallax SNR$_{\pi}=10$. For clarity, only a subset of free parameters are shown here. In each panel, the stellar parameters used to create the mock data are indicated by red squares or lines. Plotted on the right are the mock spectrum (top) and spectral energy distribution (bottom) along with 1000 models randomly sampled from the posteriors (blue lines). Above the mock spectrum, we also plot the relative residual for the 1000 posterior draws (blue lines).}
    \label{fig.examplecmockcorner}
\end{figure*}

\subsubsection{Likelihood}

The likelihood probability function within \MS\, is computed by combining the individual likelihoods of an observed spectrum and/or measured photometric magnitudes given our model. This procedure starts with a given set of EEP, \feh$_i$, \afe$_i$, and $M_i$.  Atmospheric parameters (\teff, \logg, \feh, \afe) are then predicted for the appropriate \texttt{MIST} model. Spectral and photometric models are computed for these parameters using our trained neural network. The synthetic spectrum is then convolved with a broadening function that is composed of two kernels: an instrument line profile characterized by a resolving power ($R$), and broadening from the star itself (\Vstar) which includes the stellar projected rotation velocity and large-scale macroturbulent velocity.  We then apply the a continuum normalization to the model, and calculate the likelihood for observed spectrum and photometry.

The likelihood for a spectrum with flux ($f_{o}$) over a wavelength range ($\lambda$) given a predicted flux ($f_{m}$) from our model is:

\begin{equation}
    \mathcal{L}_{\rm spec} = \prod_{\lambda}\frac{1}{\sqrt{2\pi}\sigma_{\lambda}}
    \exp{-\frac{(f_{\lambda,o}-f_{\lambda,m})^{2}}{2\sigma_{\lambda}^{2}}},
\end{equation}

\noindent
where $\sigma_{\lambda}$ is the measured flux error in the observed spectrum. In our procedure the model is always interpolated onto the observed wavelength grid.

Similarly, the likelihood of a set of observed photometry given predicted synthetic photometry is:

\begin{equation}
    \mathcal{L}_{phot} = \prod_{i}\frac{1}{\sqrt{2\pi}\sigma_{i}}
    \exp{-\frac{(m_{i,o}-m_{i,m})^{2}}{2\sigma_{i}^{2}}},
\end{equation}

\noindent
where the product is over all filters $i$, with observed and model magnitudes $m_{o}$ and $m_{m}$, respectively, and measured photometric errors $\sigma_{i}$. We could also have calculated this likelihood probability using observed fluxes instead of magnitudes. However, we choose to work in magnitudes as the dominant source of error for the majority of the bright targets we will fit in later sections are the photometric zeropoints which are approximately Gaussian in magnitude and not in flux.

Multiplying these two likelihood probabilities together gives us the total likelihood probability of observing a spectrum and photometry given a predicted model:

\begin{equation}
    \mathcal{L}_{total} = \mathcal{L}_{\rm spec} \times \mathcal{L}_{phot} \, .
\end{equation}

\noindent
The likelihood accounts for individual measurement errors in the data (assuming they are normally distributed), and can also incorporate systematic uncertainties in the data \citep[e.g., known systematic biases in the photometric calibration of surveys; see e.g.,][]{Portillo2019}. 

We also identify and handle outliers in the photometry by performing an initial $\chi^{2}$ optimization to the SED in order to find photometric data that are clearly inconsistent with this preliminary model. We classify the photometric data as an outlier if it is $>5\sigma$ away from this preliminary SED fit. Once identified, these outlier data are re-weighted by increasing the uncertainty such that they are 1$\sigma$ away from the preliminary best-fit SED model.  This procedure effectively identifies obviously erroneous photometry.  One could also consider more complex noise models for both the photometry and spectroscopy \citep[e.g.,][]{Hogg2010}.  This will be the subject of future work.

\begin{figure*}[t!]
    \centering
    \includegraphics[width=0.33\linewidth]{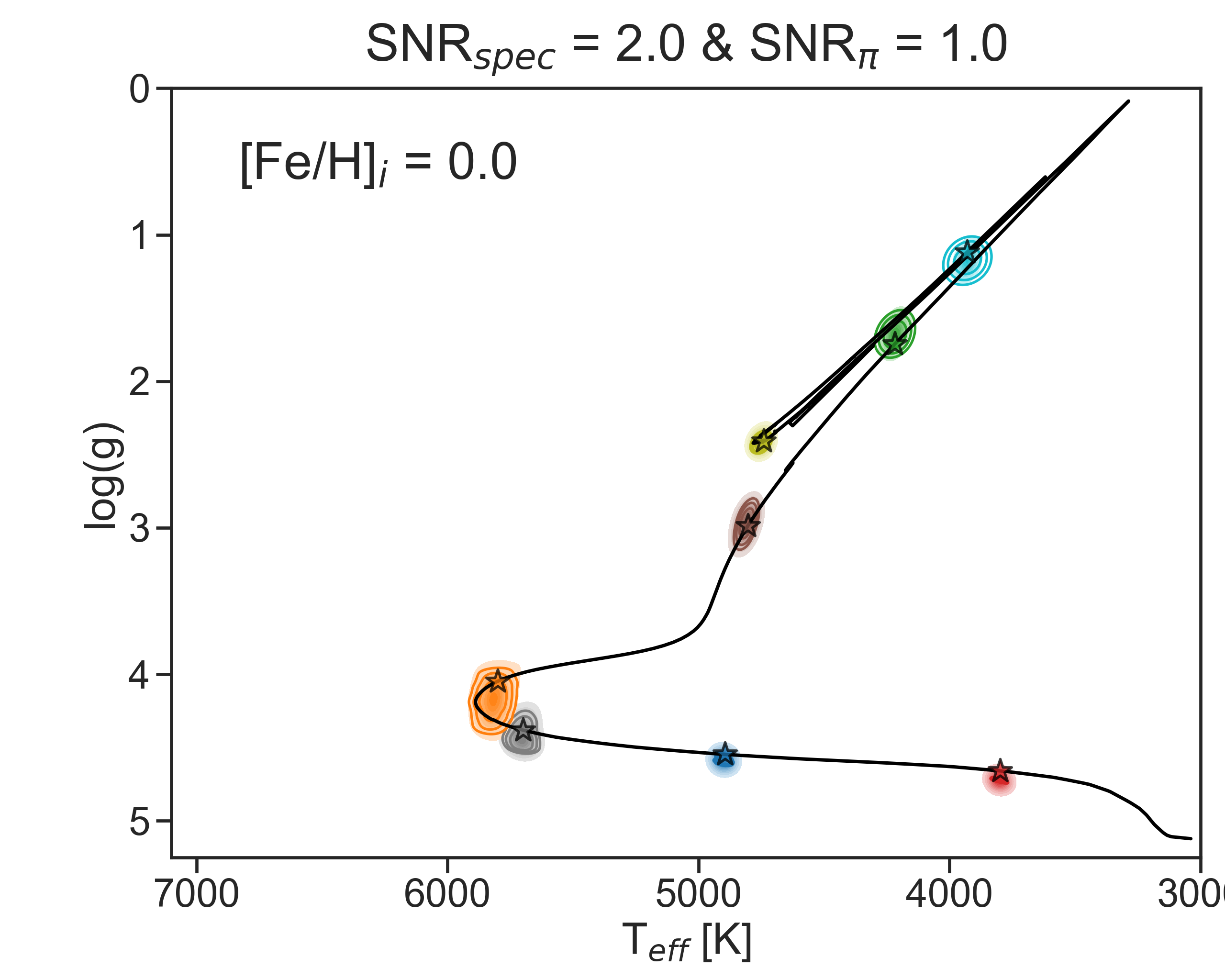}
    \includegraphics[width=0.33\linewidth]{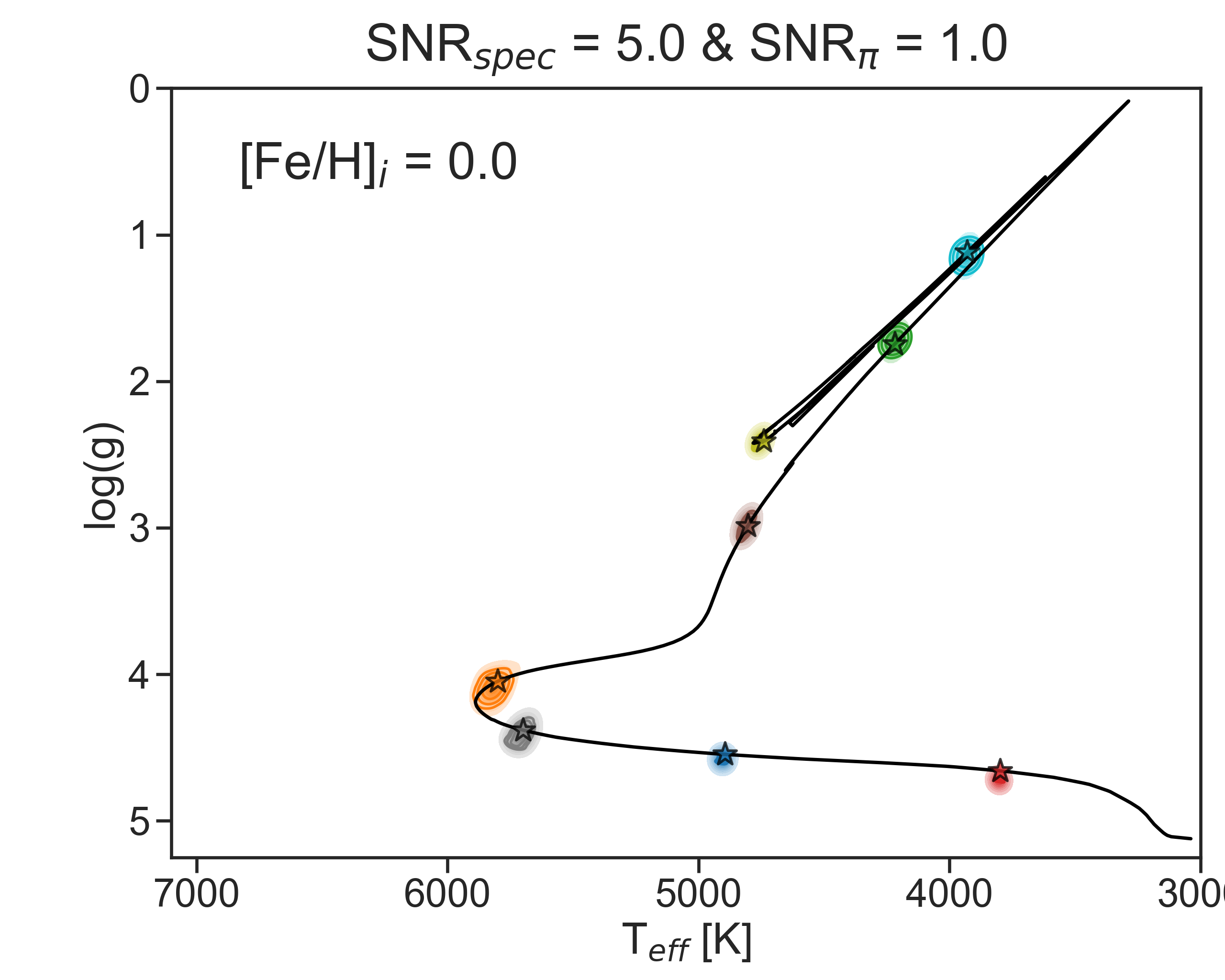}
    \includegraphics[width=0.33\linewidth]{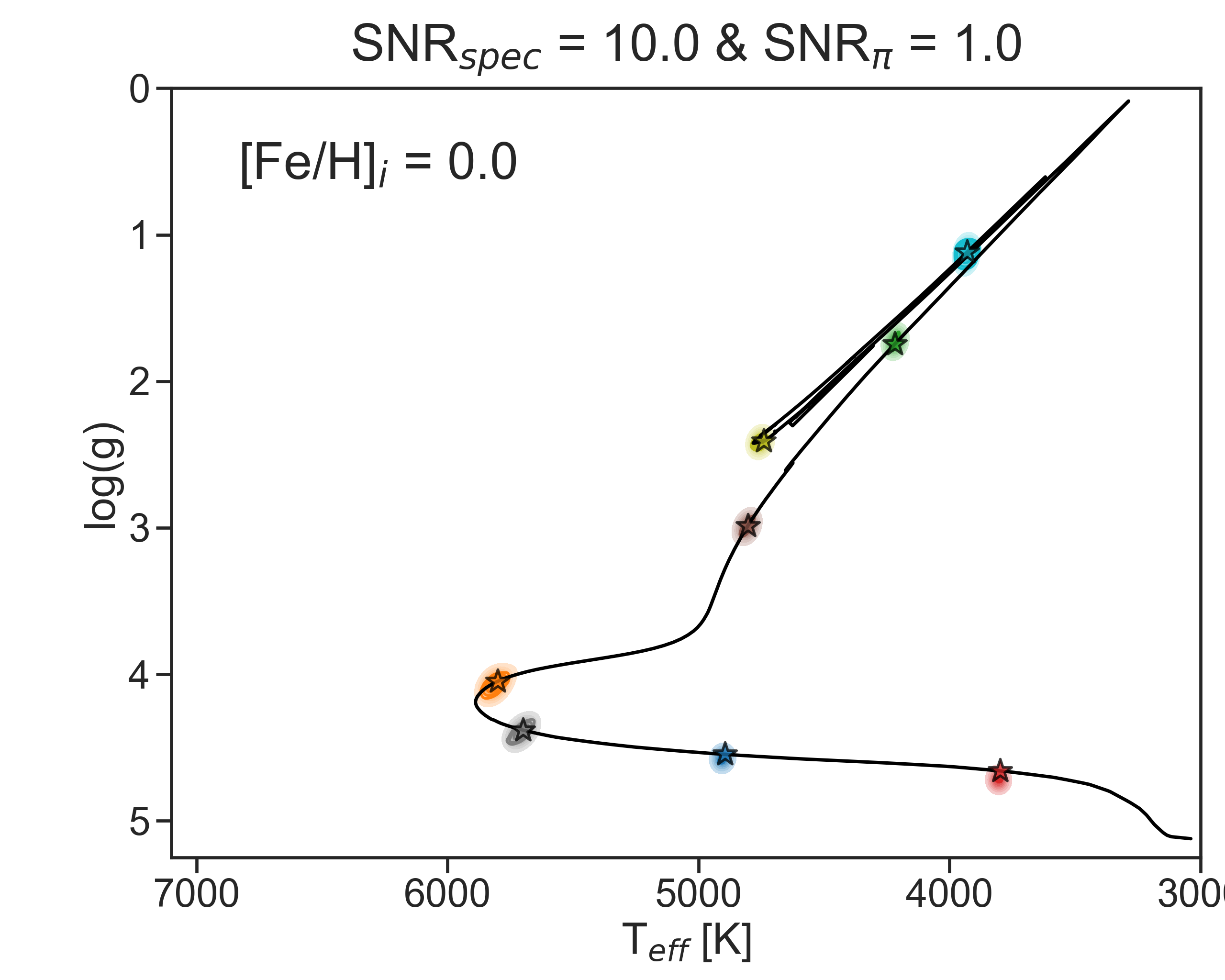}
    \includegraphics[width=0.33\linewidth]{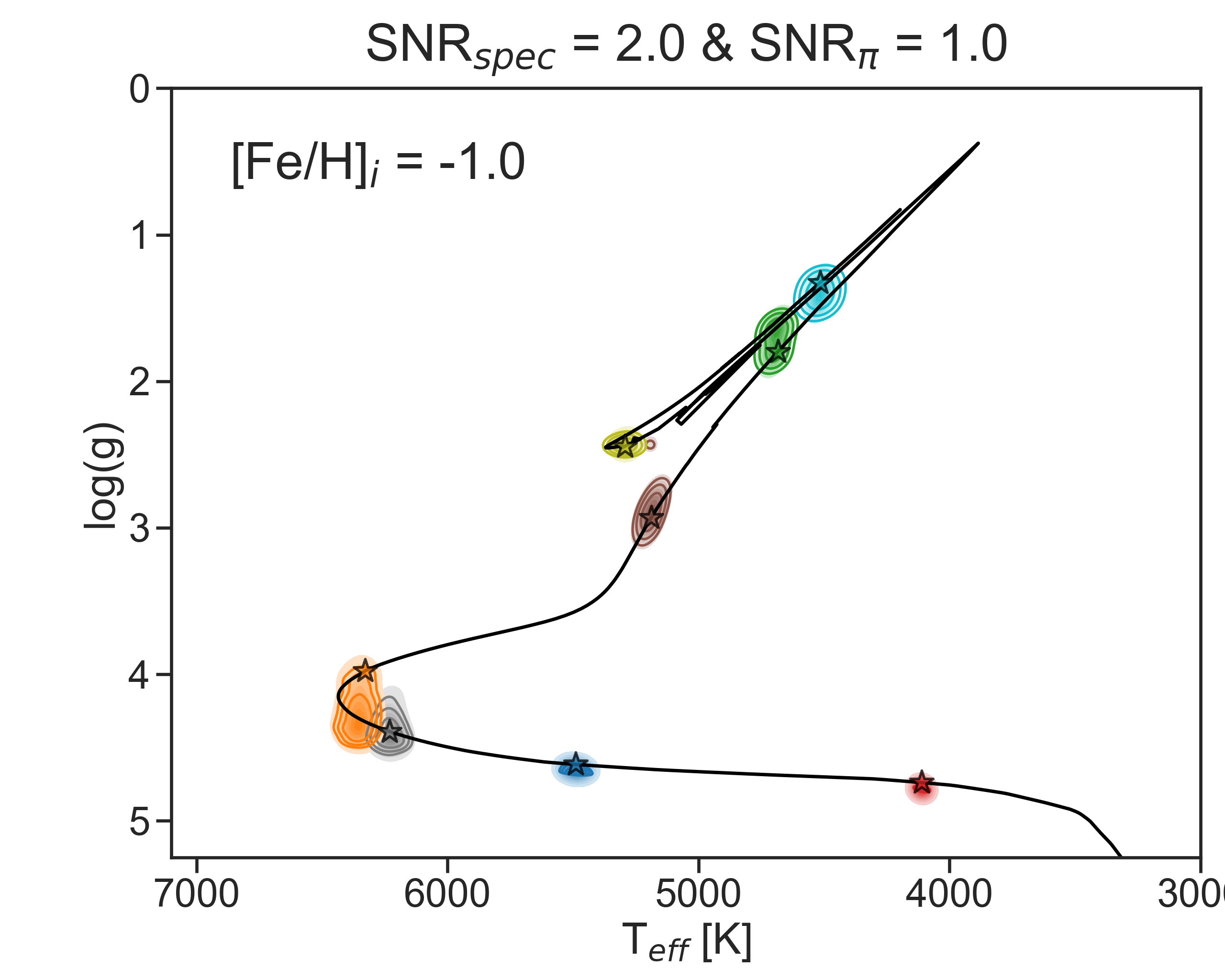}
    \includegraphics[width=0.33\linewidth]{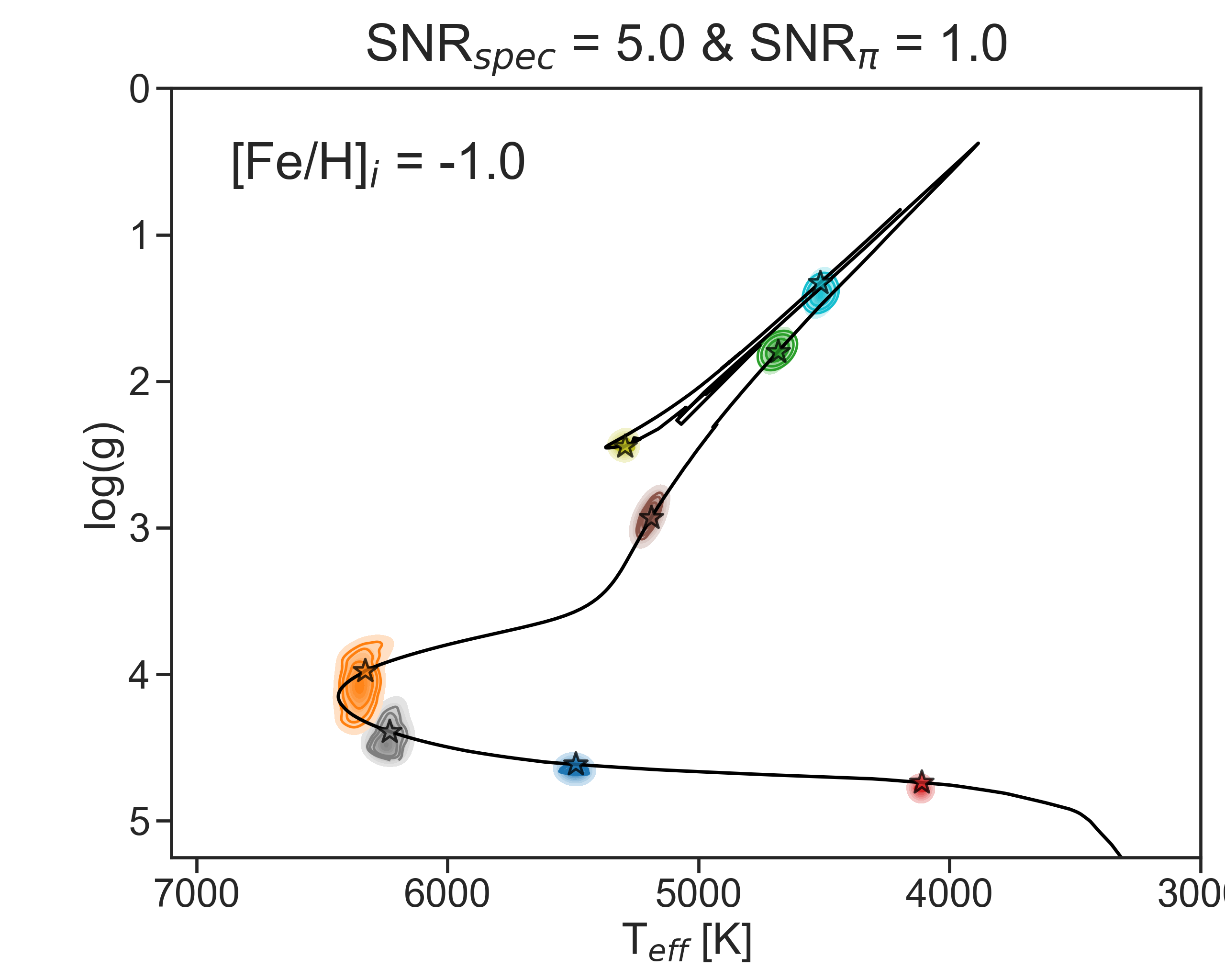}
    \includegraphics[width=0.33\linewidth]{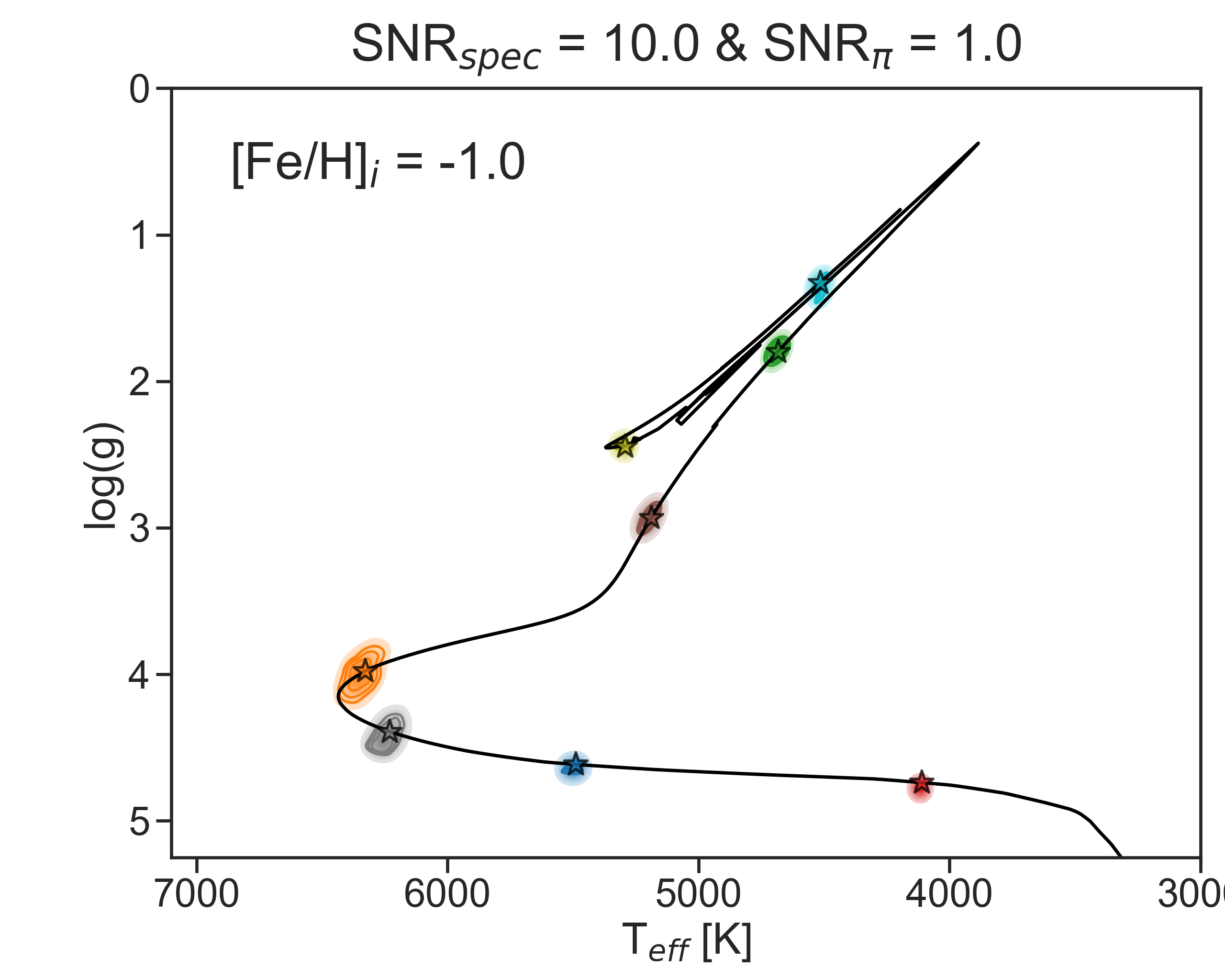}
    \includegraphics[width=0.33\linewidth]{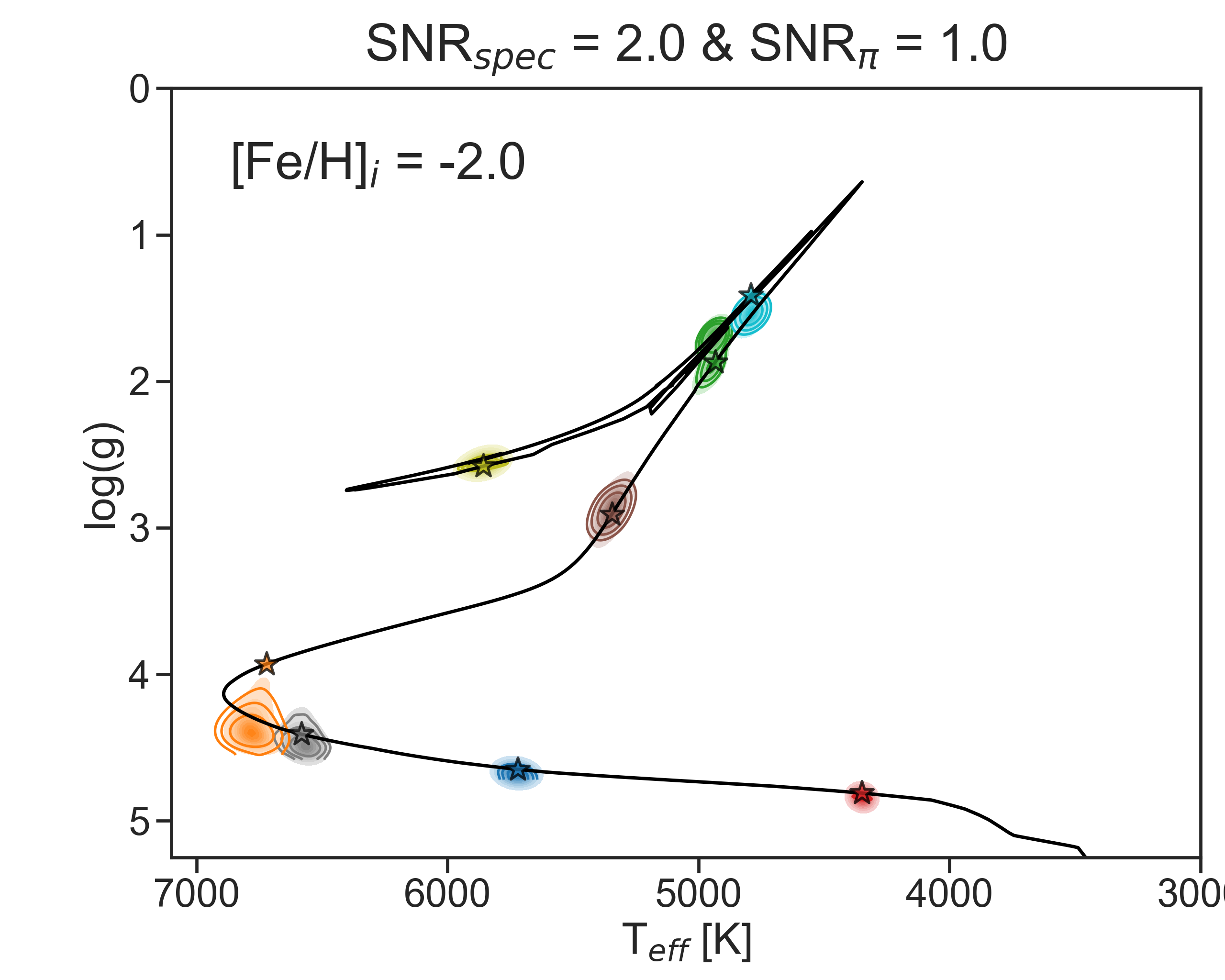}
    \includegraphics[width=0.33\linewidth]{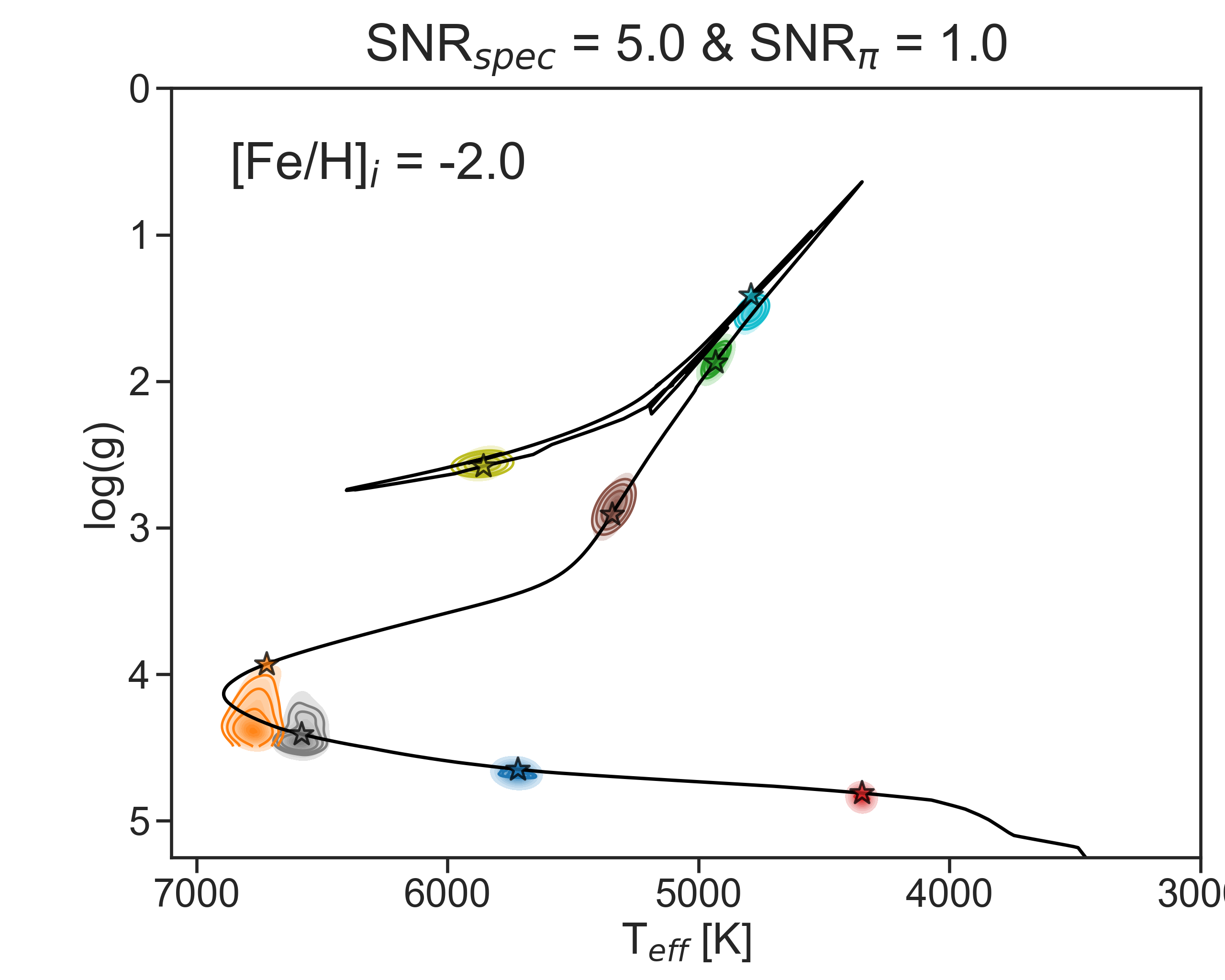}
    \includegraphics[width=0.33\linewidth]{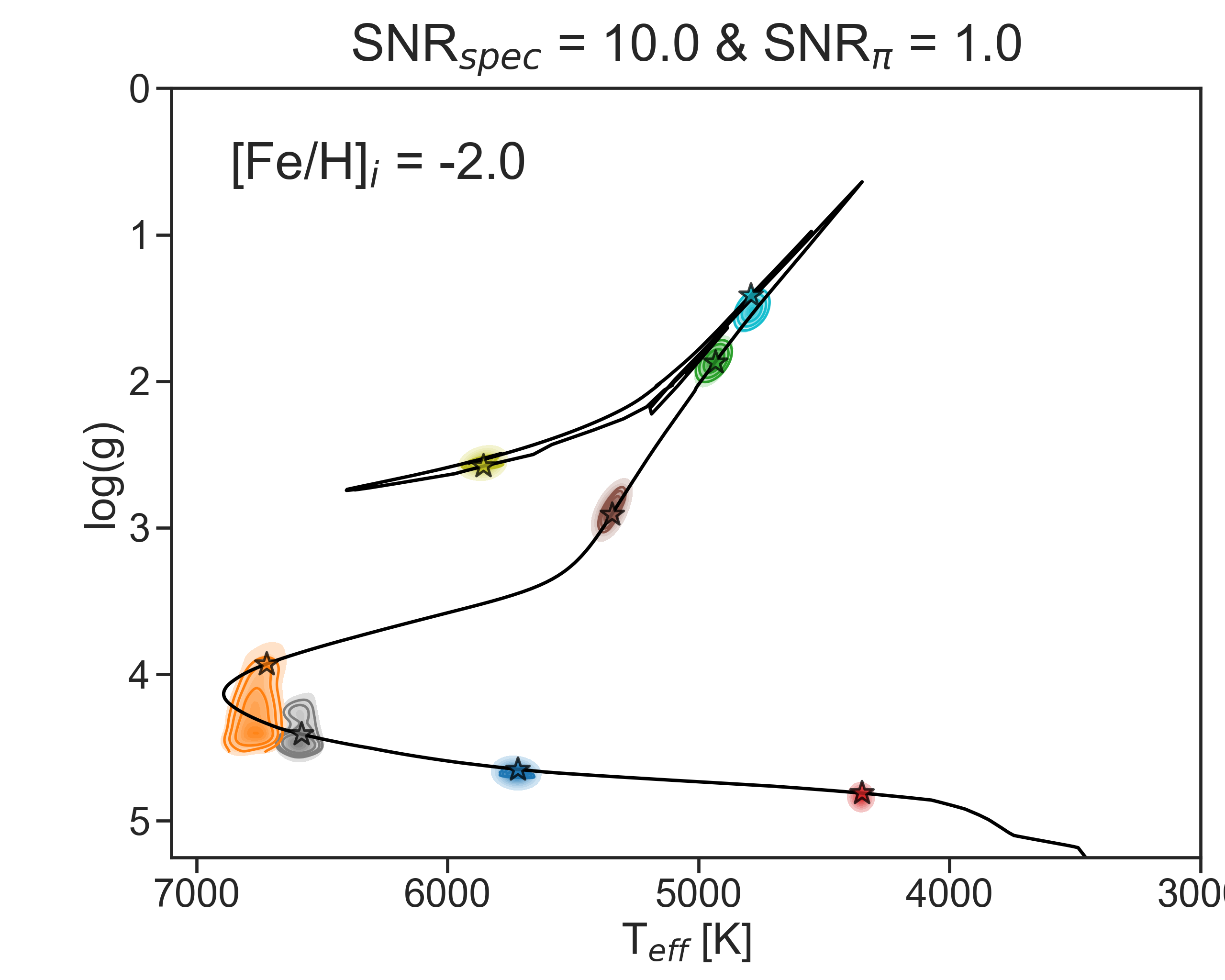}
    \caption{Kiel diagrams showing the posterior distributions resulting from modeling mock spectra and photometry with parallax SNR$_{\pi}=$1, for \feh$_i=0.0$ (top row), \feh$_i=-1.0$ (middle row), and \feh$_i=2.0$ (bottom row). Mocks were fit with photometry and a spectrum with spectral SNR$_{\rm spec}=2$ pix$^{-1}$ (left column), spectral SNR$_{\rm spec}=5$ pix$^{-1}$ (middle column), and SNR$_{\rm spec}=10$ pix$^{-1}$ (right column). Colors indicate the evolutionary state of the mock star ranging from lower main sequence through the turnoff, first ascent giant branch, horizontal branch, and asymptotic giant branch.  Contours indicate 1, 2, and 3$\sigma$ credible intervals.  True isochrone positions of these mock stars are given by the star symbols.}
    \label{fig.examplecmockkiel}
\end{figure*}

\subsubsection{Priors}

\MS\, allows for a wide range of prior probability distributions of spectroscopic and photometric parameters.  There are two types of parameters in \MS, those that are explicitly sampled, and extra parameters that are specified by the sampled parameters but not explicitly fit.  These can have priors assigned to them (e.g., a uniform age prior). In our setup the sampled parameters are $M$, EEP, \feh$_i$, \afe$_i$, distance, \Vrad, \Vstar, \Av, and four Chebyshev coefficients (T$_{n}$). Examples of extra parameters include the age and parallax.  These parameters and the associated priors are listed in Table \ref{tab.priors}. In addition, the \MS code also outputs posterior predictions for any other "derivative" stellar parameters predicted by the MIST models (stellar radius, bolometric luminosity, CNO abundance predictions, etc.)

For the stellar mass, we adopt a prior given by the \citet{Kroupa2001} initial mass function (IMF). The parameters \feh$_i$ and \afe$_i$ are uniformly sampled over the available grid range, namely $-4.0<$\feh$_i<+0.5$ and $-0.2<$\afe$_i<+0.6$. For the EEPs we adopt a uniform prior between the zero-age main sequence (EEP$=200$) and the end of the early-AGB phase (EEP$=808$).  Because the relation between EEP and age is complex and non-linear, drawing uniformly in EEP will imprint a complicated age-bias (generally in favor of short evolutionary phases).  We remove this bias by applying a prior ($\propto \frac{\mathrm d Age}{\mathrm d EEP}$) at any given stellar mass and metallicity that results in the probability of drawing any age equally likely. By removing this age-bias in this way, we then are able to apply more general age priors (e.g., uniform over the age of the Universe) without preferentially being biased towards these ages of rapid stellar evolution.

When fitting mock data and the star cluster data, we adopt a uniform prior on the distance from $0-50$ kpc. For the H3 data, we incorporate a more physical distance prior based on a simple Galactic number density model similar to other studies \citep[][]{Schonrich2014,Bailer-Jones2018,Anders2019}. Details of our specific model are provided in Speagle et al. (2020,submitted), but in short, we use a 3-D Galactic model that includes three components: thin-disk, thick-disk, and halo populations. Each of these components have specific shapes and relative fractional contributions based on \citet[][]{BlandHawthorn2016} for the thin and thick disks, and \citet[][]{Xue2015} for the Galactic halo. In a similar way, we place a prior on stellar age on the H3 data based on this Galactic density model, with the thin-disk having a uniform prior from 4 to 14 Gyr, a Gaussian prior for the thick disk centered at 10 Gyr with a sigma of 2 Gyr, and a Gaussian prior for the Halo centered at 12 Gyr with a sigma of 2 Gyr. For the mocks and cluster modeling, we use uniform priors given in Table \ref{tab.priors}. We use the measured {\it Gaia} DR2 parallaxes as a prior for the cluster and H3 data. We assume the {\it Gaia} parallax measurements are normally distributed about the true parallax \citep[][]{Luri18}.  

We place a strong prior on the radial velocity, \Vrad, by first performing a least-squares optimization to find a best-fit velocity offset between the observed spectrum and a model spectrum for a star with a \teff\ and \feh\, estimated from an initial photometry-only least-squares fit to the stellar SED. We then adopt a uniform prior that is centered at this initial velocity $\pm$25 km s$^{-1}$. 

The stellar broadening kernel is determined by two parameters, \Vstar, and the instrumental resolution with resolving power $R$.  For \Vstar\, we adopt a sigmoid prior function with an exponential probability decay at 7 and 10 \kms\ for giants and dwarfs, respectively.  We typically choose to fix the instrumental resolving power to a value determined through a fitting a line-spread-function to Th-Ar or solar-twilight calibration spectra. In this way, we reduce the dimensionality of the model by one free, highly-degenerate parameter.

The Chebyshev coefficients are given Gaussian priors with unit mean and widths of $1.0, 0.25, 0.1$, and $0.1$.  These were determined based on experimentation from fitting the various datasets described below.

And finally, the prior on the total extinction value \Av\, when we fit  the mocks is a Gaussian distribution centered at the input "truth" value of 0.1 mag. For the cluster data, we use a less informative, uniform prior of \Av $=0-1.5$ mag. For the H3 data we adopt a Gaussian prior with mean determined by the \citet{Schlegel1998} dust maps with a width of 15\% \citep[we have adopted the 14\% lower normalization of the dust maps provided by][]{Schlafly2011}. 

\subsubsection{Inference of Stellar Parameters}

We approximate the posterior distribution function in \MS\, using the dynamic nested sampling package \texttt{dynesty} \citep[][]{Speagle2020a}. Nested sampling is particularly well suited for fitting data in \MS, as it is more efficient in sampling complex, multi-modal, and/or highly covariant posterior distributions compared to other popular Markov Chain Monte Carlo samplers. Nested sampling also directly computes the Bayesian evidence, and has a well defined stopping criteria based on evidence estimation. In this study, we use a nested sampler with 750 live points and a stopping criteria of $\Delta \log{Z} = 0.01$ where $Z$ is the Bayesian evidence. We have tested these choices to insure that our results are not sensitive to their exact values. With these \texttt{dynesty} settings, for a single star we can sample a full posterior distribution inferred from a medium resolution spectrum with several thousand pixels and $5-10$ bands of photometry in $3-4$ CPU hours.

\begin{figure*}[t!]
\vspace{0.2cm}
    \includegraphics[width=0.33\linewidth]{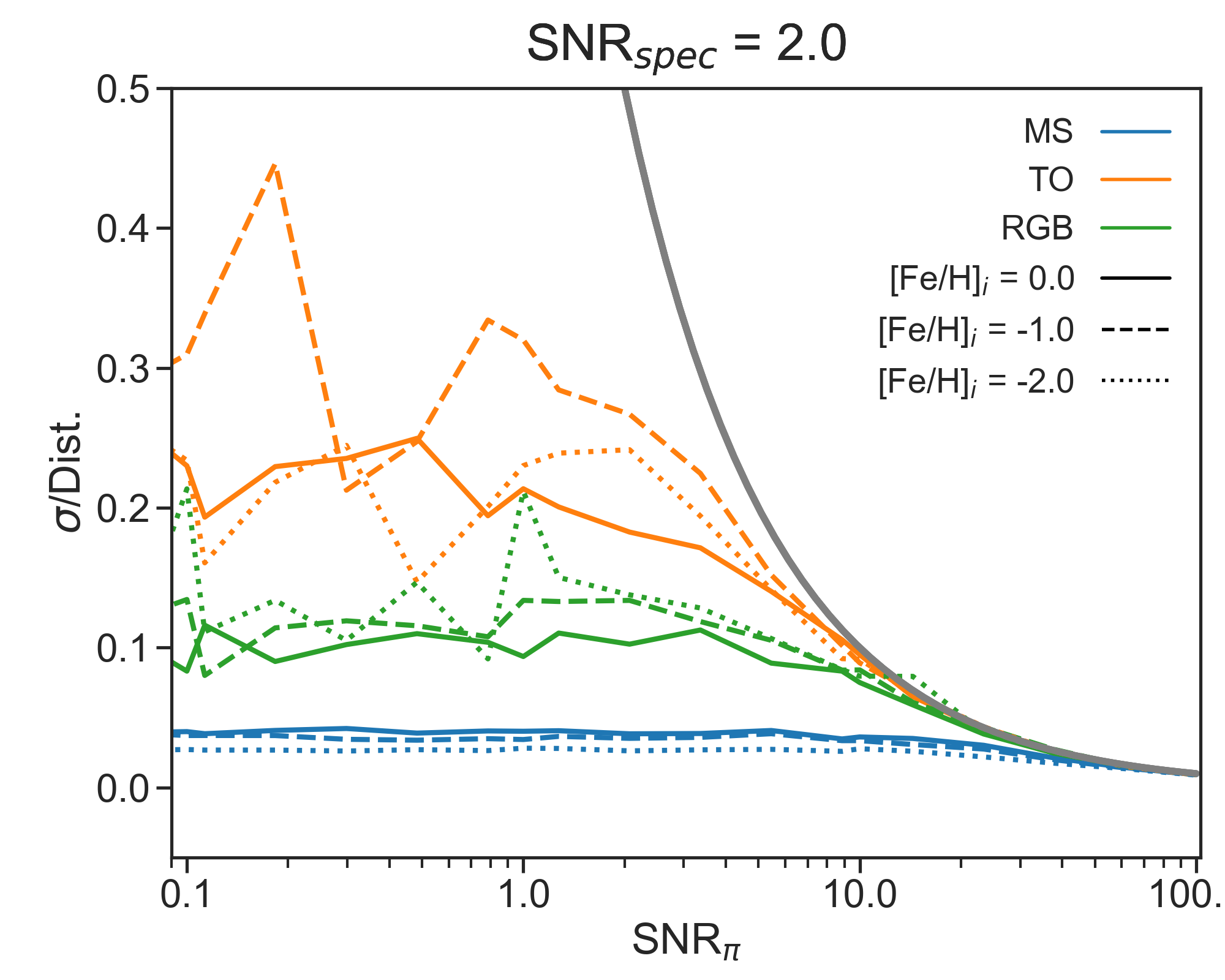}
    \includegraphics[width=0.33\linewidth]{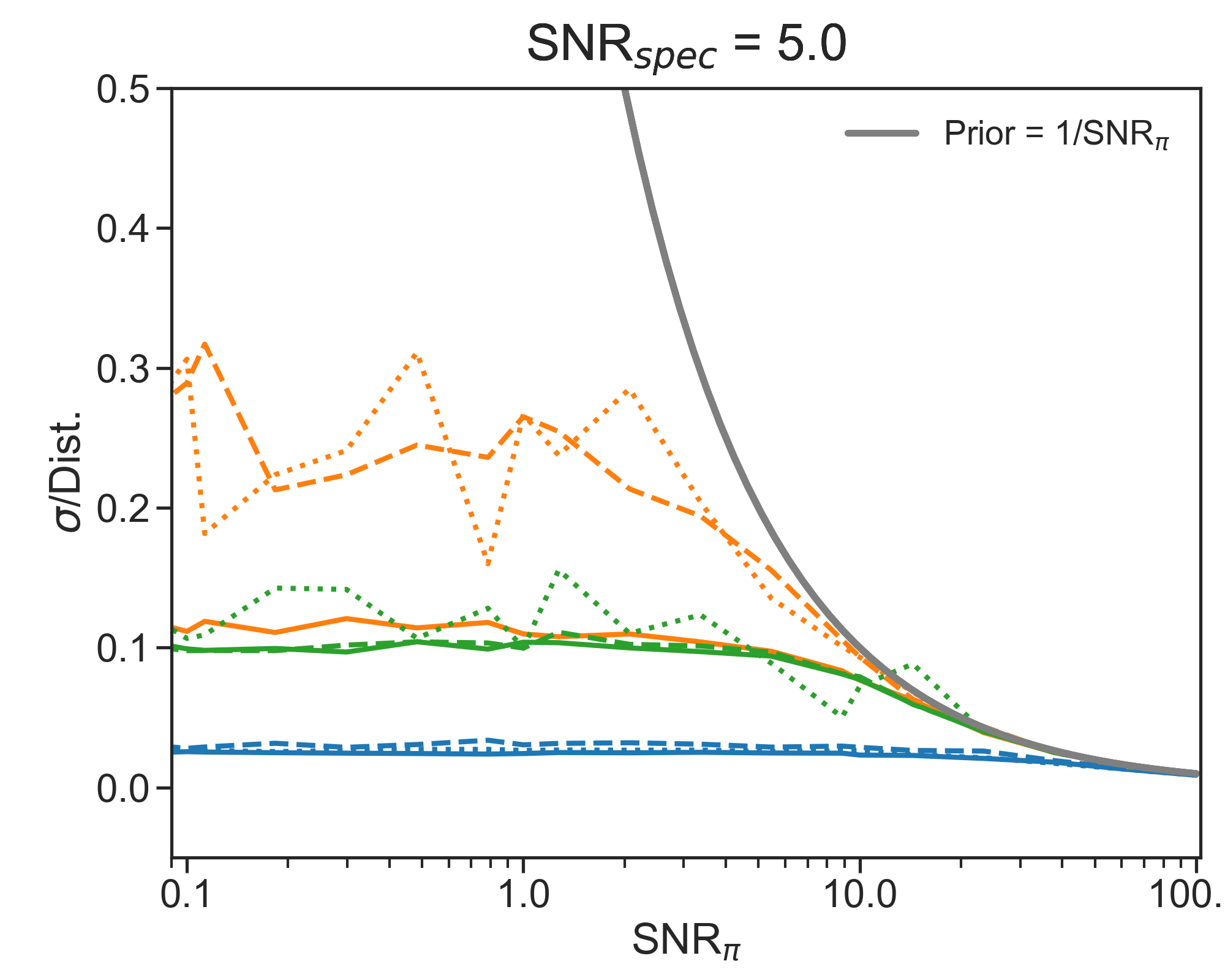}
    \includegraphics[width=0.33\linewidth]{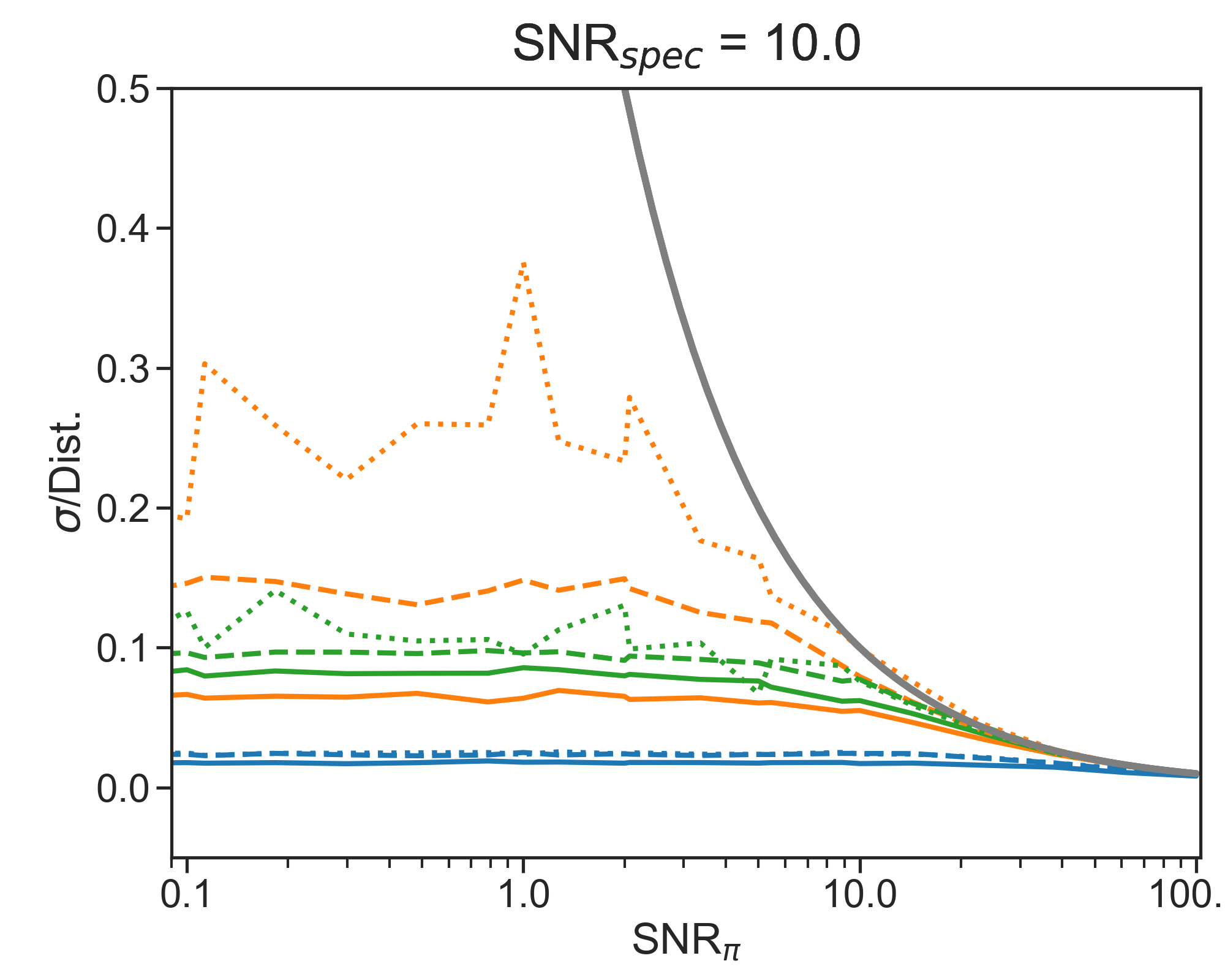}
    \includegraphics[width=0.33\linewidth]{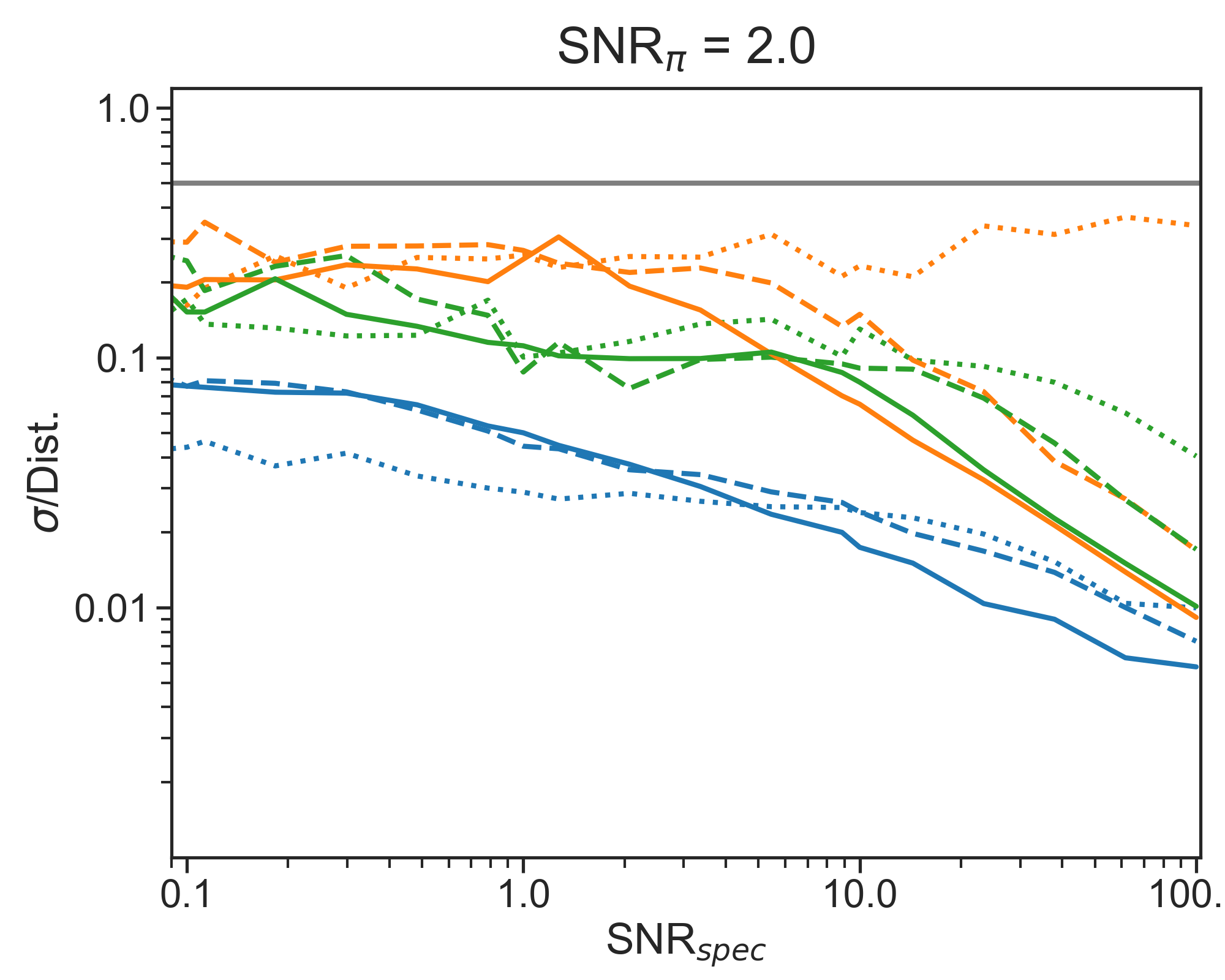}
    \includegraphics[width=0.33\linewidth]{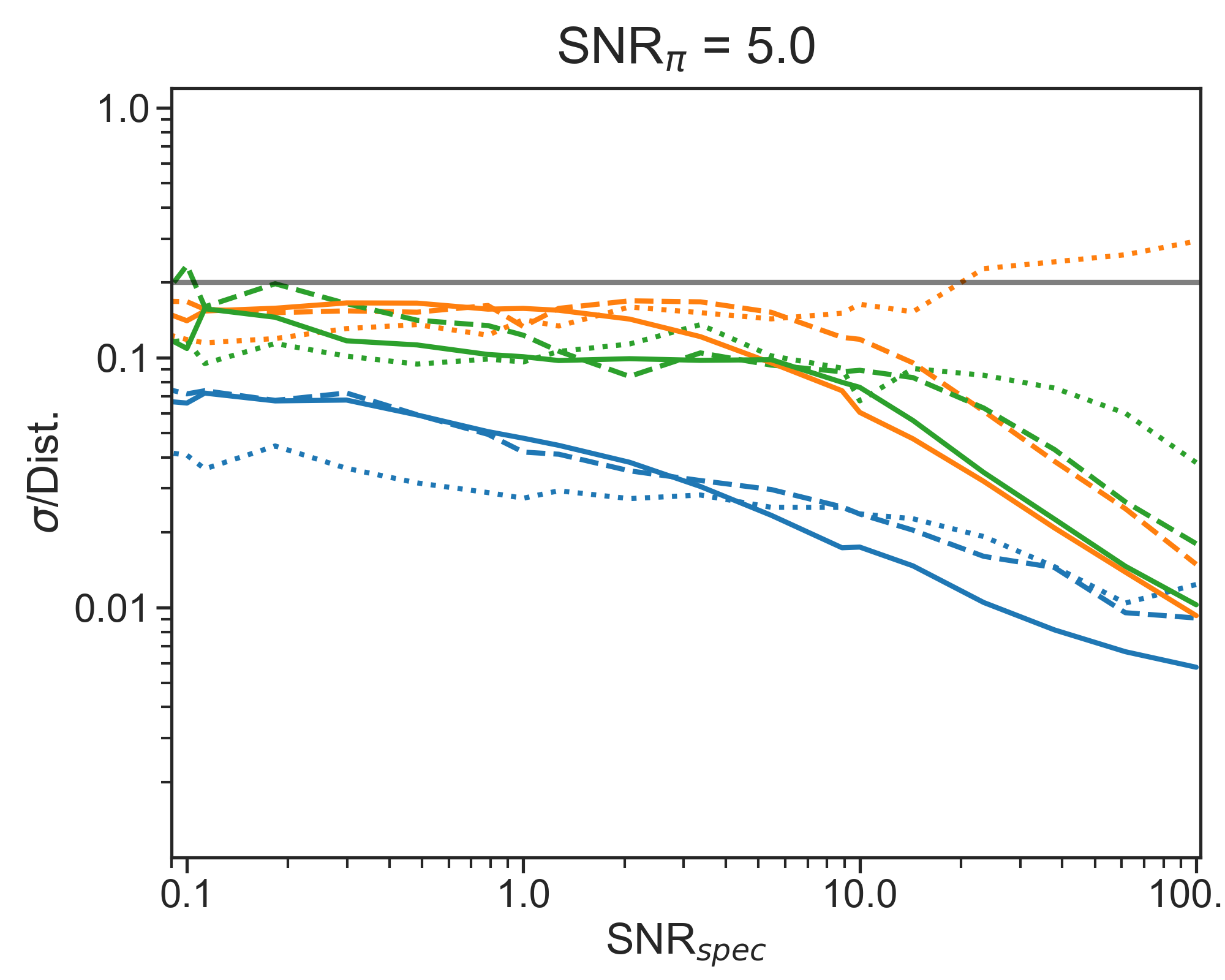}
    \includegraphics[width=0.33\linewidth]{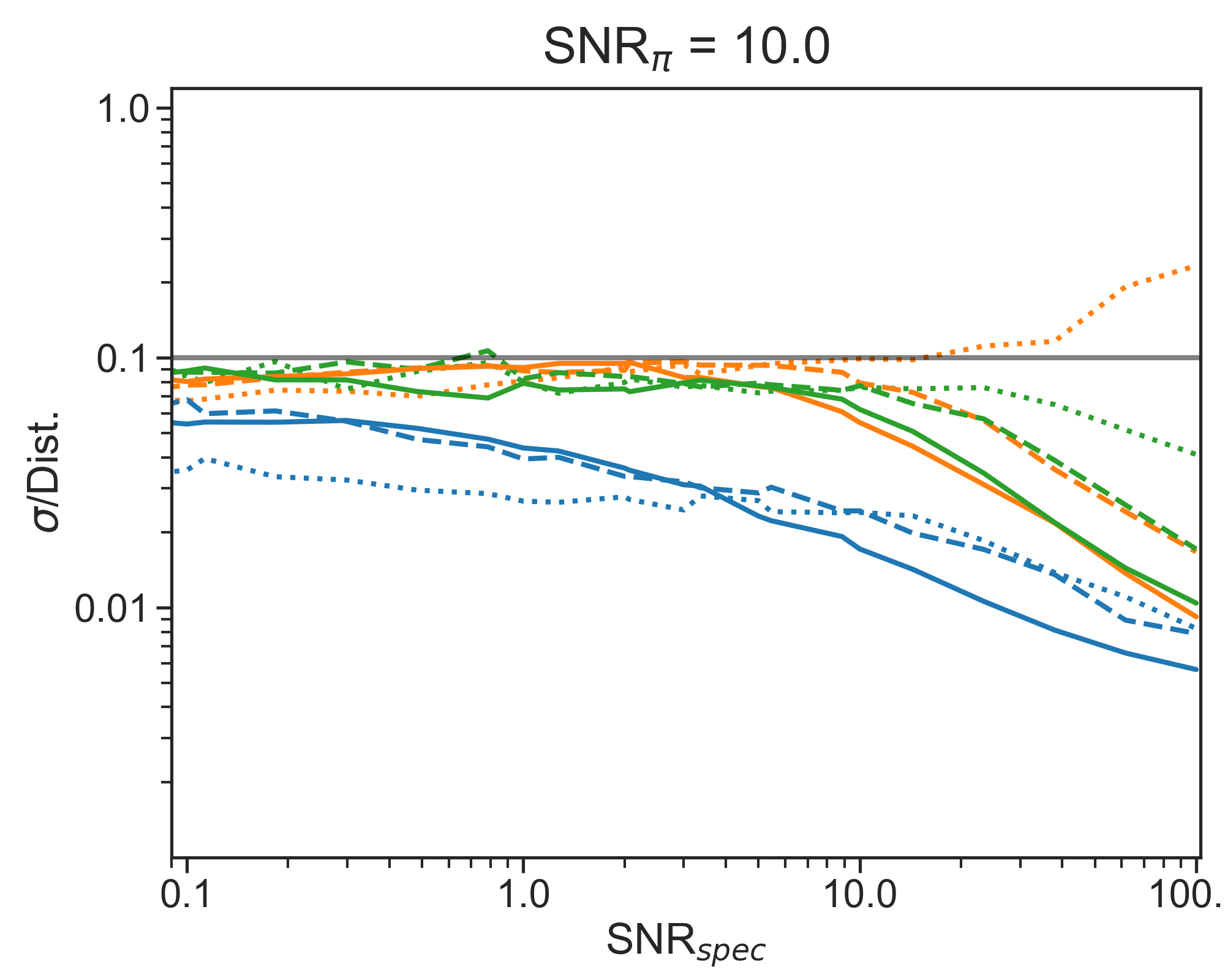}
    \vspace{0.3cm}
    \caption{Fractional distance uncertainty as a function of parallax SNR (top panels) and spectrum SNR (bottom panels) for three evolutionary phases: main-sequence (MS), turn-off (TO), and red giant branch (RGB), each at three metallicities: \feh$_i=-2.0, -1.0, 0.0$.  The parallax SNR prior is shown as a grey line in each panel.  Distances and uncertainties are determined for each mock based on the median and $1\sigma$ credible interval, respectively.}
    \label{fig.reldisterr}
\end{figure*} 

\begin{figure*}[hbt!]
    \centering
    \vspace{0.3cm}
    \includegraphics[width=0.24\linewidth]{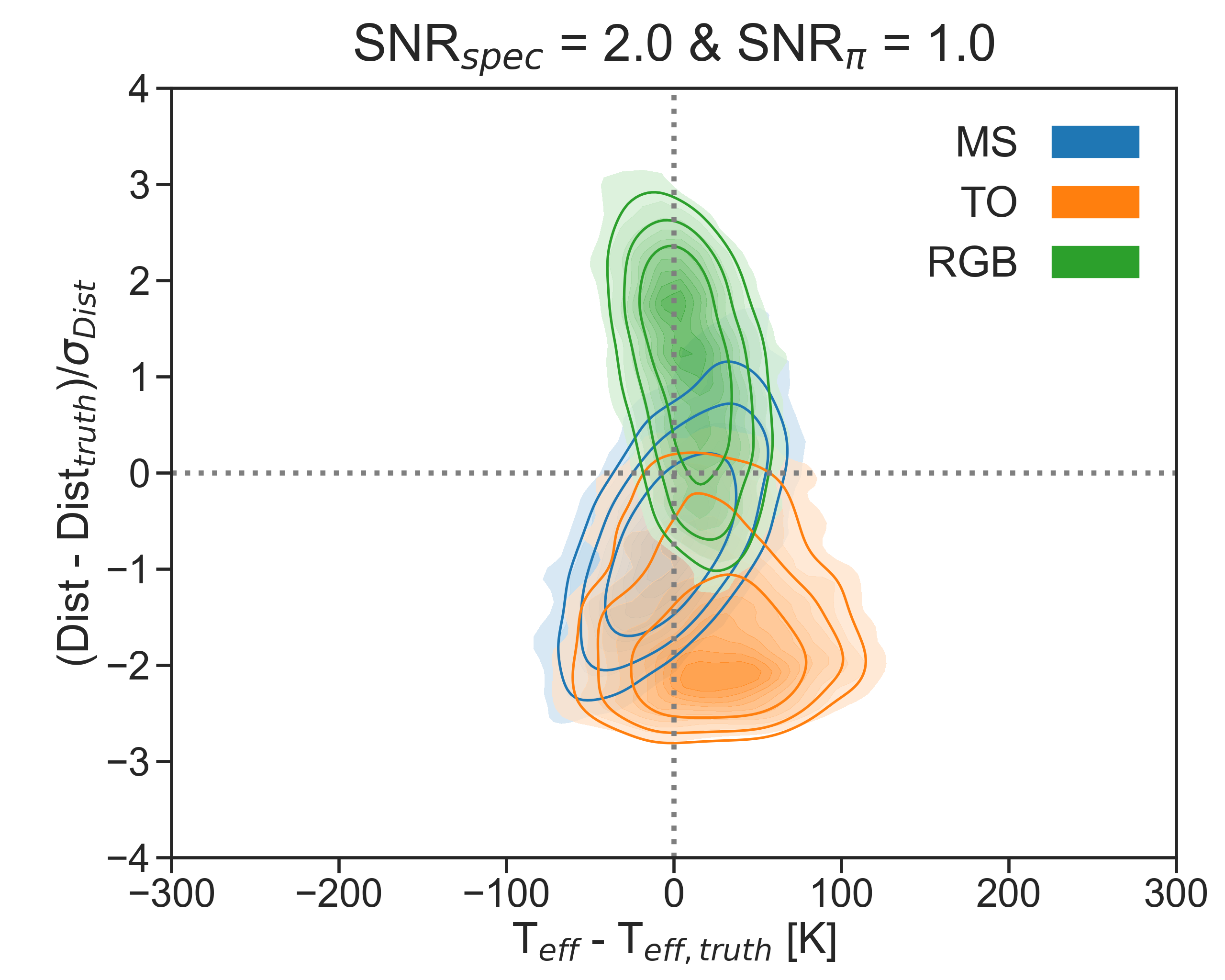}
    \includegraphics[width=0.24\linewidth]{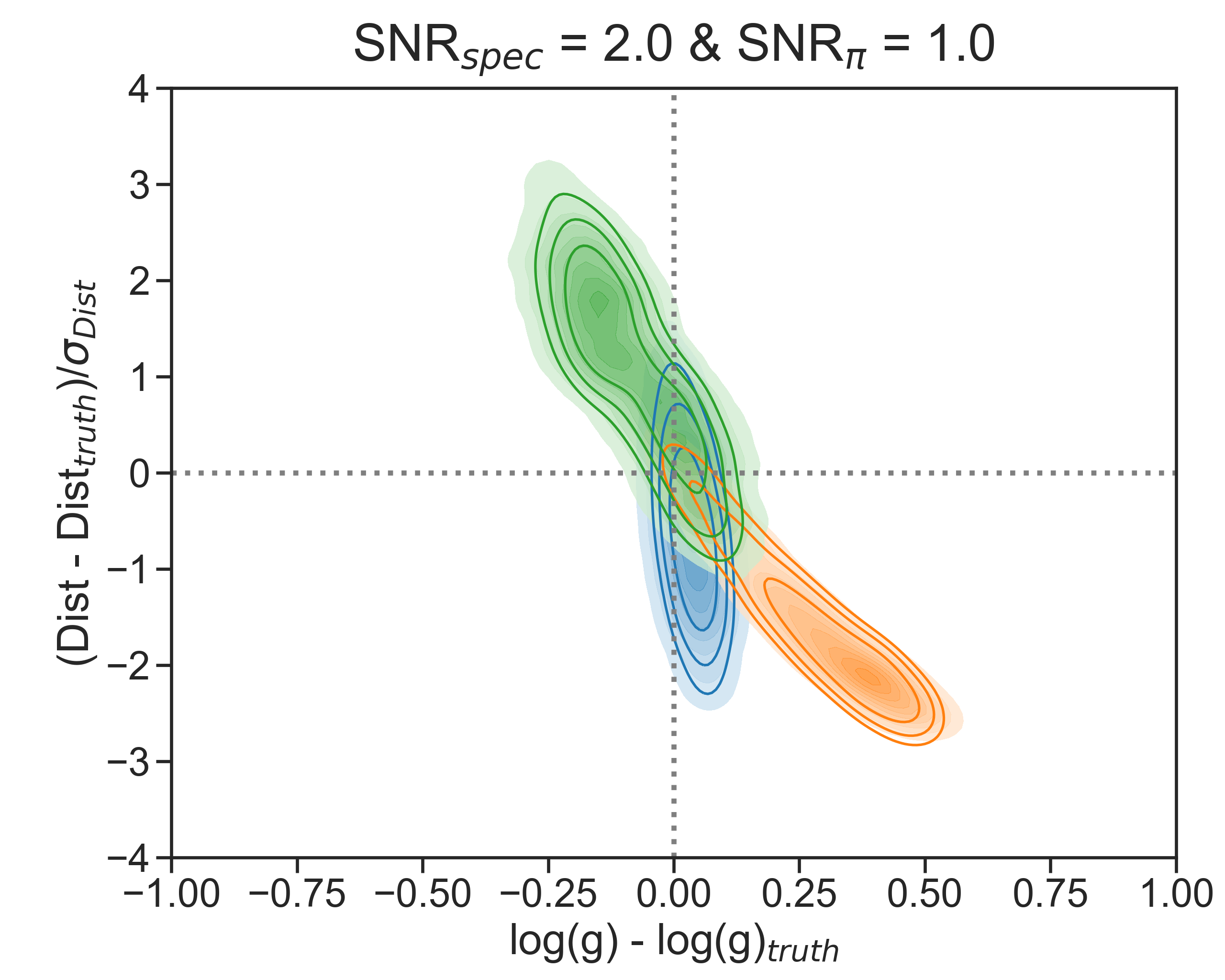}
    \includegraphics[width=0.24\linewidth]{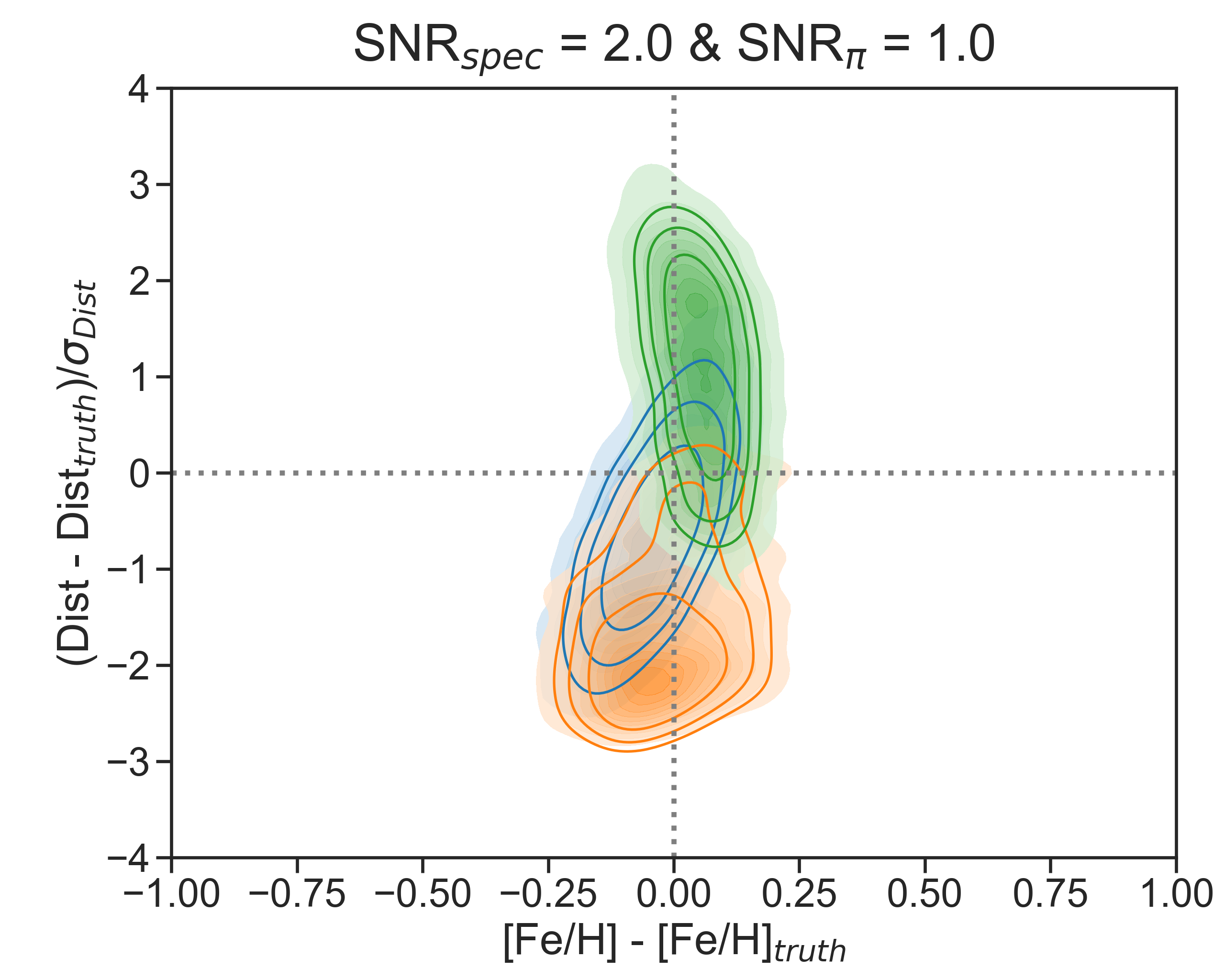}
    \includegraphics[width=0.24\linewidth]{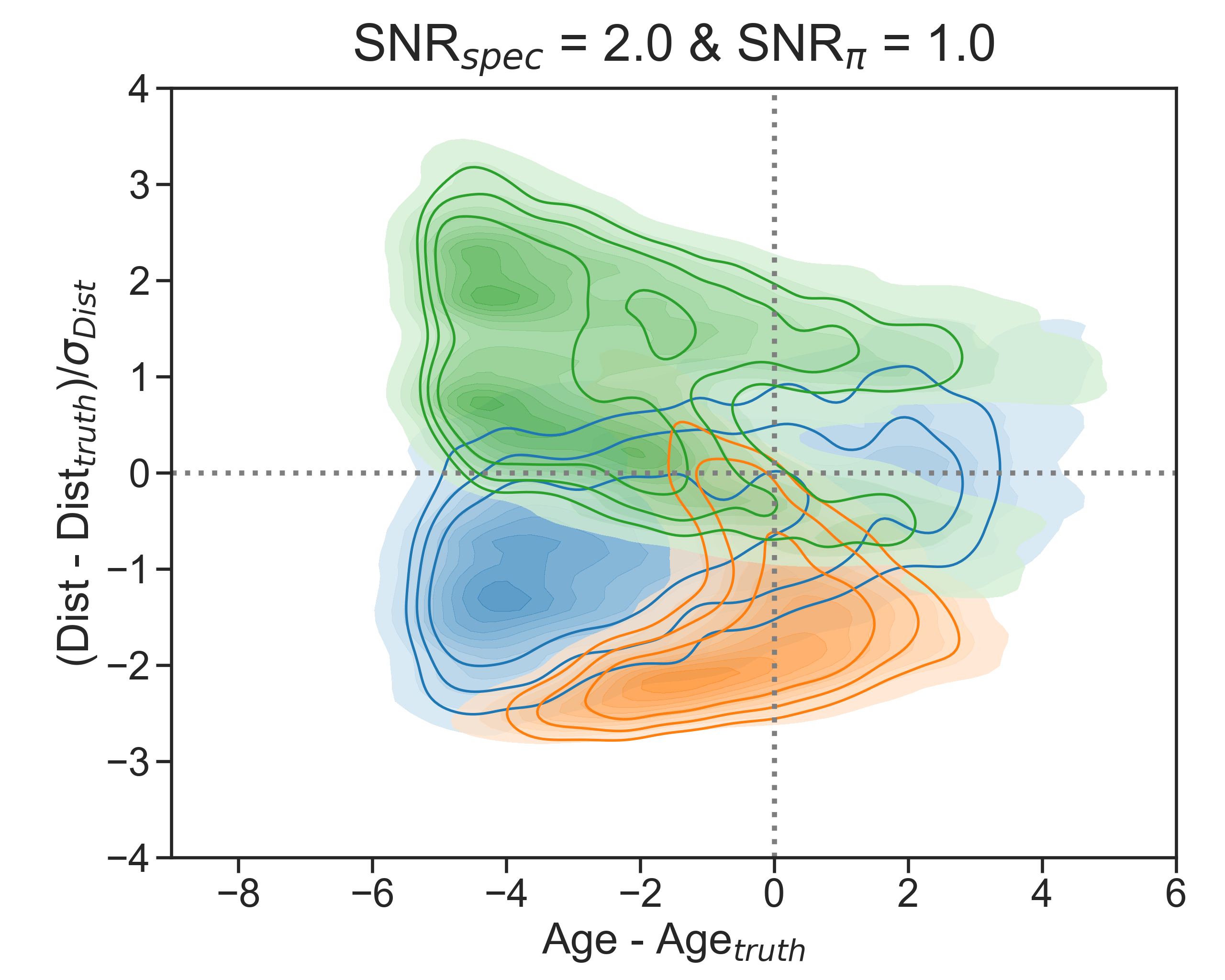}
    \includegraphics[width=0.24\linewidth]{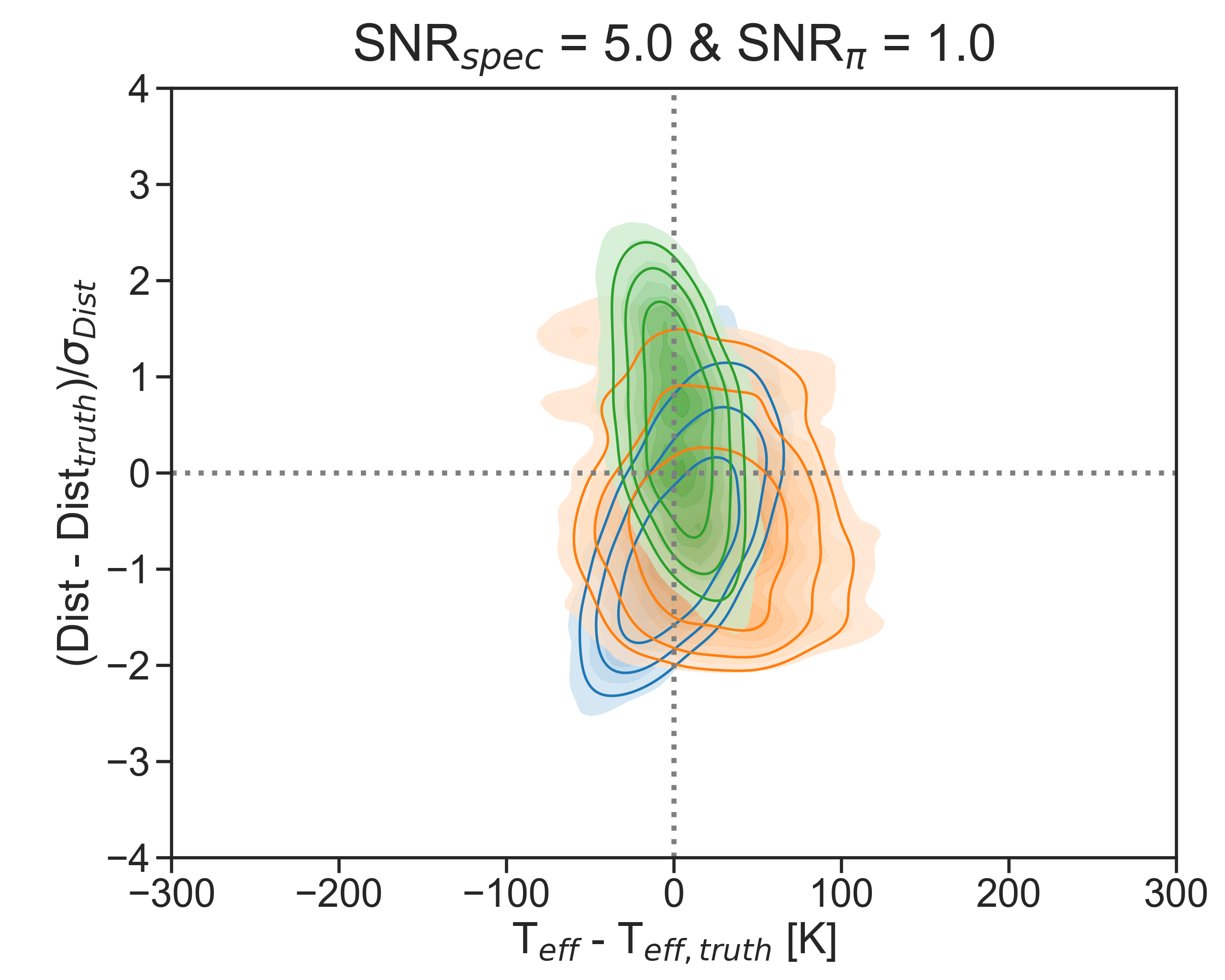}
    \includegraphics[width=0.24\linewidth]{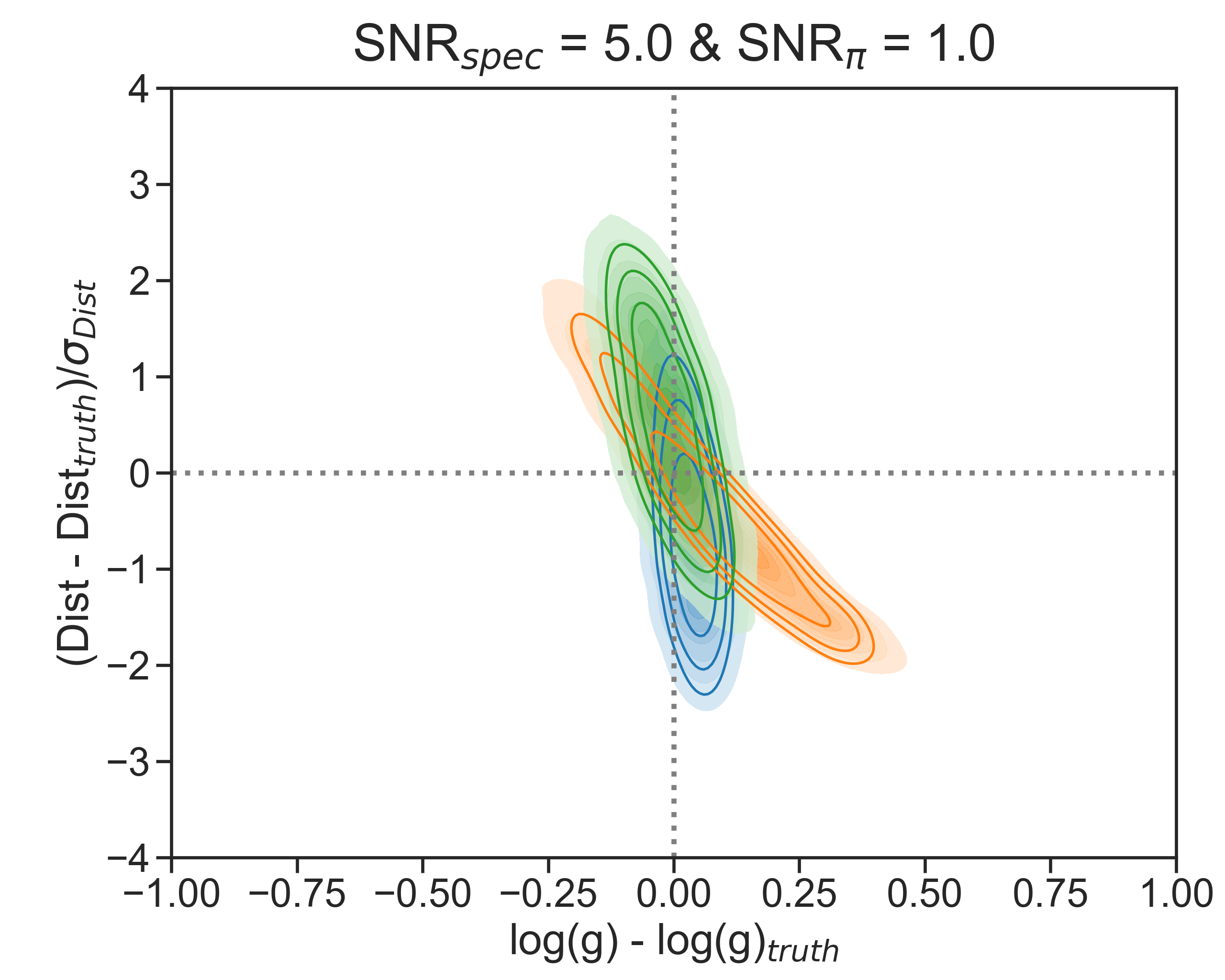}
    \includegraphics[width=0.24\linewidth]{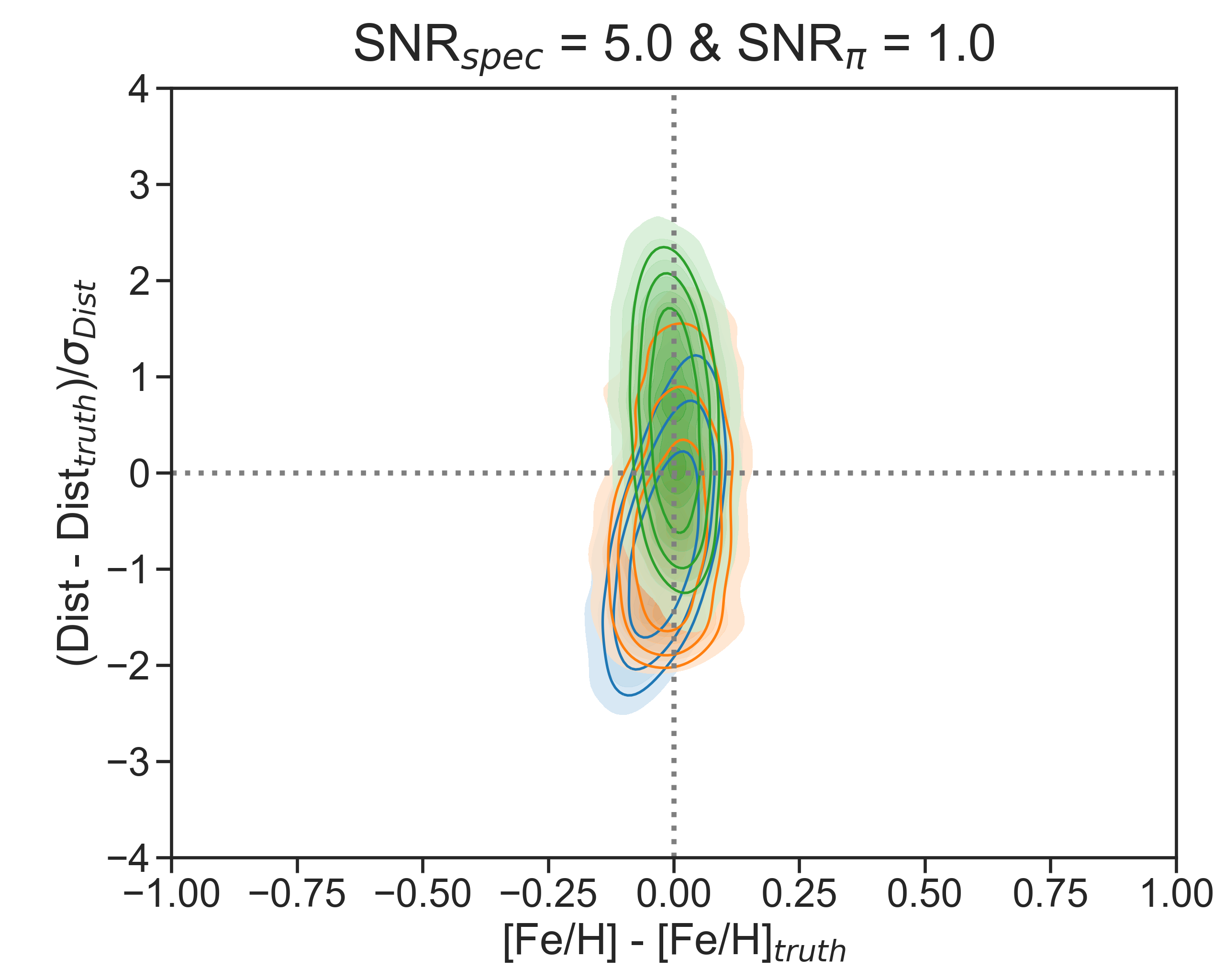}
    \includegraphics[width=0.24\linewidth]{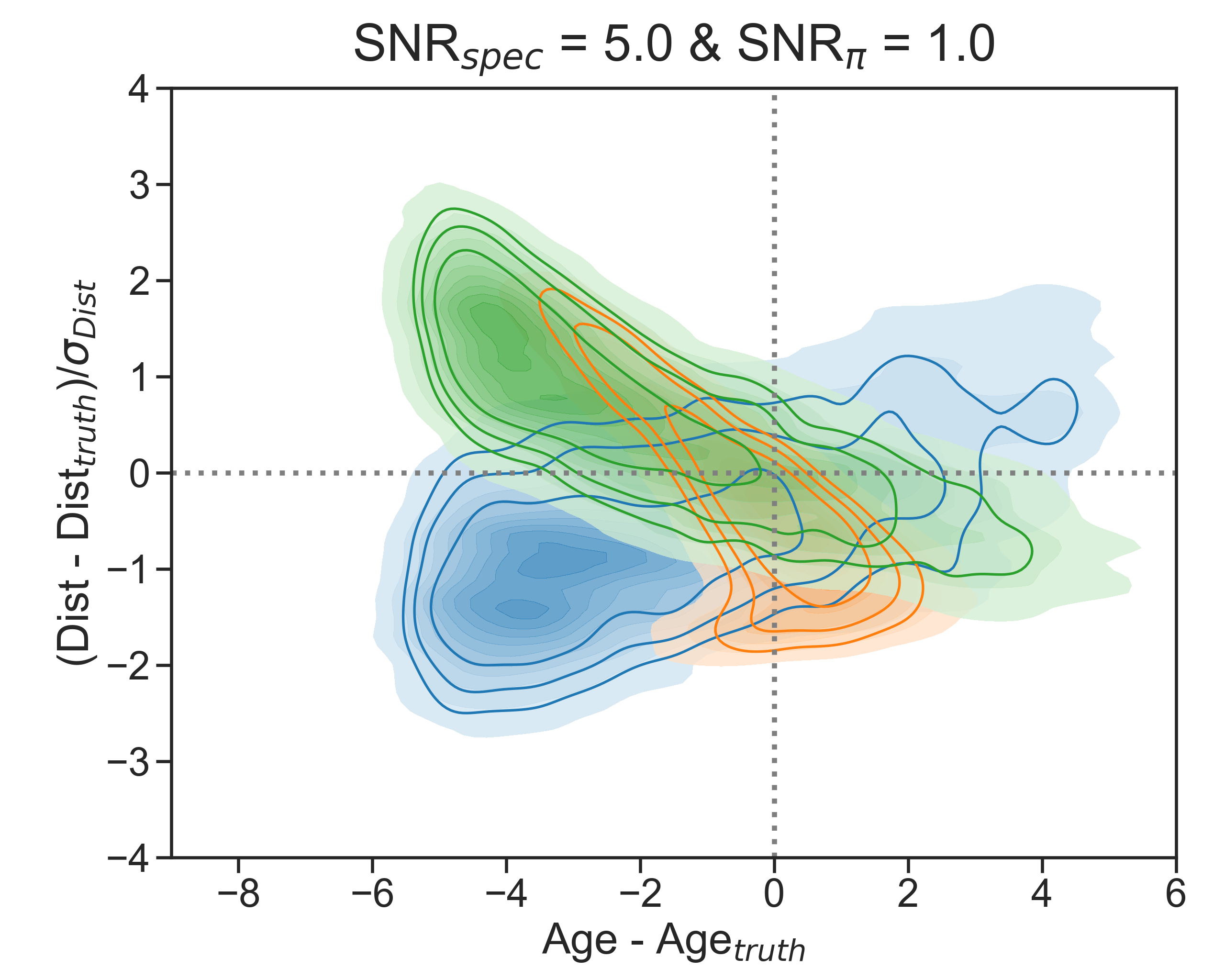}
    \includegraphics[width=0.24\linewidth]{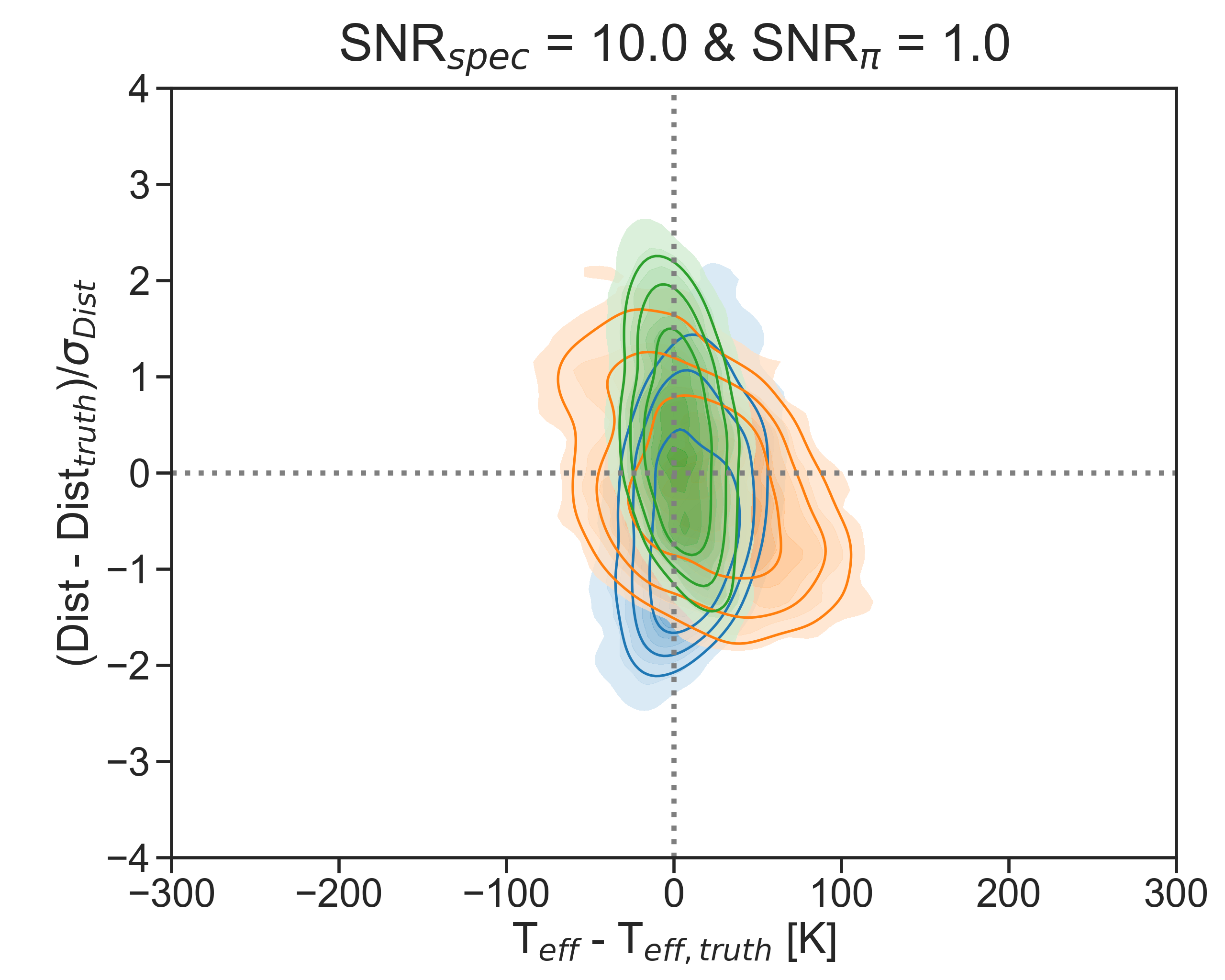}
    \includegraphics[width=0.24\linewidth]{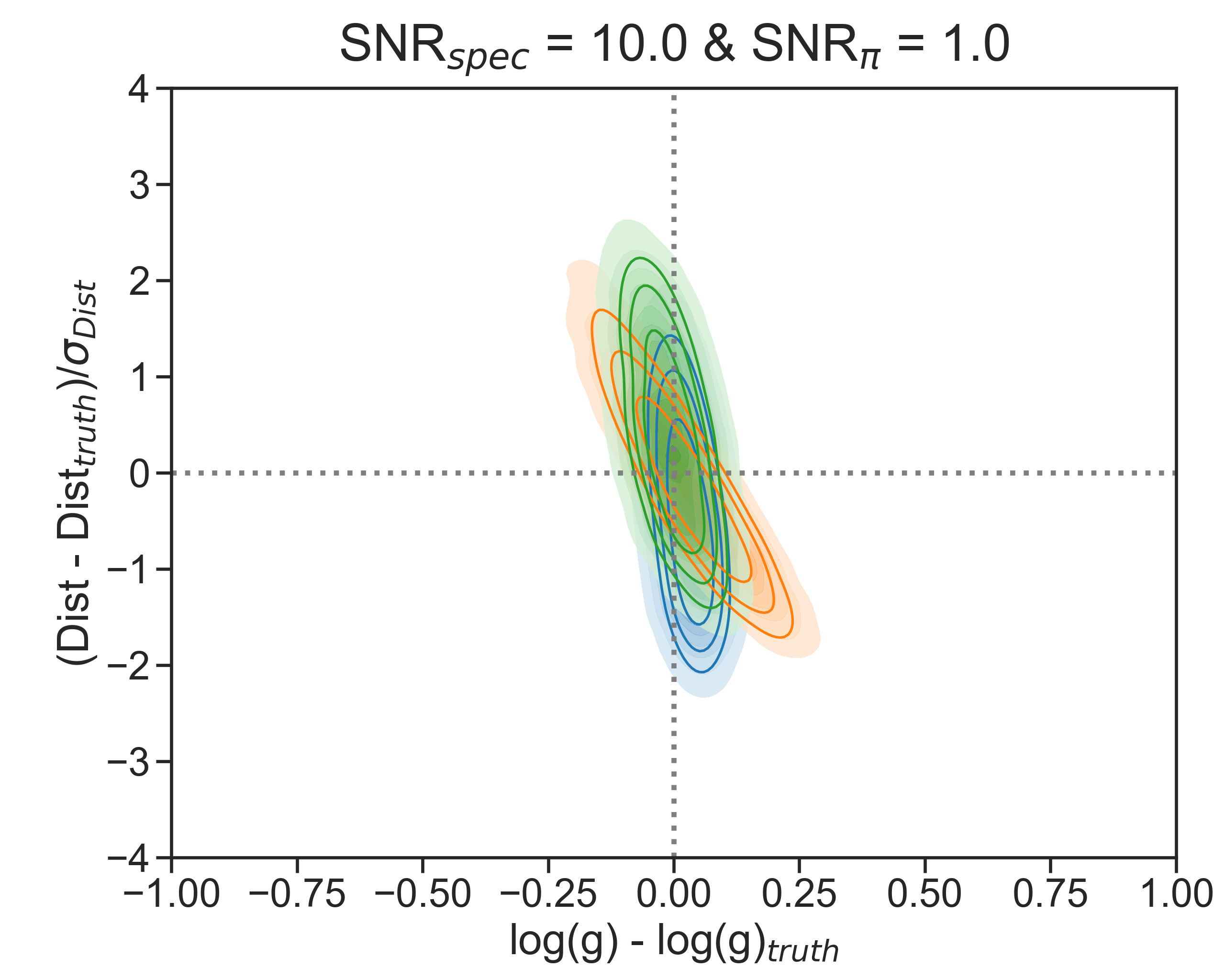}
    \includegraphics[width=0.24\linewidth]{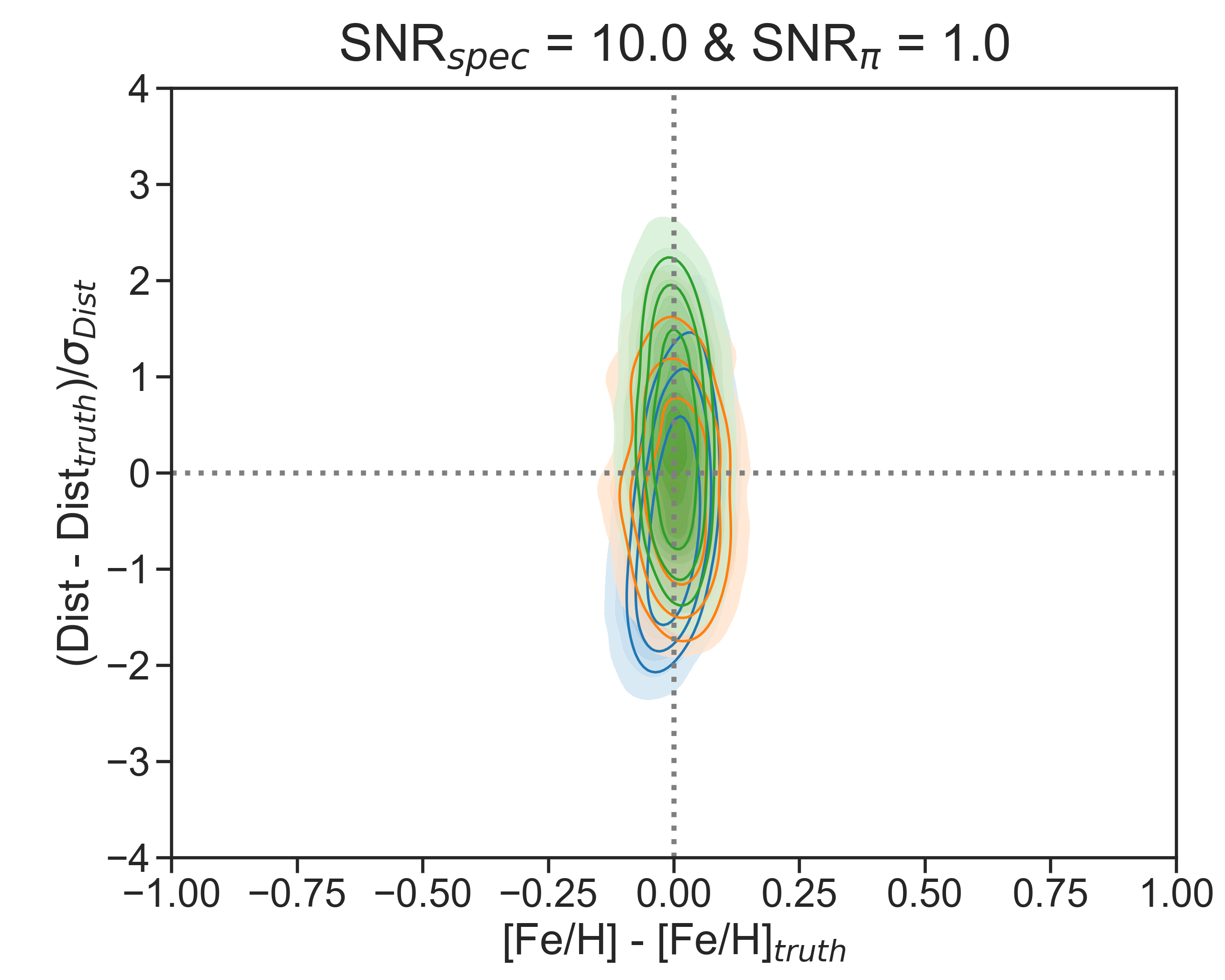}
    \includegraphics[width=0.24\linewidth]{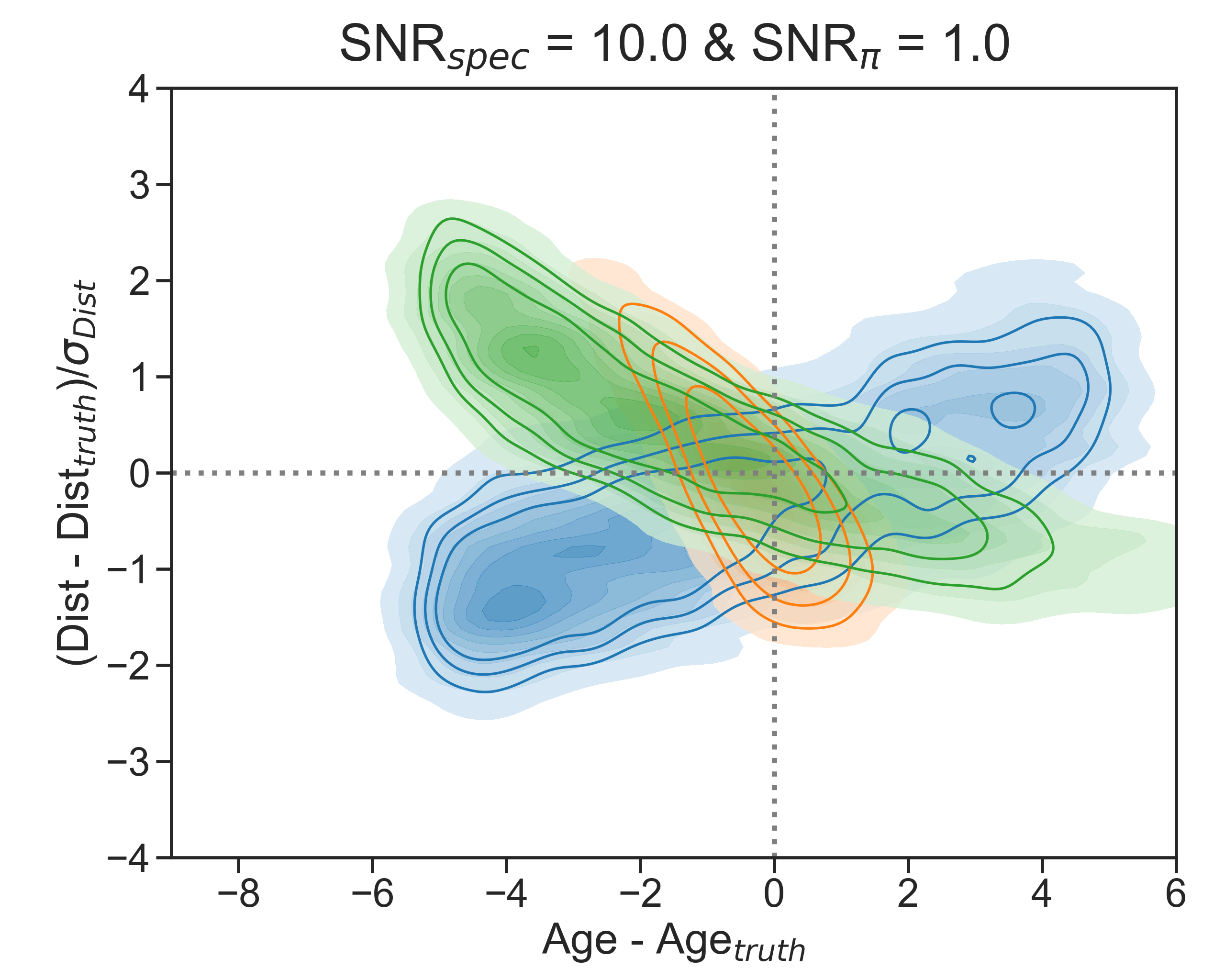}
    \caption{Constraints on stellar parameters relative to the true input values for three evolutionary phases (MS, TO, RGB) at \feh$_i=-1.0$ and parallax SNR$_\pi=1.0$.  Rows show results for SNR$_{\rm spec}=$ 2 (top), 5 (midde) and 10 (bottom).}
    \label{fig.delta_mock_params}
\end{figure*}

\begin{figure*}[t!]
    \includegraphics[width=0.33\linewidth]{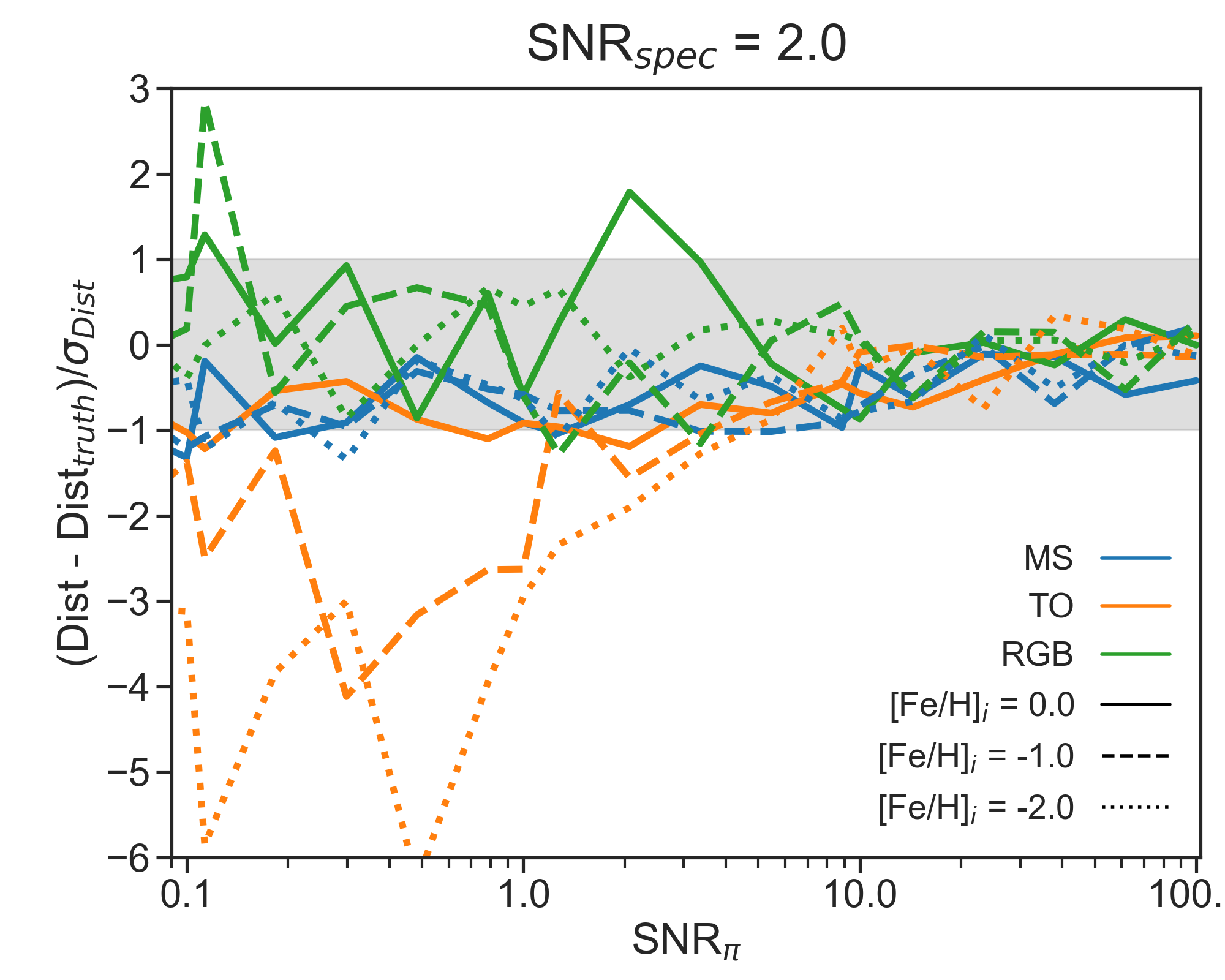}
    \includegraphics[width=0.33\linewidth]{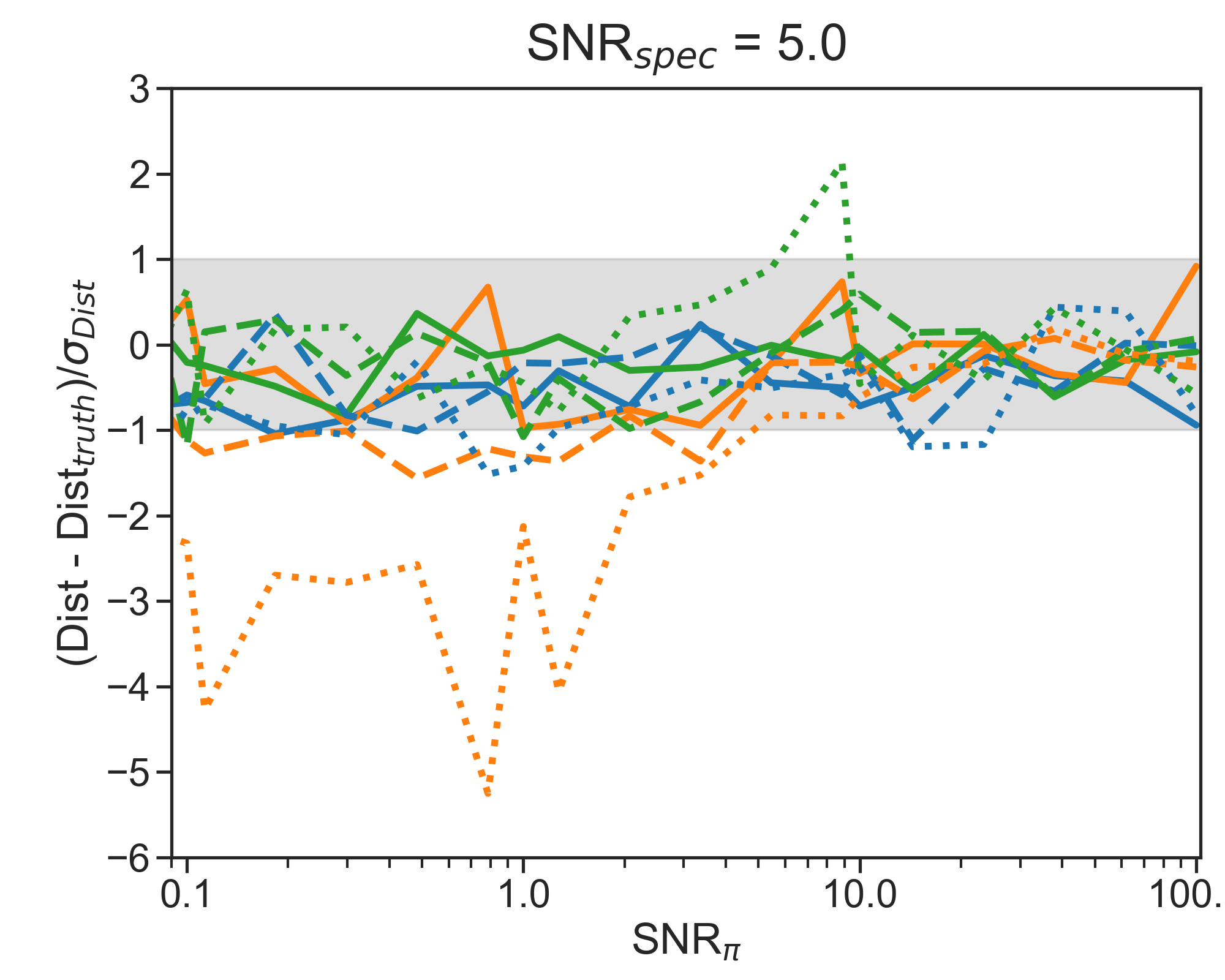}
    \includegraphics[width=0.33\linewidth]{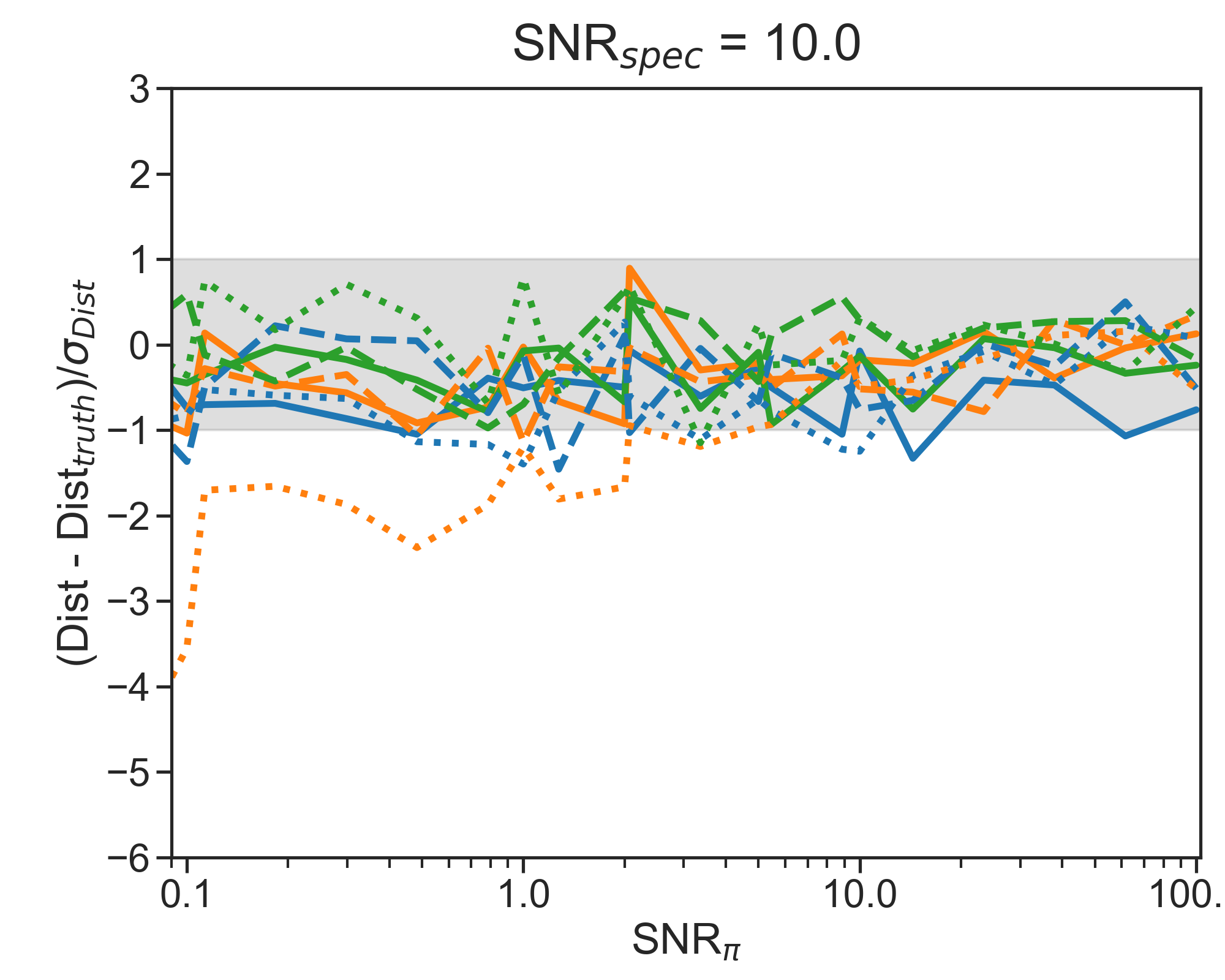}
    \includegraphics[width=0.33\linewidth]{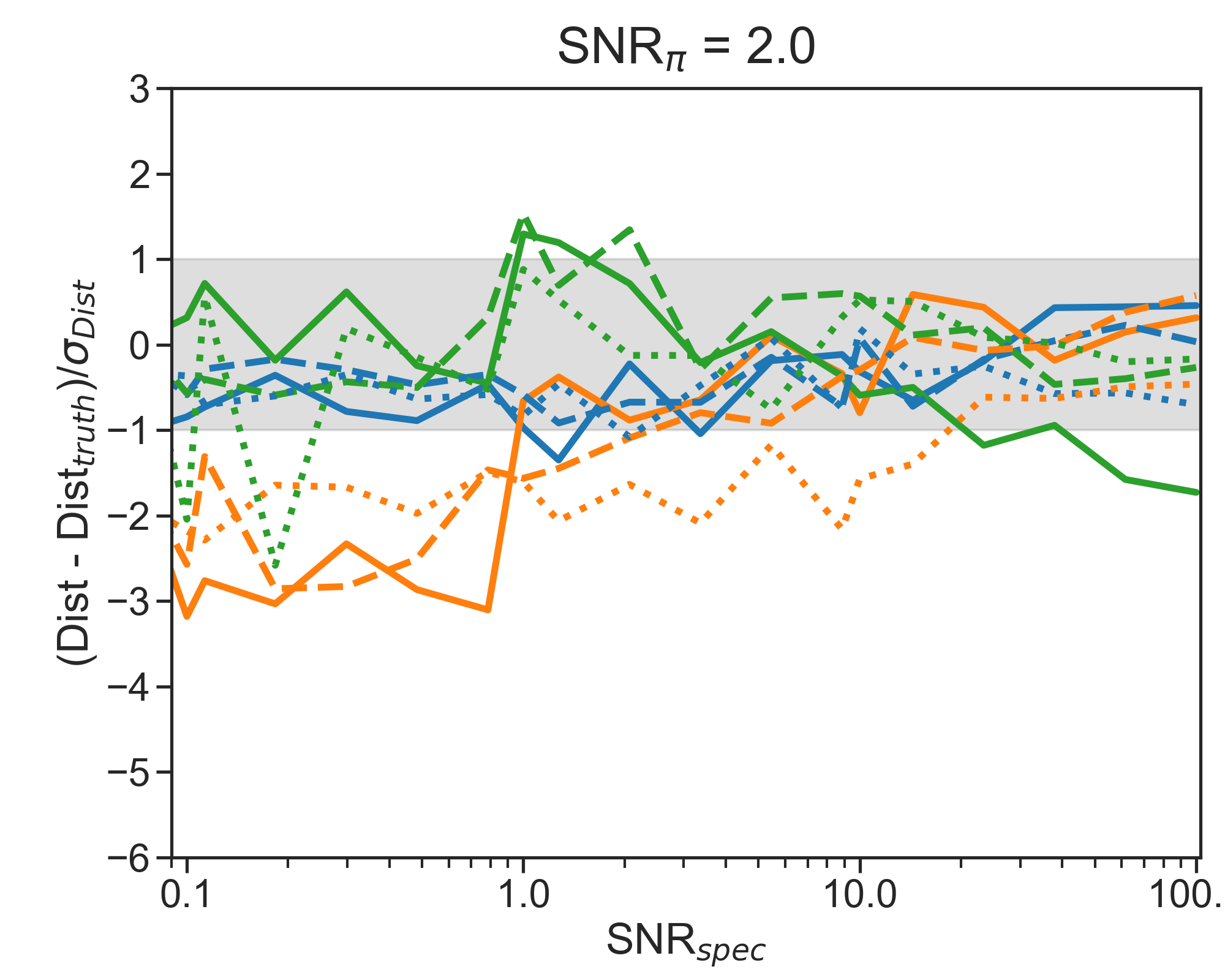}
    \includegraphics[width=0.33\linewidth]{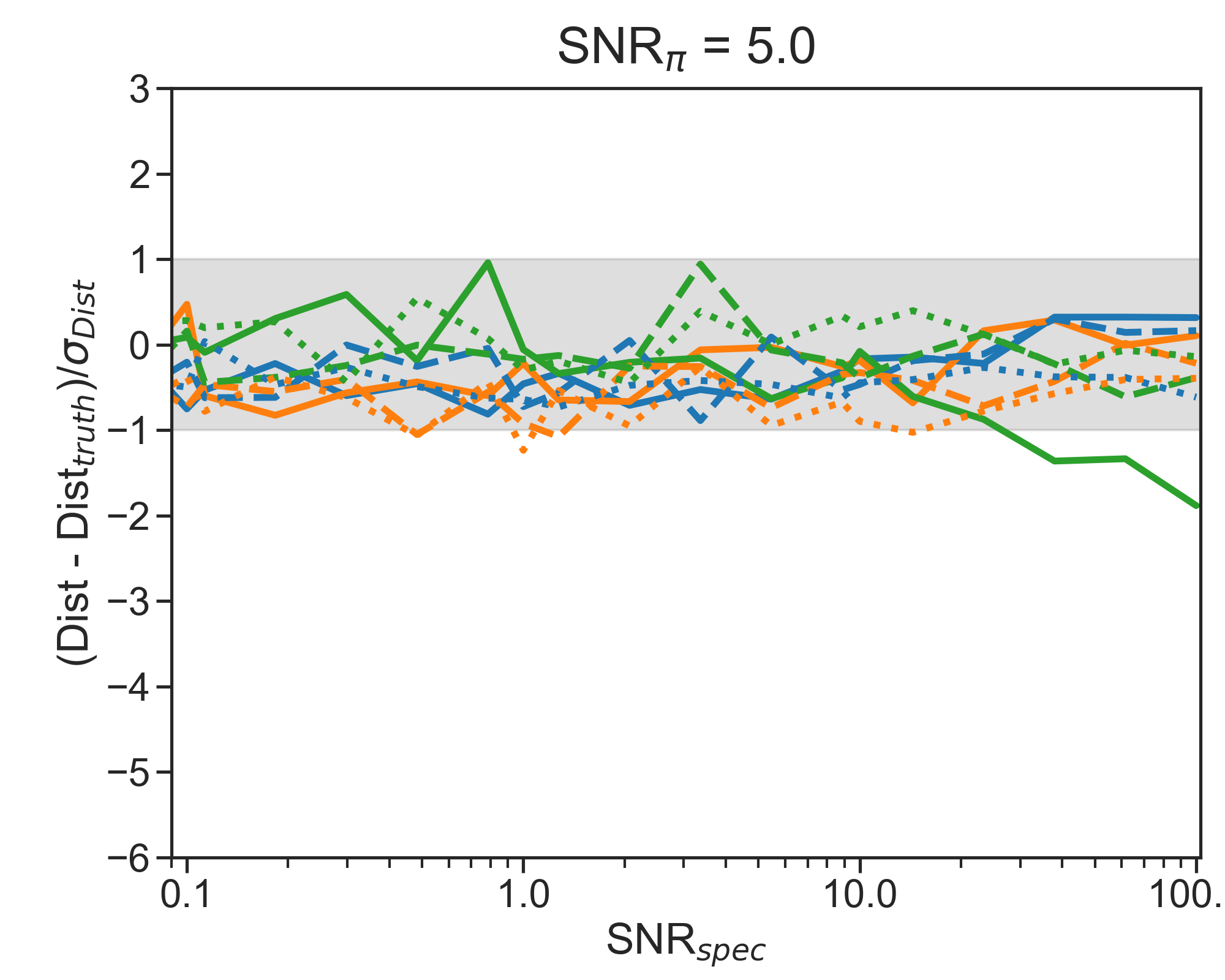}
    \includegraphics[width=0.33\linewidth]{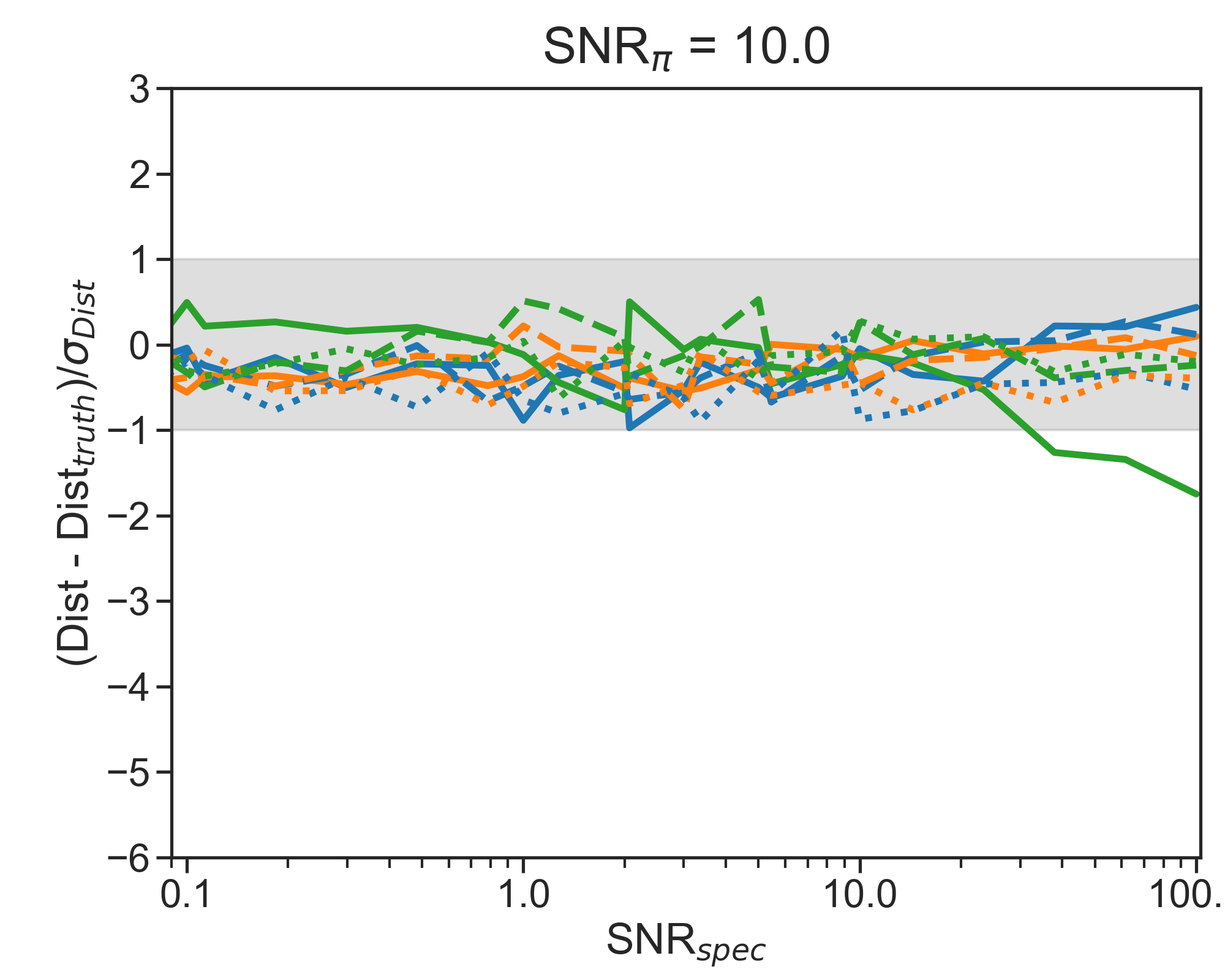}
    \caption{Accuracy of recovered distances and reported uncertainties of mock data.  The y-axis shows the best-fit distance minus the true distance in units of the measurement uncertainty. Each panel shows three evolutionary phases (MS, TO, RGB) at three metallicities: \feh$_i=-2.0, -1.0, 0.0$.  The top panels show results as a function of parallax SNR for three SNR$_{\rm spec}$, while the bottom panels show results as a function of SNR$_{\rm spec}$ for three choices of parallax SNR.  In general the measured distances agree with the true distances to within $\approx1\sigma$.  Significant deviations are limited to low metallicity TO stars at low SNR.}
    \label{fig.absdisterr}
\end{figure*}

\section{Model Validation}
\label{sec.modtest}

In this section we present a variety of tests of the \MS-derived parameters.  We begin in Section 3.1 with a suite of mock tests that demonstrate the measurement precision of derived parameters as a function of SNR, metallicity, and stellar evolutionary phase.  Turning to data, in Section \ref{sec.modtest.benchmarkstars} we fit high SNR spectra of well-studied benchmark stars that have independently-determined stellar parameters.  In Section \ref{sec.modtest.clusters} we present fits to six star clusters spanning a wide range in metallicity, and in Section \ref{sec.modtest.h3} we fit data from the H3 Survey and compare \MS-derived distances to parallax-based distances.

Throughout our validation tests, we consider photometry from Pan-STARRS griz, {\it Gaia} G, BP, and RP, 2MASS JHK$_{\textrm{s}}$, and WISE W1 W2, which together provide coverage across the bulk of the stellar SED for the stellar types considered here.  For the mock tests we synthesize photometry in these systems.  Where relevant, we include a minimum error floor of $\sigma_{\rm Pan-STARRS}=$ 0.02, $\sigma_{\rm 2MASS}=$ 0.05, and $\sigma_{\rm WISE}=$ 0.04 mag.  In the case of the cluster and H3 data we include SDSS photometry where available, with an error floor of $\sigma_{\rm SDSS}=$ 0.02.  

For spectra, we focus on a spectral wavelength range and resolution delivered by the MMT Hectochelle instrument \citep{Szentgyorgyi2011} with the RV31 order-blocking filter, which provides a spectral resolution of $R=32,000$ across the wavelength range  $5150-5350$\AA. This wavelength range contains the \ion{Mg}{1} triplet, commonly used to determine the surface gravity of main-sequence and giant stars.  The H3 and cluster data were acquired with this setup.  For simplicity and consistency, the mock spectra are also generated with this wavelength range and resolution, and the high SNR high resolution benchmark spectra are smoothed and downsampled to mimic this configuration.

\begin{figure*}[t!]
    \centering
    \includegraphics[width=0.33\linewidth]{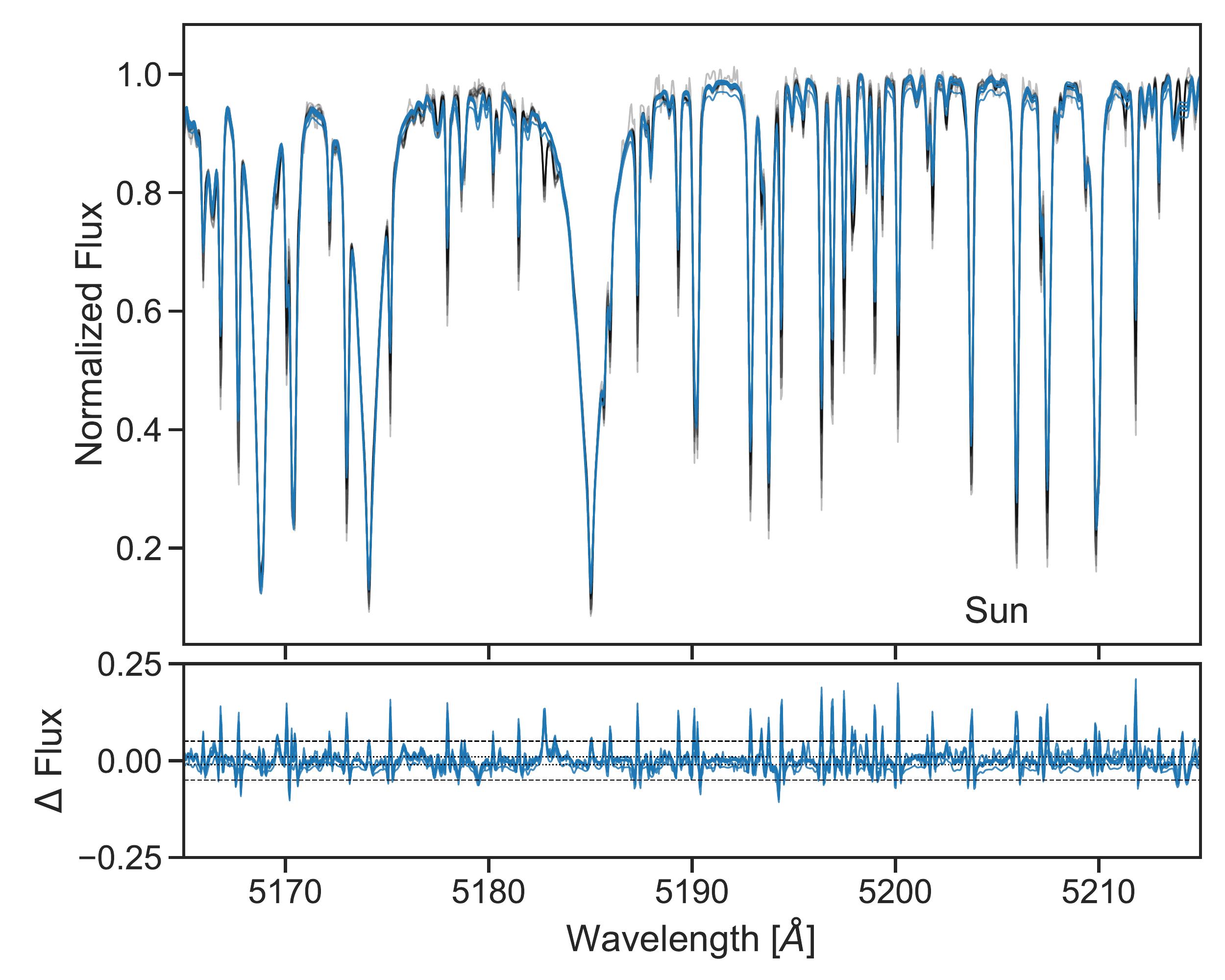}
    \includegraphics[width=0.33\linewidth]{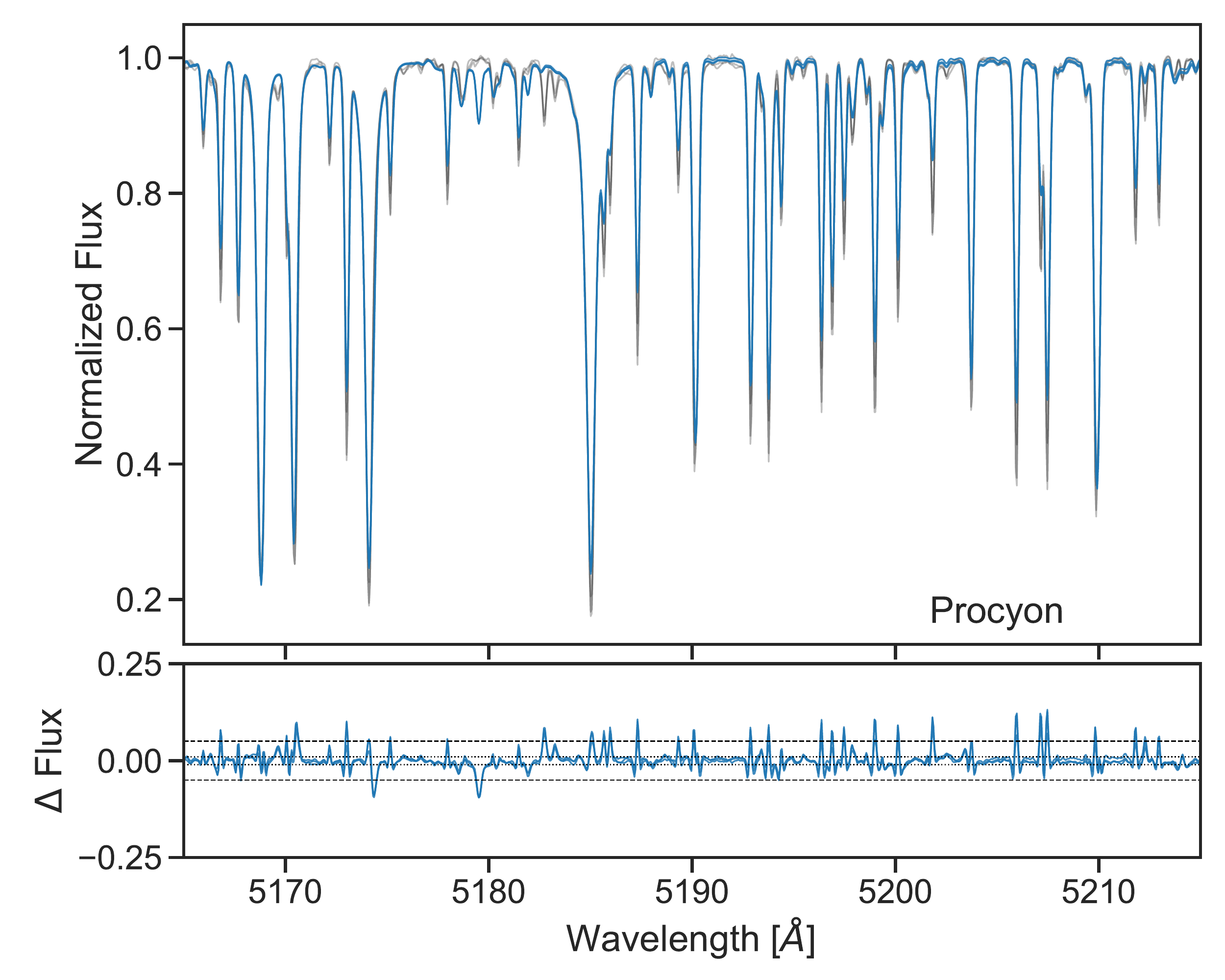}
    \includegraphics[width=0.33\linewidth]{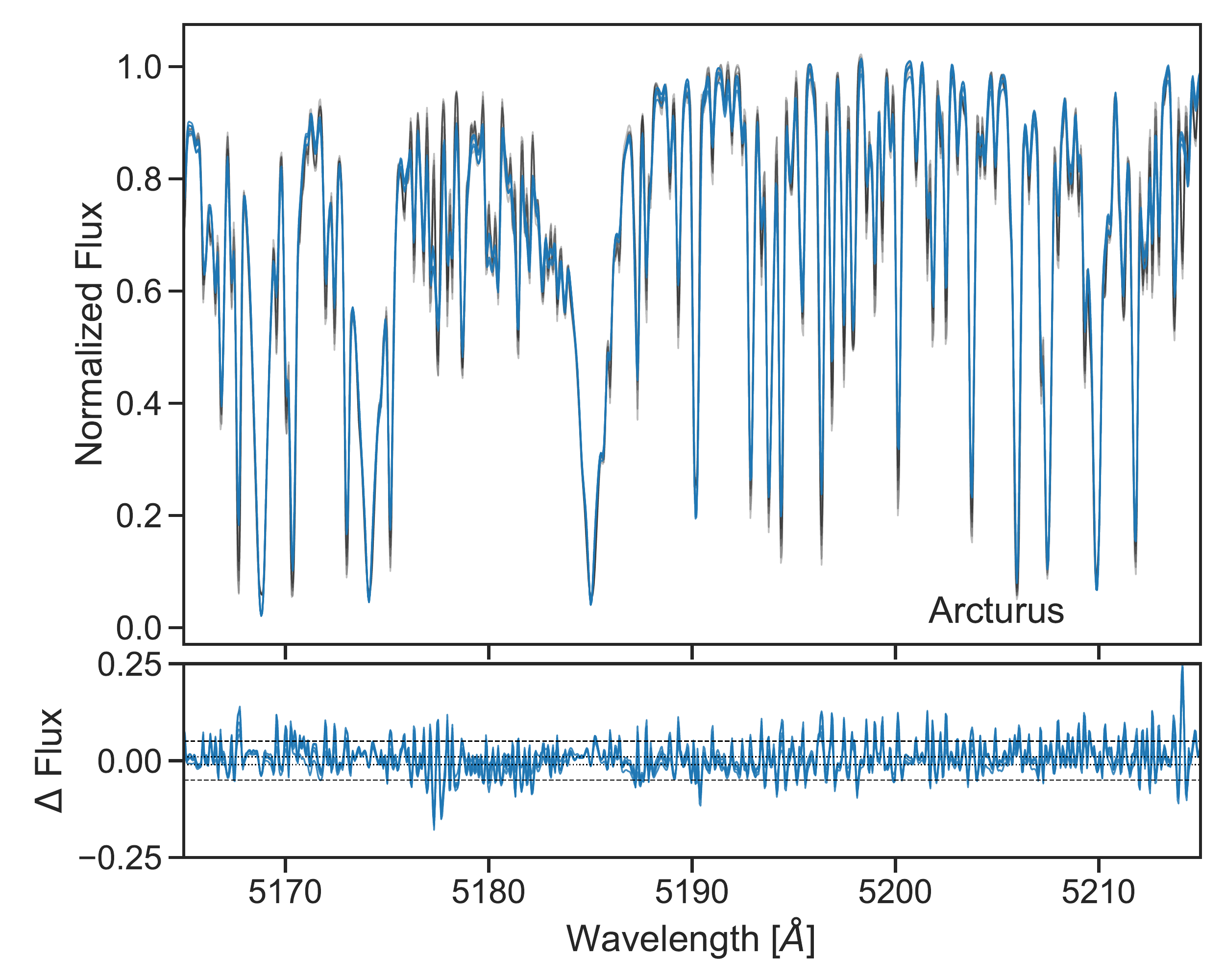}
    \caption{{\it Gaia} FGK benchmark spectra and best-fit models (top panels) with residuals (bottom panels) for the Sun (left), Procyon (middle), and Arcturus (right). Spectra were smoothed to a common resolution of $R=32,000$ before fitting.  Parameters inferred from \MS\ modeling of these spectra are given in Table \ref{tab.benchmarkstars}. Also displayed with the residuals are lines indicating 1\% (dotted) and 5\% (dashed) residuals.}
    \label{fig.Bestfit_Sun_Procyon_Arcturus}
\end{figure*}

\begin{figure*}[t!]
    \centering
    \includegraphics[width=1.\linewidth]{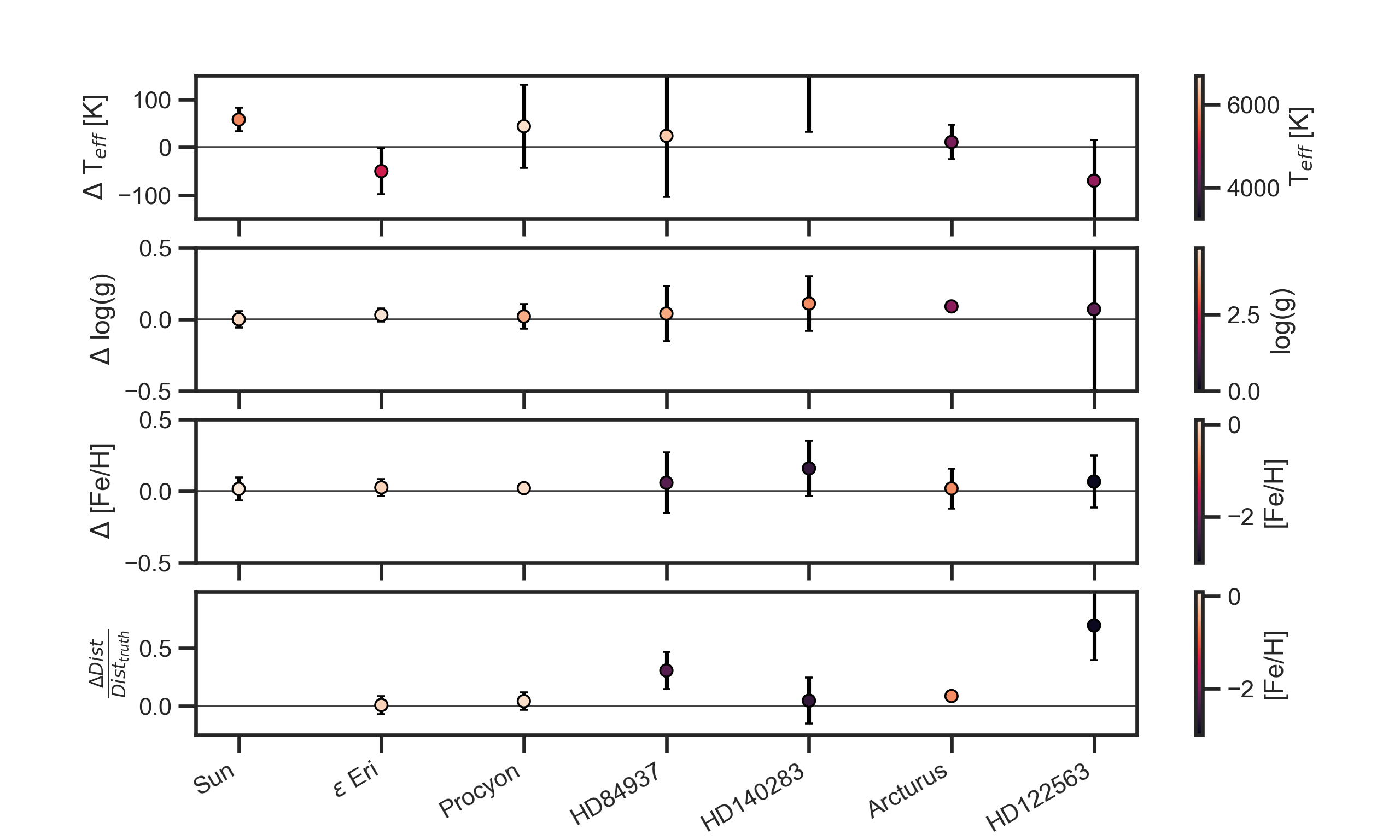}
    \caption{Comparison of \MS\, parameters and literature values for a set of benchmark stars. The differences are \MS\, parameters minus the literature value.  The error bars represent the \MS\, uncertainties added in quadrature with the literature uncertainties, and in some cases are smaller than the symbol size.  See Table \ref{tab.benchmarkstars} for literature references. For HD~140283 the difference in \teff\, is 338 K and is therefore beyond the range of the figure.}
    \label{fig.benchmarks.comp}
\end{figure*}

\begin{deluxetable*}{llllllllll}
    \tabletypesize{\scriptsize} 
    \tablecaption{\MS\, Parameters for Benchmark Stars}
    \tablehead{ 
        \colhead{Star} & \colhead{\teff} & \colhead{\logg} & \colhead{\feh} & \colhead{\afe} & \colhead{$M$} & \colhead{$L$} & \colhead{Age} & \colhead{Dist.} & \colhead{$N_{\rm spec}$\tablenotemark{a}} \\
        \colhead{}     & \colhead{(K)}   & \colhead{}      & \colhead{}     & \colhead{}     & \colhead{($M_{\odot}$)}   & \colhead{($L_{\odot}$)} & \colhead{(Gyr)} & \colhead{(pc)} & \colhead{phot.\tablenotemark{b}} } 
\startdata
{\bf Sun} \\
\emph{this work} & $5829\pm 25$ & $4.45\pm0.06$ & $0.08\pm0.02$ & $ 0.03\pm0.02$ & $ 1.04\pm0.01$ & $    1.06\pm  0.10$ & $ 2.88\pm 2.50$ & $    ...\pm  ...$ &  9 \\
 \emph{this work}\tablenotemark{c} & $5812\pm 22$ & $4.42\pm 0.02$ & $0.07\pm 0.04$ & $0.03\pm 0.02$ &  $1.03\pm 0.03$ & $1.11\pm 0.09$ & $4.56\pm 0.04$ & $...\pm ...$ & 9 \\
\emph{lit.} & $5772\pm 1$ & $4.44\pm... $ & $0.00\pm... $ & $ 0.00\pm... $ & $ 1.00\pm... $ & $         1.00\pm  ... $ & $ 4.57\pm 0.01$ & $    ...\pm  ...$ &  ...\\ 
{\bf $\epsilon$ Eri} \\
\emph{this work} & $5026\pm 37$ & $4.59\pm0.04$ & $-0.10\pm0.03$ & $ 0.02\pm0.04$ & $ 0.81\pm0.04$ & $    0.37\pm  0.01$ & $ 6.07\pm 7.05$ & $   3.25\pm 0.25$ &  4 \\
\emph{lit.} & $5076\pm 30$ & $4.60\pm0.03$ & $-0.09\pm0.06$ & $ 0.01\pm0.12$ & $ 0.80\pm0.06$ & $         0.32\pm  0.01$ & $ 0.4  - 0.9 $ & $   3.20\pm 0.01$ & G,2M\\
{\bf  Procyon} \\
\emph{this work} & $6598\pm 24$ & $3.93\pm0.05$ & $-0.01\pm0.02$ & $ 0.04\pm0.01$ & $ 1.52\pm0.04$ & $    7.86\pm  1.10$ & $ 1.89\pm 0.07$ & $   3.66\pm 0.26$ &  6 \\
\emph{lit.} & $6554\pm 84$ & $4.00\pm0.02$ & $-0.04\pm0.08$ & $0.08\pm0.11$ & $ 1.50\pm0.07$ & $          6.88\pm  0.36$ & $ 1.5 -  2.5 $ & $   3.51\pm 0.01$ & 2M\\
{\bf HD~84937} \\
\emph{this work} & $6380\pm 84$ & $3.89\pm0.09$ & $-2.30\pm0.06$ & $ 0.52\pm0.03$ & $ 0.81\pm0.02$ & $    3.98\pm  0.79$ & $10.31\pm 0.95$ & $  95.17\pm11.18$ &  5 \\
\emph{lit.} & $6356\pm 97$ & $4.06\pm0.04$ & $-2.09\pm0.08$ & $ 0.33\pm0.14$ & $ 0.75\pm0.04$ & $         2.09\pm  0.03$ & $11.0 -  13.5$ & $  72.52\pm 4.64$ & 2M\\
{\bf HD~140283} \\
\emph{this work} & $5860\pm287$ & $3.59\pm0.19$ & $-2.59\pm0.16$ & $ 0.44\pm0.05$ & $ 0.77\pm0.03$ & $    5.57\pm  1.51$ & $12.97\pm 1.90$ & $  64.98\pm12.40$ &  5 \\
\emph{lit.} & $5522\pm105$ & $3.58\pm0.11$ & $-2.43\pm0.10$ & $ 0.10\pm0.12$ & $ 0.68\pm0.17$ & $         4.65\pm  0.09$ & $ \ge 12.0    $ & $  62.06\pm 0.28$ & G,2M\\
{\bf Arcturus} \\
\emph{this work} & $4297\pm  9$ & $1.66\pm0.03$ & $-0.64\pm0.02$ & $ 0.41\pm0.00$ & $ 1.28\pm0.08$ & $  229.92\pm 21.33$ & $ 3.65\pm 0.63$ & $  12.24\pm 0.49$ &  5 \\
\emph{lit.} & $4286\pm 35$ & $1.64\pm0.09$ & $-0.53\pm0.08$ & $ 0.37\pm0.14$ & $ 1.03\pm0.21$ & $       198.05\pm  5.06$ & $ 4.0 - 10.0 $ & $  11.28\pm 0.05$ & 2M\\
{\bf HD~122563} \\
\emph{this work} & $4517\pm 61$ & $1.15\pm0.32$ & $-2.91\pm0.07$ & $ 0.50\pm0.12$ & $ 1.17\pm0.40$ & $  907.52\pm349.47$ & $ 3.19\pm 9.18$ & $ 493.53\pm87.28$ &  5 \\
\emph{lit.} & $4587\pm 60$ & $1.61\pm0.07$ & $-2.74\pm0.22$ & $ 0.39\pm0.24$ & $ 0.86\pm0.03$ & $       343.66\pm 14$ & $     ...     $ & $ 290.36\pm 5.22$ & G,2M,W\\
\enddata
\tablecomments{Literature parameters are adopted from \citet[][]{Jofre2014}, \citet{Jofre2015}, \citet[][]{Heiter2015} and \citet[][]{Sahlholdt2019}, except for distances, which are derived directly from parallaxes given in {\it Gaia} DR2.  Luminosities have also been updated to reflect the DR2 parallaxes.  The values given for \MS\ are the weighted mean and standard deviation measured from the set of fitted spectra.}
\tablenotetext{a}{Number of spectra analyzed.} 
\tablenotetext{b}{Photometric data used in the fitting: G:{\it Gaia DR2}, 2M: 2MASS, and W: WISE.}
\tablenotetext{c}{Parameters include a strong age prior based on meteorite data: Age$=$4.567$\pm$0.01 Gyr \citep{Connelly2012}.}
\label{tab.benchmarkstars}
\end{deluxetable*}

\subsection{Mock Data} \label{sec.modtest.mocks}

We first test \MS\, using simulated data that are drawn from the same models used to fit the data. These mock data allow us to determine the precision of derived parameters as a function of SNR, in the limit where the models are a perfect representation of the data. We create mock stars residing on a 9 Gyr isochrone at three metallicities: \feh$_i=-2.0,-1.0$, and 0.0.  The stars are placed at a distance of 10 kpc.  For most of the analysis in this section we focus on three archetypal stellar types: a warm main sequence (MS) star, a turnoff (TO) star, and a first ascent red giant branch (RGB) star.  These three types allow us to explore the parameter recovery for three very different regimes within old stellar populations.


Figure \ref{fig.examplecmockcorner} shows an example set of posterior distributions resulting from \MS\, modeling of a 9 Gyr solar metallicity RGB star. The spectrum for this mock has errors corresponding to a signal-to-noise (SNR$_{\rm spec}$) of 10 per pixel.  Note that we are not adding noise to the actual spectrum. The parallax prior for this mock assumes a parallax signal-to-noise (SNR$_{\pi} = \pi/\sigma_{\pi}$) of 1.0. 

The posteriors show that we recover all of the input stellar parameters. Figure \ref{fig.examplecmockcorner} also shows significant correlations between parameters, resulting in non-symmetric posteriors, as well as non-linear covariances.  These include ``trivial" covariances, e.g., between log $R$, distance, and log $L$, that are a consequence of their definitions.  Other correlations are more interesting, such a between age and distance, and between \feh, \logg, and \teff.  The latter set are correlated in such a way as to preserve the line-depths in the spectrum (higher temperatures go with higher metallicities and higher \logg).


Figure \ref{fig.examplecmockkiel} shows recovered stellar parameters in Kiel diagrams (\logg$-$\teff) for eight evolutionary phases ranging from lower MS through the TO, RGB, horizontal branch (HB) and asymptotic giant branch (AGB).  Rows show results for three metallicities: \feh$_i=-2.0,-1.0,0.0$.  Columns show results for SNR$_{\rm spec}=2,5,10$.  In all cases we have assumed a parallax SNR (SNR$_\pi$) of 1.0, and broadband photometry from Pan-STARRS, {\it Gaia}, 2MASS, and WISE are included in the fitting.  Also shown is the 9 Gyr isochrone from which the mock data were drawn.

In general the posteriors agree very well with the true values (shown as star symbols). It is noteworthy that this remains true even at SNR$_{\rm spec}=2$ at low metallicity.  \MS\ struggles in the case of the TO star, especially at low SNR$_{\rm spec}$ and low metallicity.  The primary reason for this is the age prior.  The TO is a relatively short-lived phase, and so the prior strongly prefers the higher-\logg\ solution at the same temperature.  High SNR and/or high metallicity is required for the spectrum to favor the lower-\logg\ TO solution. 

The parameter space explored here was tailored to the stellar populations and data quality relevant for the H3 Survey (see Section \ref{sec.modtest.h3} below), and so one should be careful not to generalize these results too broadly.  In particular, the combined use of photometry and spectra is essential here, as the former strongly constrains \teff, which in turn allows the spectrum to place stronger constraints on \logg\ and \feh\ even at low SNR$_{\rm spec}$. 


We now focus on the precision with which \MS\, recovers distances.  Figure \ref{fig.reldisterr} shows the fractional distance uncertainties for the MS, TO, and RGB stars as a function of metallicity, SNR$_\pi$, and SNR$_{\rm spec}$.  The parallax prior is also shown as a grey line in each panel. 

The top panels show the fractional distance uncertainties as a function of parallax SNR. From these plots it is clear that the parallax prior dominates the distance uncertainties for SNR$_\pi\gtrsim10$.  At lower SNR$_\pi$ the uncertainties are governed by the spectrum (and photometry, which is also included in the fit).  At low SNR$_{\rm spec}$ (left panel), the distance precision depends strongly on stellar type but not on metallicity, while at high SNR$_{\rm spec}$ (right panel) there is a strong metallicity-dependence for the TO star (see also the right panels of Figure \ref{fig.examplecmockkiel}).  The distances are well constrained for the MS star in all cases owing to the nature of stellar isochrones: if the star can be determined to be a dwarf, not near the TO, with a well-measured \teff\ from photometry, then its location in the HR diagram can be determined quite precisely.

The bottom panels of Figure \ref{fig.reldisterr} show the fractional distance uncertainties as a function of SNR$_{\rm spec}$.  It is noteworthy that the distance precision for the RGB star does not begin to significantly improve with increasing SNR$_{\rm spec}$ until SNR$_{\rm spec}\gtrsim10$.  For the TO star the distance precision starts improving for SNR$_{\rm spec}\gtrsim2$, except for the lowest metallicity, which shows no discernible improvement with SNR$_{\rm spec}$.  Finally, the MS star shows a more continuous increase in distance precision with SNR$_{\rm spec}$ for all metallicities.


In Figure \ref{fig.delta_mock_params}, we show for several stellar properties the output posterior distributions from our \MS\, modeling of the mock data with SNR$_{\rm spec} = 2$, 5 and 10 pix$^{-1}$ and SNR$_\pi=1.0$. Here we focus on the MS, TO, and RGB stars with \feh$_i=-1.0$.

The posteriors for the three stellar types have different correlations and systematic offsets as a result of the information content of the spectrum and photometry, as well as the shape of the isochrone at the three different evolutionary phases.  In many cases, increasing SNR$_{\rm spec}$ has the expected effect of reducing the posterior regions.  Interesting exceptions to this rule include the distance constraint for the TO star (as noted earlier), and the age constraint for the MS and RGB star (right panels).  For the TO star, while the distance uncertainty is not reduced, the systematic bias is removed at higher SNR.  In the case of the MS and RGB stars, they have weak age constraints for the well-known reason that the isochrones at those locations are nearly independent of age for the range of ages considered here.  We also note that in all cases \feh\, is recovered with no bias and small uncertainties, even at SNR$_{\rm spec}=2.0$.


Figure \ref{fig.absdisterr} explores the accuracy of the recovered distances and uncertainties.  In this figure we show the measured distance minus the true distance in units of the reported uncertainty.  This quantity should lie within $\pm1$ if the measured distances and uncertainties are accurate.  We show this quantity for the MS, TO, and RGB stars for metallicities \feh$_i=-2.0,-1.0,0.0$ and as a function of SNR$_{\rm spec}$ and SNR$_\pi$.  

In most cases the reported distances and uncertainties are within the expected range (demarcated by the grey band), indicating that these quantities are well-measured by \MS.  There are two outlier cases.  First are the TO stars at low SNR, especially at \feh$=-2.0$.  The distances deviate by several sigma compared to the true values, as noted earlier in this section.  The reason for this is that at low SNR the prior plays an increasingly important role, and the TO star is a less likely phase to find a star compared to the higher-\logg\, solution at the same \teff.  The distances in this case are therefore systematically under-estimated.  This bias is removed for SNR$_{\rm spec}>2$ except for \feh$_i=-2.0$, where the bias is lifted for SNR$_{\rm spec}>10$.

The second outlier case is at very high SNR$_{\rm spec}$ for the solar metallicity RGB star.  In this case the issue is related to the posterior solutions becoming poorly sampled, as the uncertainties are dropping to below 1\%.  We have added a noise floor of 1\% in the distance uncertainties to partially alleviate this issue.  In any event, this issue is not relevant in the analysis of real data, as systematic uncertainties dominate the error budget at the percent level.

\begin{deluxetable*}{lcrrrrrrrr}
\tabletypesize{\scriptsize}
\tablecaption{Cluster Parameters \label{tab.clusters}}
\tablehead{
\colhead{Cluster} & \colhead{Number} & \multicolumn{4}{c}{Lit.} & \multicolumn{4}{c}{\MS}\\
\colhead{}  & \colhead{of} & \colhead{Age}   & \colhead{Dist.} & \colhead{\feh\tablenotemark{a}} & \colhead{\afe\tablenotemark{b}} & \colhead{Age}   & \colhead{Dist.} & \colhead{\feh} & \colhead{\afe} \\
\colhead{}  & \colhead{Stars} & \colhead{[Gyr]} & \colhead{[kpc]} & \colhead{}     & \colhead{}     & \colhead{[Gyr]} & \colhead{[kpc]} & \colhead{}     & \colhead{}
}
\startdata
M92  & 38  & 13.25 & 17.7 & $-2.23$, $-2.31$ &  0.14--0.47 & ...             &   $9.26\pm2.46$ & $-2.13\pm0.15$ & $0.29\pm0.12$\\
M13  & 80  & 12.9  & 12.5 & $-1.50$, $-1.53$ &  0.13--0.26 & ...             &   $8.33\pm1.33$ & $-1.44\pm0.06$ & $0.19\pm0.05$\\
M3   & 82  & 12.6  & 10.5 & $-1.40$, $-1.50$ &  0.15--0.25 & ...             &  $10.66\pm1.22$ & $-1.34\pm0.08$ & $0.17\pm0.04$\\
M107 & 40  & 12.75 & 6.75 & $-1.01$, $-1.02$ &  0.24--0.51 & ...             &   $5.80\pm1.07$ & $-0.97\pm0.09$ & $0.35\pm0.10$\\
M71  & 124 & 12.5  & 4.44 & $-0.68$, $-0.78$ &  0.21--0.49 &  $9.92\pm1.37$  &   $4.27\pm0.65$ & $-0.72\pm0.13$ & $0.35\pm0.09$\\
M67  & 178 & 3.58  & 0.88 & $-0.01$          & -0.05--0.05 &  $3.92\pm0.37$  &   $0.80\pm0.08$ & $ 0.04\pm0.06$ & $0.06\pm0.05$\\
\enddata
\tablecomments{Uncertainties from \MS\, modeling are based on the 68th percentile scatter weighted by the maximum posterior probability ($\propto$ maximum likelihood) for the individual clusters.  They therefore include both the measurement uncertainty as well as  any systematic bias.  The formal error on the mean values are much smaller and so are not reported here.}
\tablenotetext{a}{Literature values of globular clusters for \feh\, are listed twice, from \citet{Meszaros2015} and \citet{Carretta2009}.}
\tablenotetext{b}{Literature values for \afe\, are quoted as a range encompassing the results of \citet{Meszaros2015}, \citet{Carretta2009}, and the literature compilation of \citet{Pritzl2005}.}
\end{deluxetable*}

\subsection{FGK Benchmark Stars} \label{sec.modtest.benchmarkstars}

We now turn to validation of \MS\, with real data.  In this section we model seven well-studied "benchmark" stars: the Sun, $\epsilon$ Eri, Procyon A (hereafter Procyon), HD 84937, HD 140283, Arcturus, and HD 122563.  These stars were chosen because they contain multiple spectra in the {\it Gaia} Benchmark Stars online database \citep{Blanco-Cuaresma2014}, and because they span a range in [Fe/H], \teff, and \logg.  They are nearby and bright, and hence have precisely-determined parallax-based distances and very high SNR data.

Spectra for these stars were obtained from the {\it Gaia} Benchmark Stars online database, an archive of high resolution and high signal-to-noise spectra of FGK stars \citep{Blanco-Cuaresma2014}. We use all available spectra in the archive.  These spectra uniformly are of high resolution ($R\approx$80,000) and signal-to-noise (SNR$_{\rm spec} \ge 200$).  We limit the wavelength range to $\lambda=5150-5350$\AA, and convolve the spectra with a Gaussian instrument profile to a resolving power of $R=32,000$.  We choose this setup as it is similar to the MMT/Hectochelle data analyzed in later sections.  As is the case with many ``naked-eye" stars, these {\it Gaia} Benchmark Stars lack high-quality photometry in many modern surveys. Therefore, we include photometry only if available from three well-calibrated photometric surveys: Gaia, 2MASS, and/or WISE.  Each available spectrum plus photometry is fit with \MS.  We adopt a flat distance prior of $0-1$ kpc and a flat prior in log(age) from $0.1-14$ Gyr.  Independent parallax information is not used as a prior in the fitting.  The Sun was fit without photometry and so no distance is reported.

A summary of best-fit parameters for these stars is given in Table \ref{tab.benchmarkstars}, and a comparison between the \MS\, and literature values for \teff, \logg, \feh\, and distances is presented in Figure \ref{fig.benchmarks.comp}.  Best-fit models to the spectra of the Sun, Arcturus, and Procyon are shown in Figure \ref{fig.Bestfit_Sun_Procyon_Arcturus}.  The quoted uncertainties from this work represent the scatter amongst the fits to the independent spectra available for each star.  The formal $1\sigma$ uncertainties reported by \MS\, are always much smaller than this value due to the very high SNR$_{\rm spec}$.  The literature sources are adopted directly from the {\it Gaia} benchmark online table\footnote{\texttt{www.blancocuaresma.com/s/benchmarkstars}}, and are based on spectroscopic analysis from \citet{Jofre2014}, \citet{Jofre2015}, and \citet{Heiter2015}, and isochrone fitting from \citet{Sahlholdt2019}.   In \MS\, all of the $\alpha-$elements are varied in lockstep.  However, the available wavelength range is dominated by the \ion{Mg}{1} triplet, so we expect that \afe\, is most sensitive to [Mg/Fe].  For the literature values, we therefore adopt [Mg/Fe] as a tracer of \afe.

Overall, the \MS\, parameters are in good agreement with the parameters published in the literature for these stars.  In nearly all cases parameters agree to within $\lesssim0.1$ dex or the quoted uncertainties, whichever is larger.  It is particularly noteworthy that we are able to recover the literature parameters given the wide range of values: metallicities range from $-2.7$ to $+0.0$; \logg\, values range from 4.6 to 1.6, and temperatures from 4286 K to 6550 K.  The most significant outlier is HD 122563.  Previous work has also found this star to lie well off of standard stellar evolution models given its measured metallicity, \teff, and bolometric luminosity \citep{Creevey2012, Creevey2019}, and so the fact that this star is an outlier is not surprising.

Our analysis shows how sensitive ages for main-sequence stars like the Sun are to small changes in measured parameters (see Table \ref{tab.benchmarkstars}). For the Sun, a change of \teff\,$\approx10$ K  results in an age difference of $\approx2$ Gyr.  We have therefore re-fit the Sun with a strong prior on the age derived from meteorite data \citep{Connelly2012}, and show the resulting derived parameters for this case also in Table \ref{tab.benchmarkstars}.  The parameters are nearly unchanged with respect to the less-constrained fit.

\begin{figure*}[t!]
    \centering
    \includegraphics[width=0.45\linewidth,height=0.25\textheight]{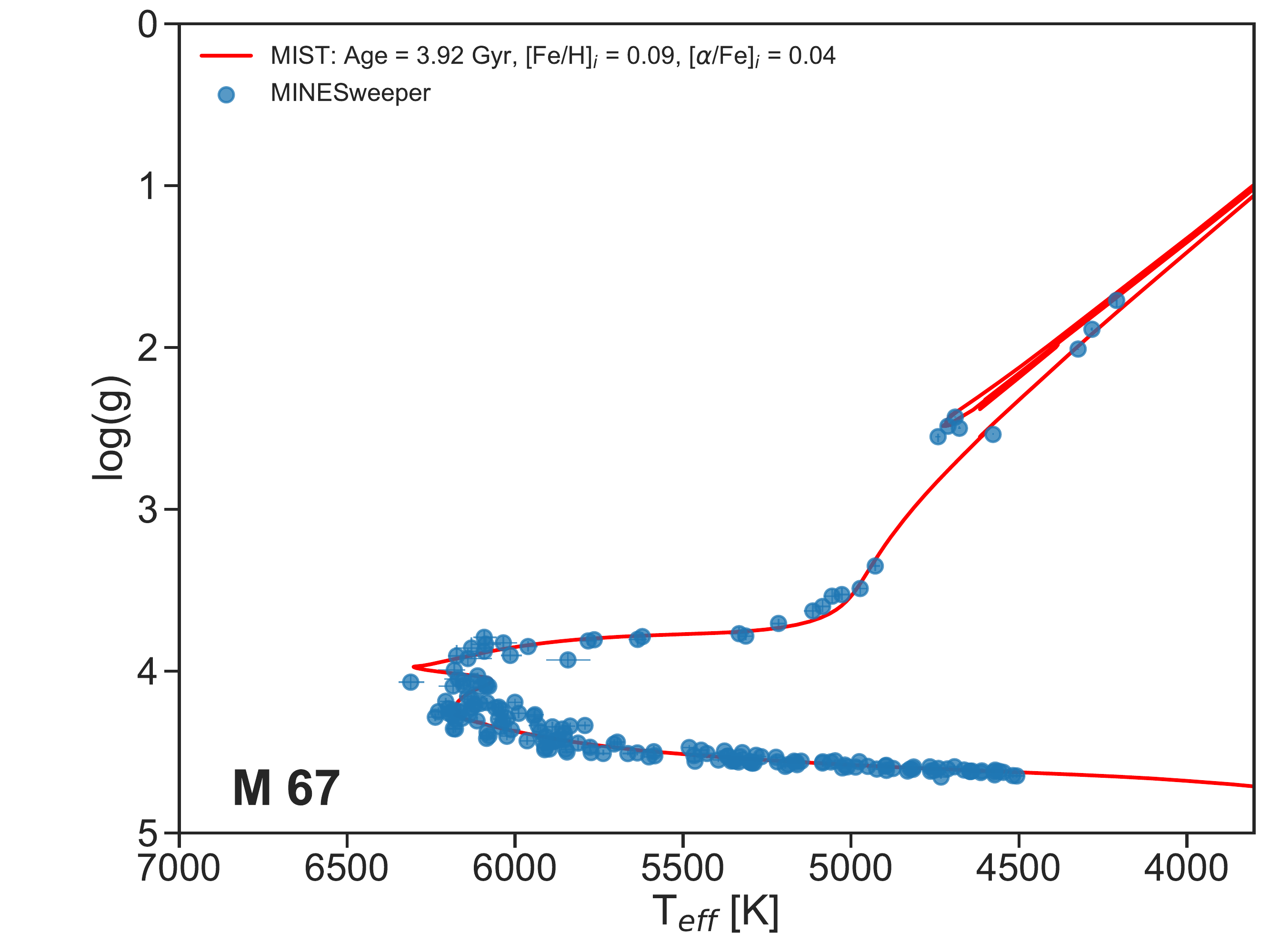}
    \includegraphics[width=0.45\linewidth,height=0.25\textheight]{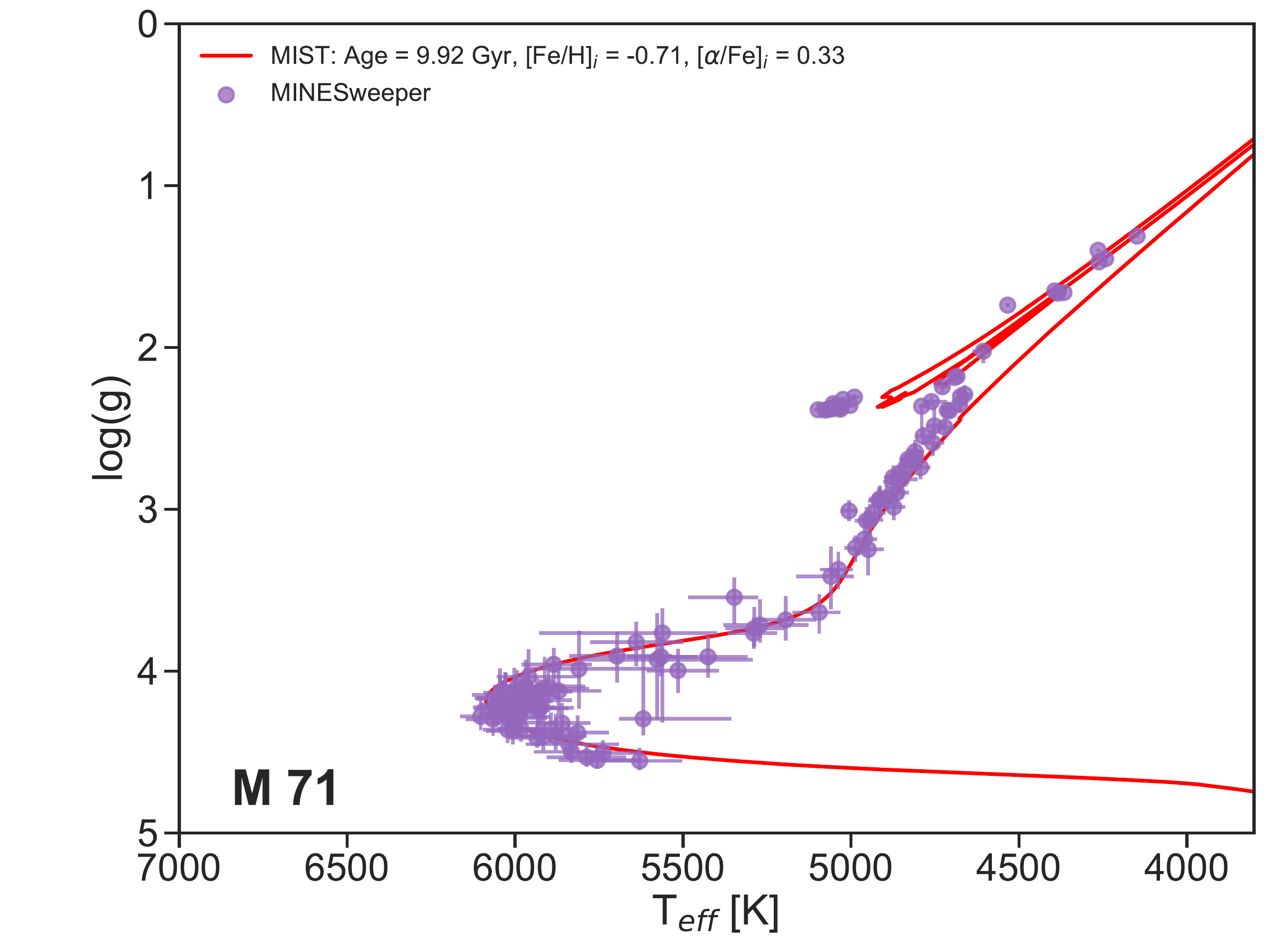}
    \includegraphics[width=0.45\linewidth,height=0.25\textheight]{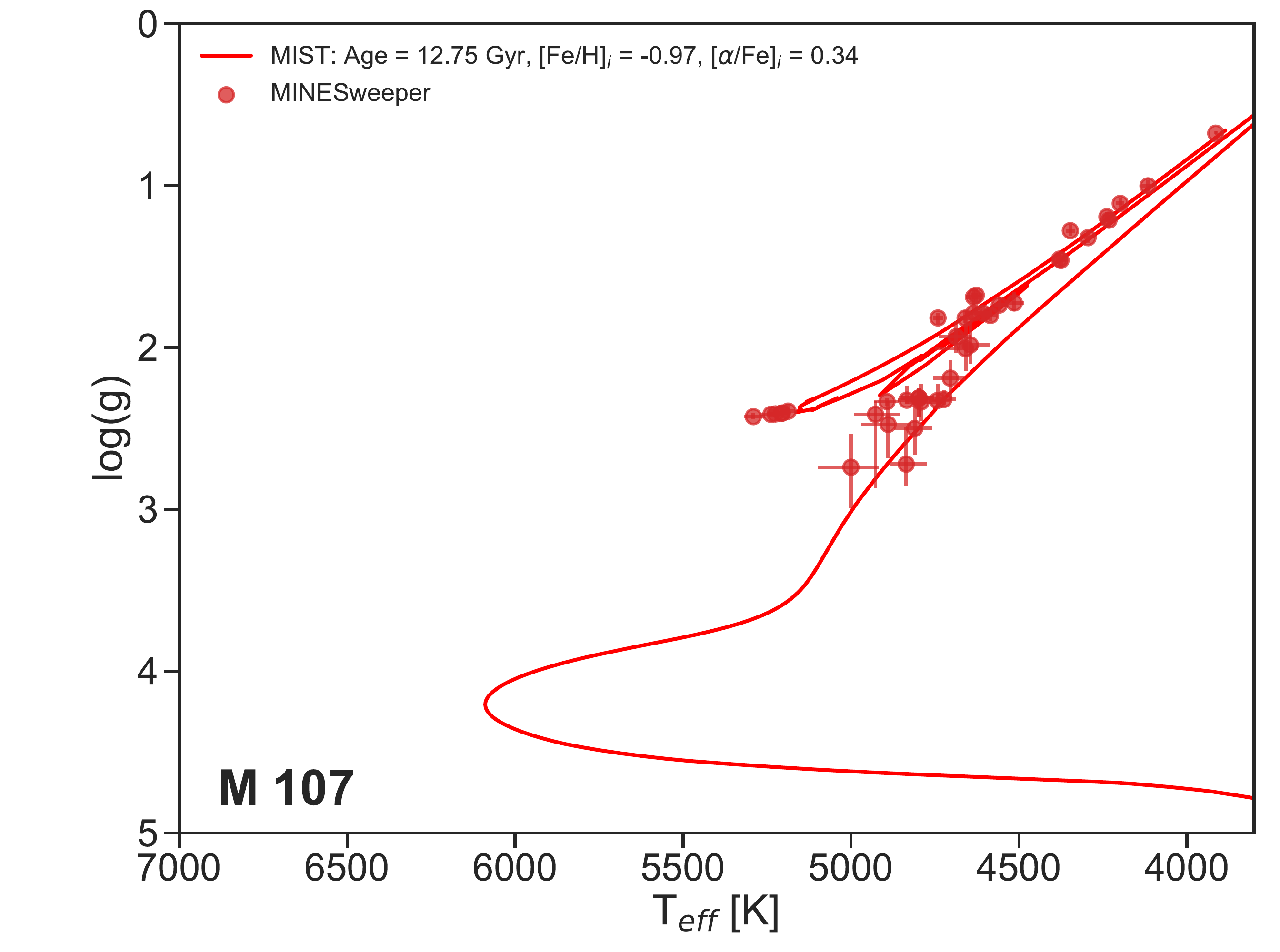}
    \includegraphics[width=0.45\linewidth,height=0.25\textheight]{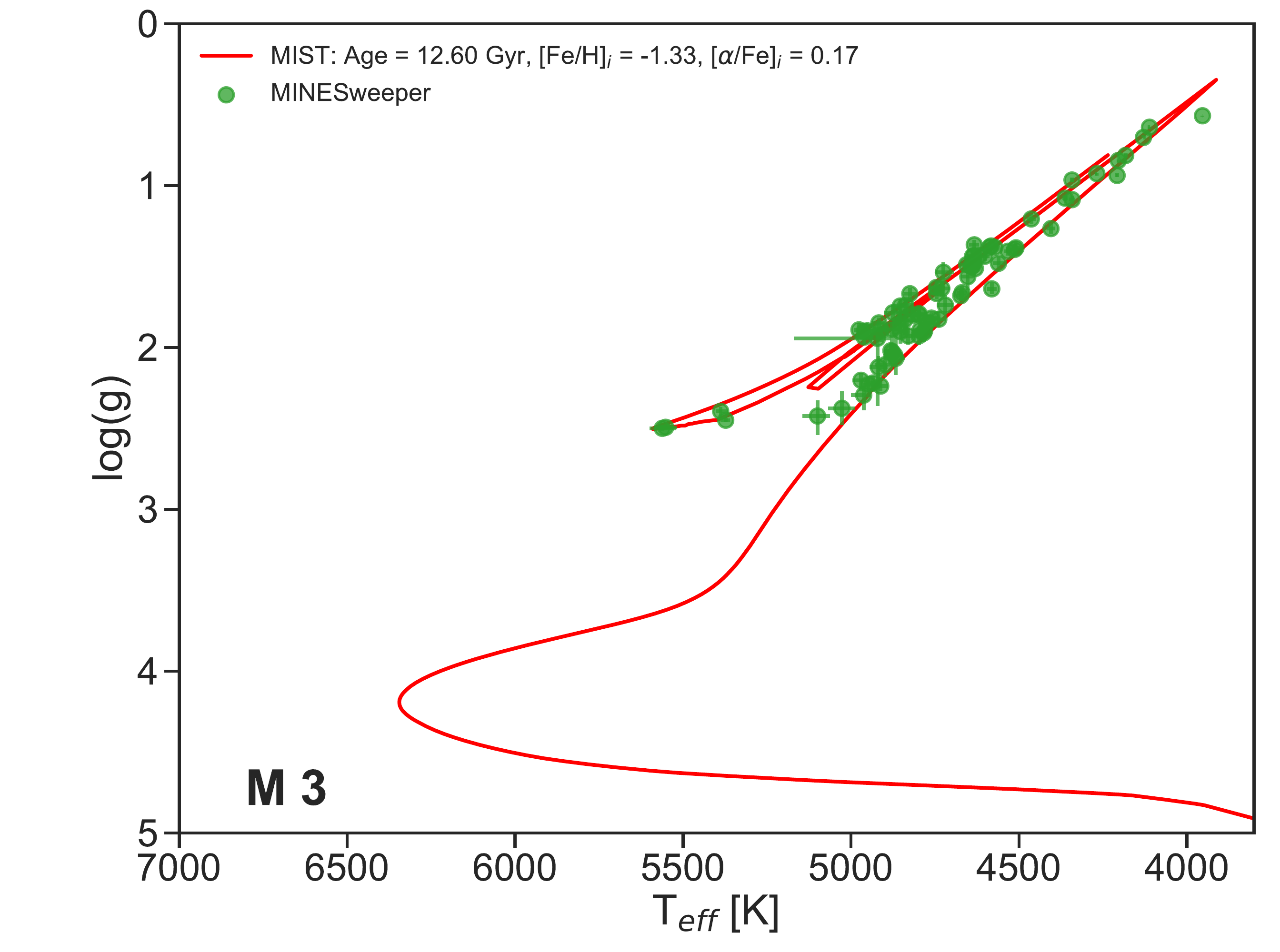}
    \includegraphics[width=0.45\linewidth,height=0.25\textheight]{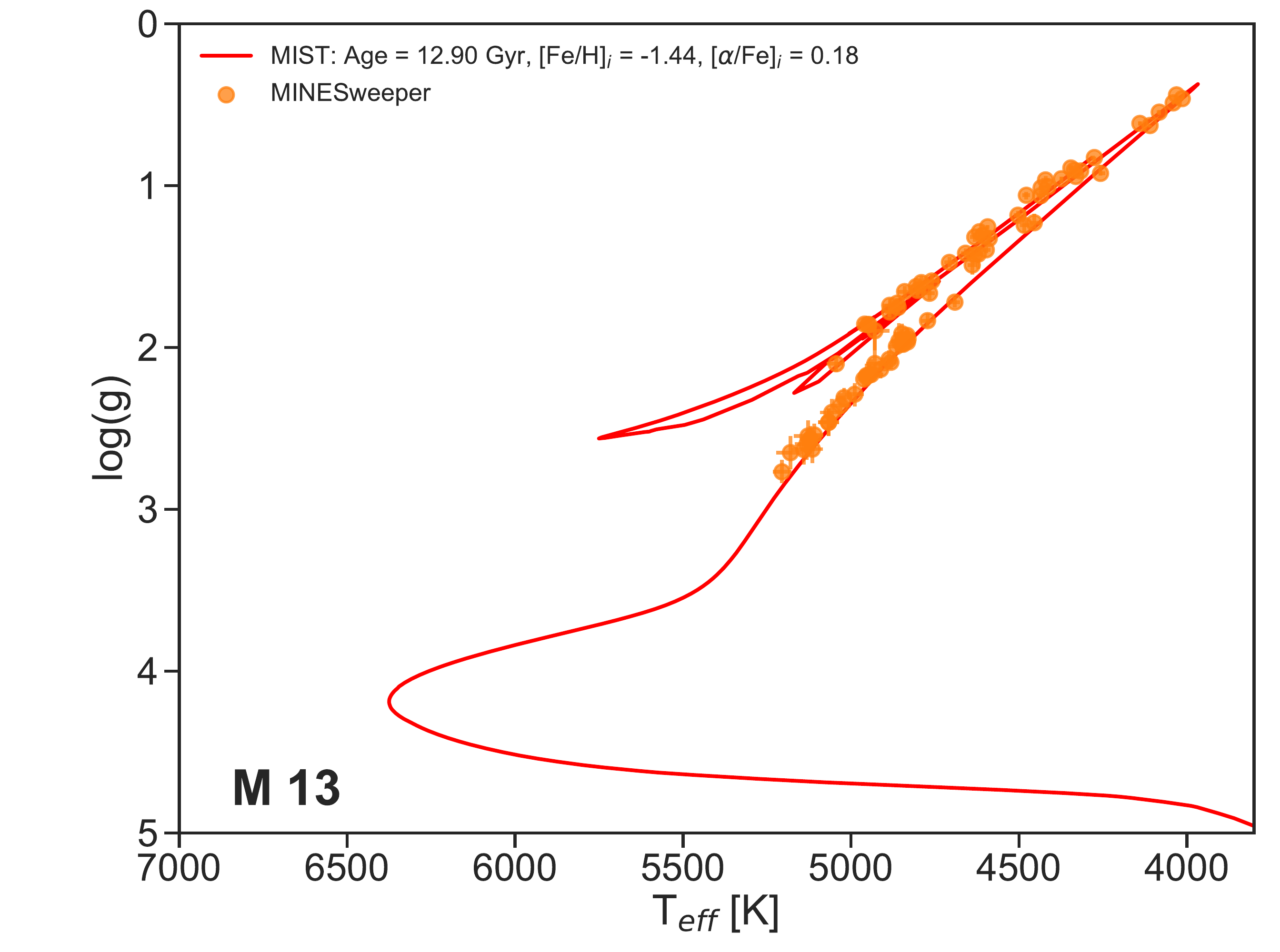}
    \includegraphics[width=0.45\linewidth,height=0.25\textheight]{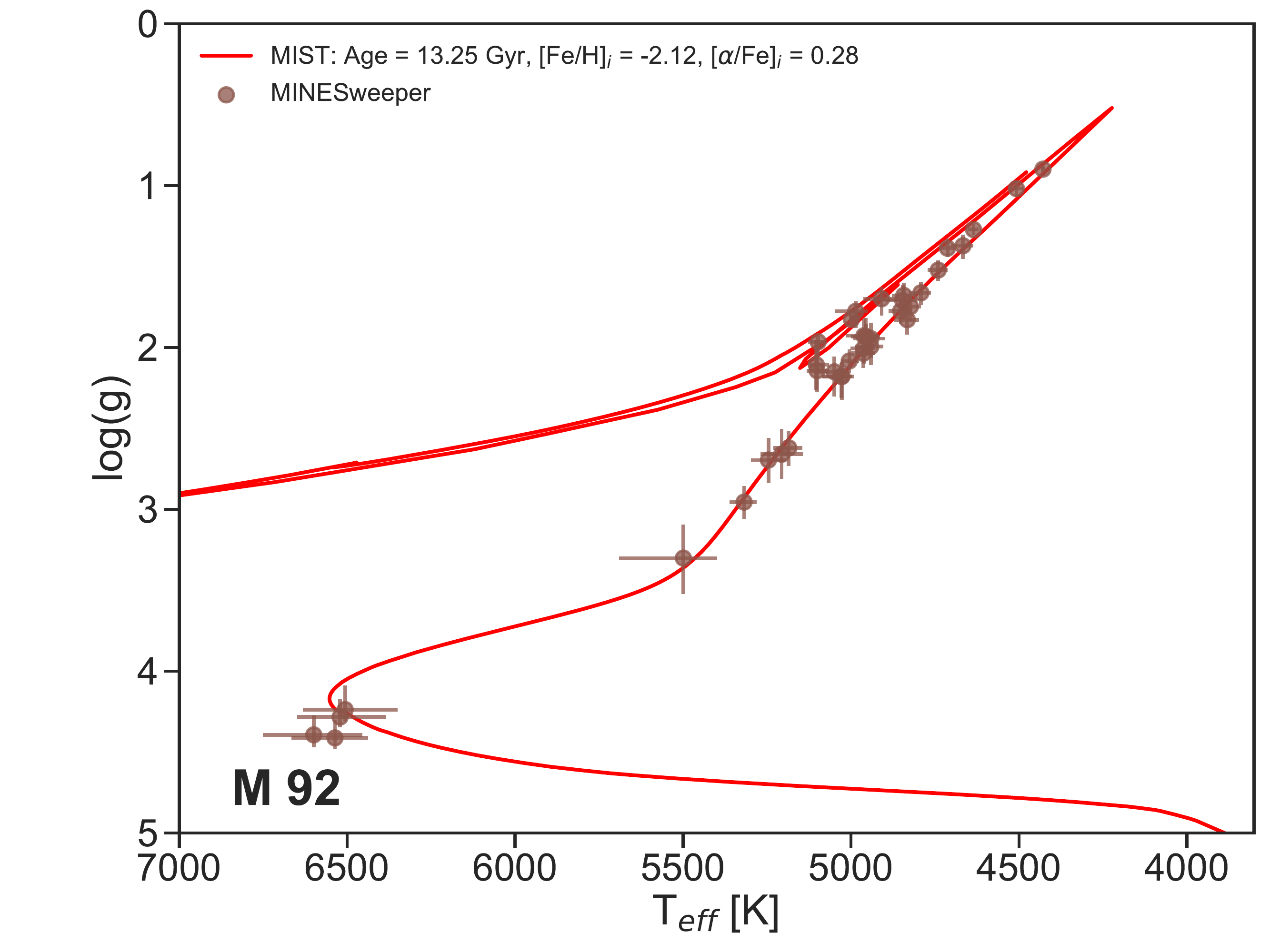}
    \caption{Kiel diagrams for clusters M67 (top left), M71 (top right), M107 (middle left), M3 (middle right), M13 (lower left), and M92 (lower right). For each cluster, we plot the \MS\ measured \logg\, and \teff\, for stars with {\it Gaia} DR2 parallax priors included in the modeling. Errors on the points give the 68th percentile credible interval for these parameters. Over-plotted are the \texttt{MIST} isochrone predictions (red lines) for estimated average cluster ages and metallicities.  In the cases of M107, M3, M13, and M92, the isochrone ages are based on literature values.}
    \label{fig.cluster.Kiel}
\end{figure*}

\begin{figure*}[t!]
    \centering
    \includegraphics[width=0.45\linewidth,height=0.25\textheight]{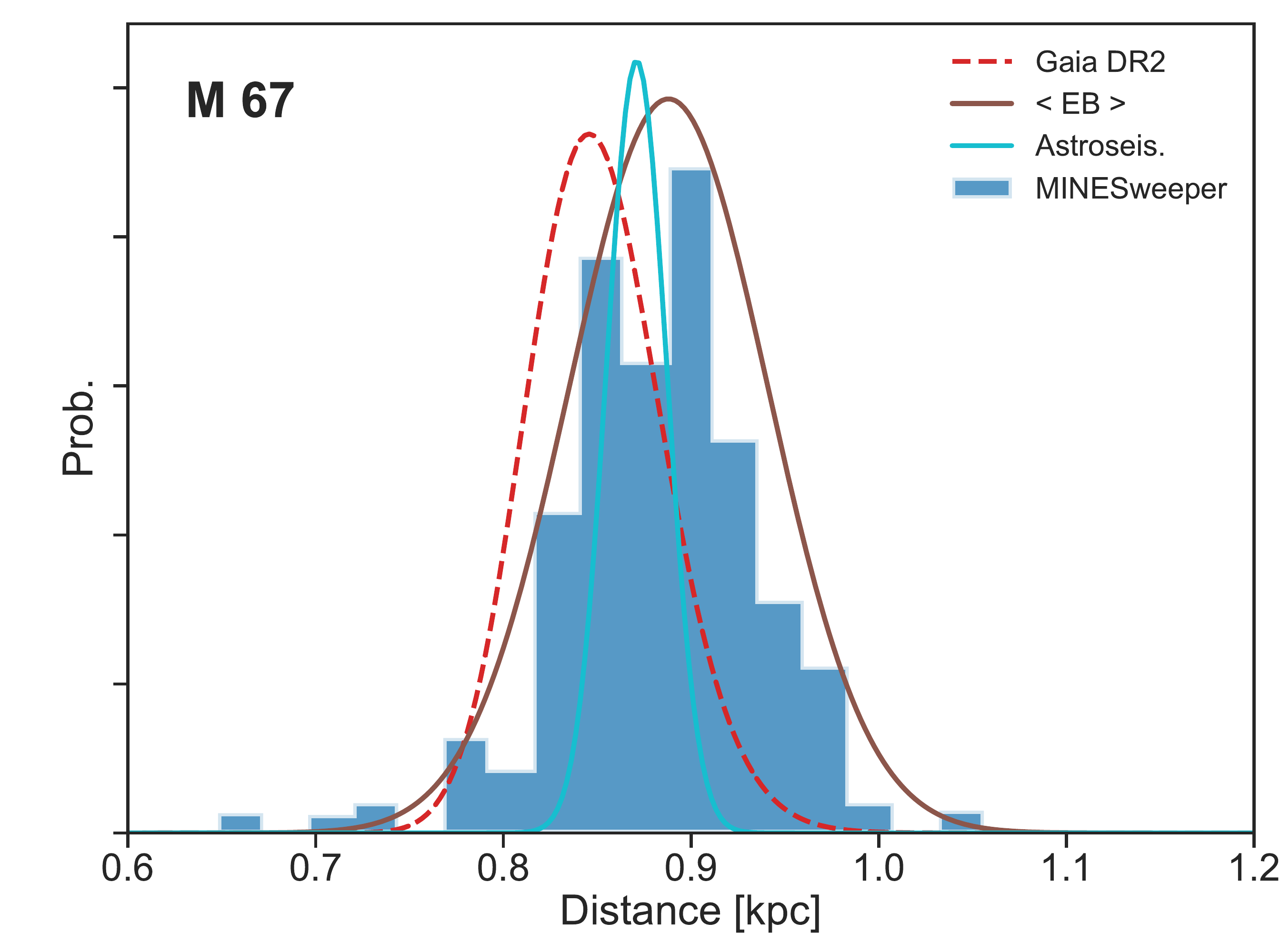}
    \includegraphics[width=0.45\linewidth,height=0.25\textheight]{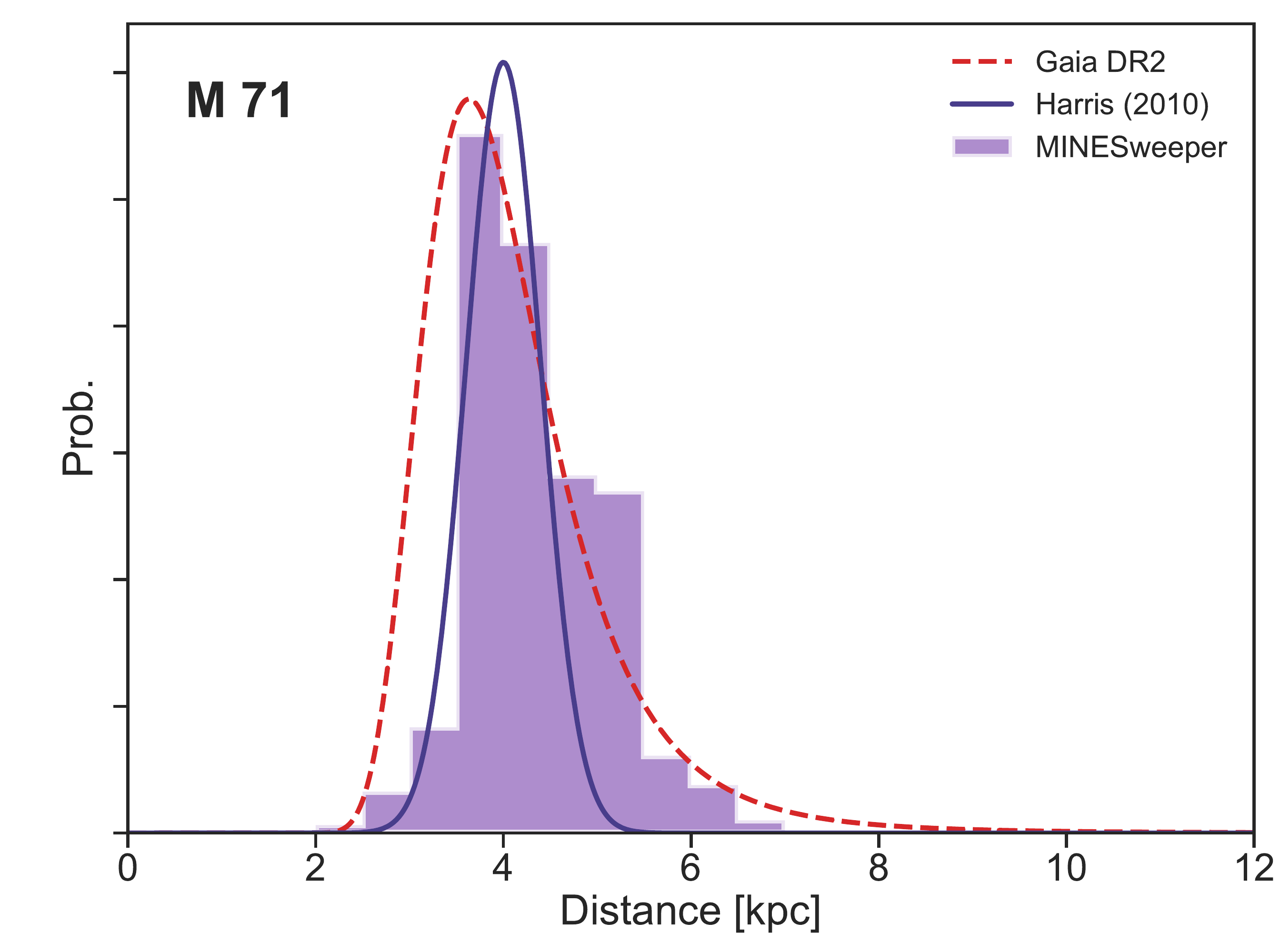}
    \includegraphics[width=0.45\linewidth,height=0.25\textheight]{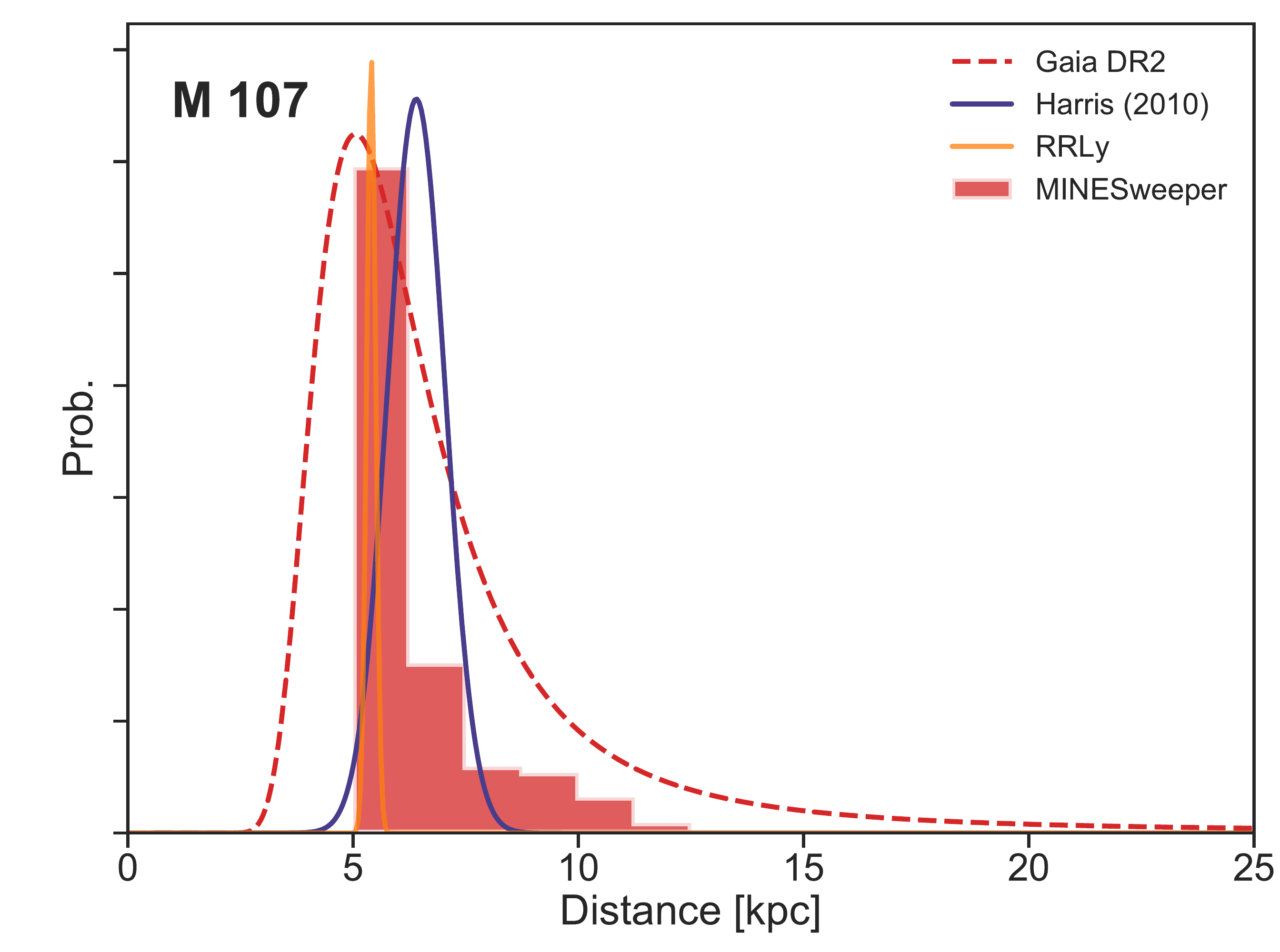}
    \includegraphics[width=0.45\linewidth,height=0.25\textheight]{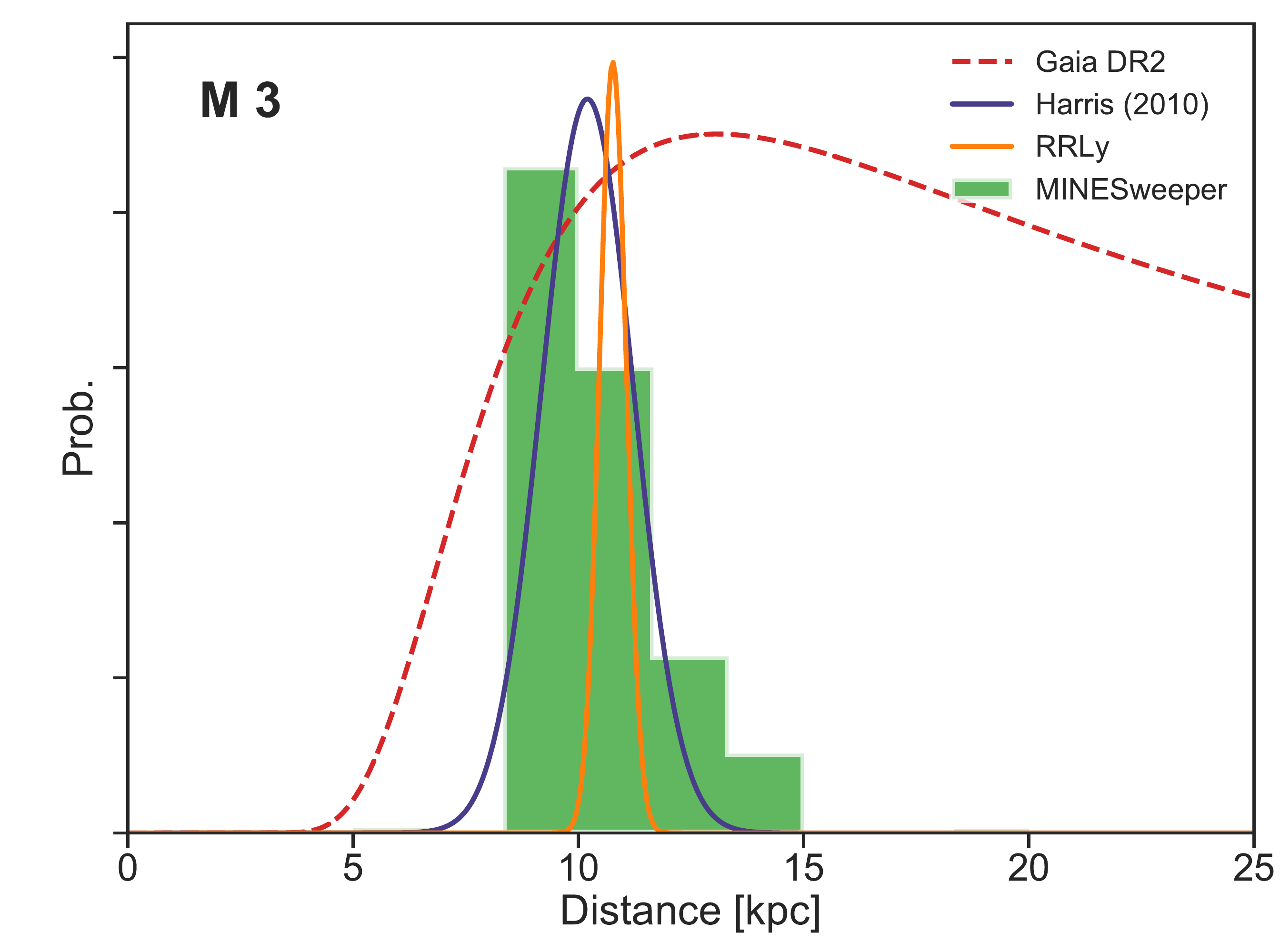}
    \includegraphics[width=0.45\linewidth,height=0.25\textheight]{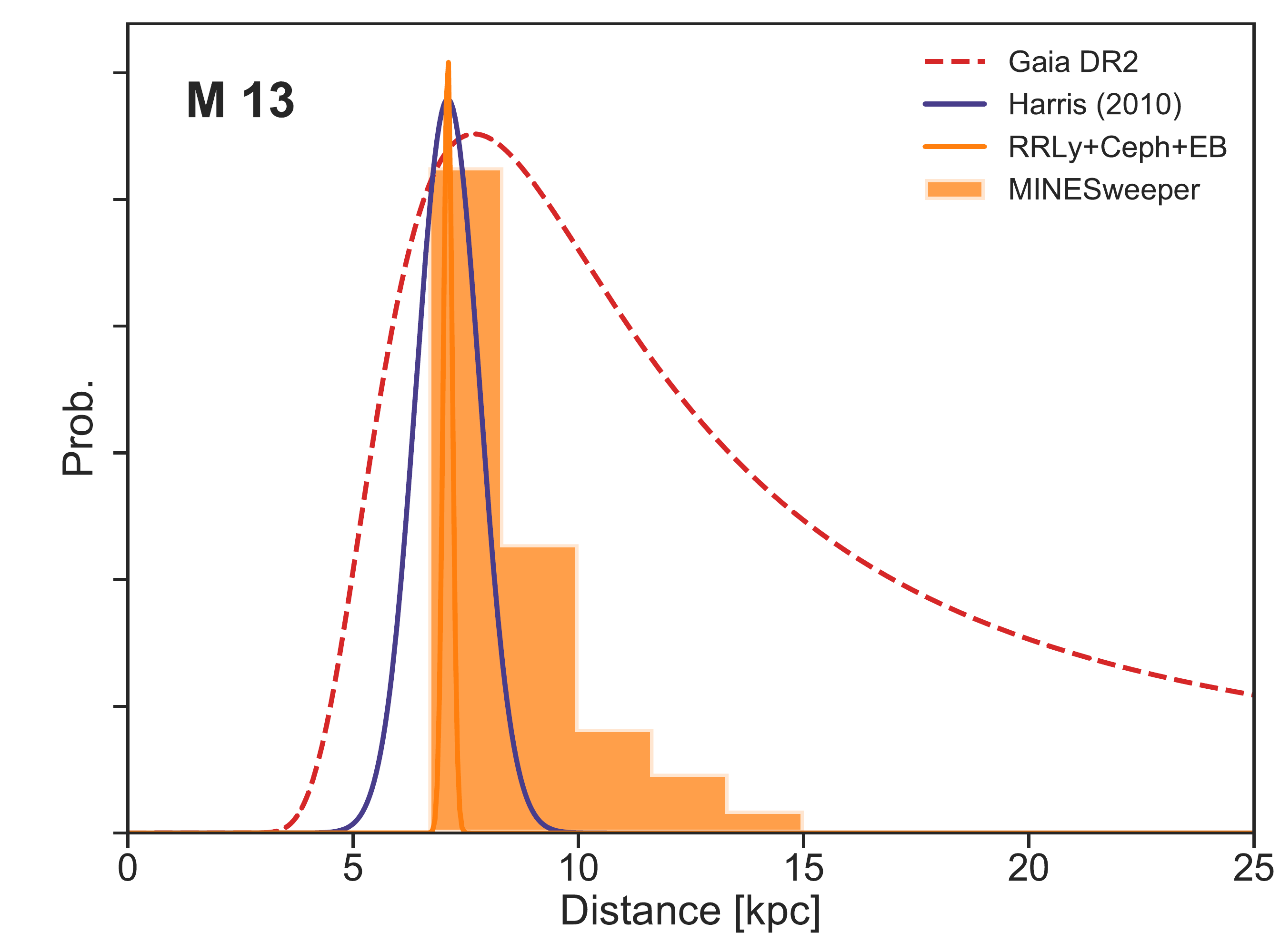}
    \includegraphics[width=0.45\linewidth,height=0.25\textheight]{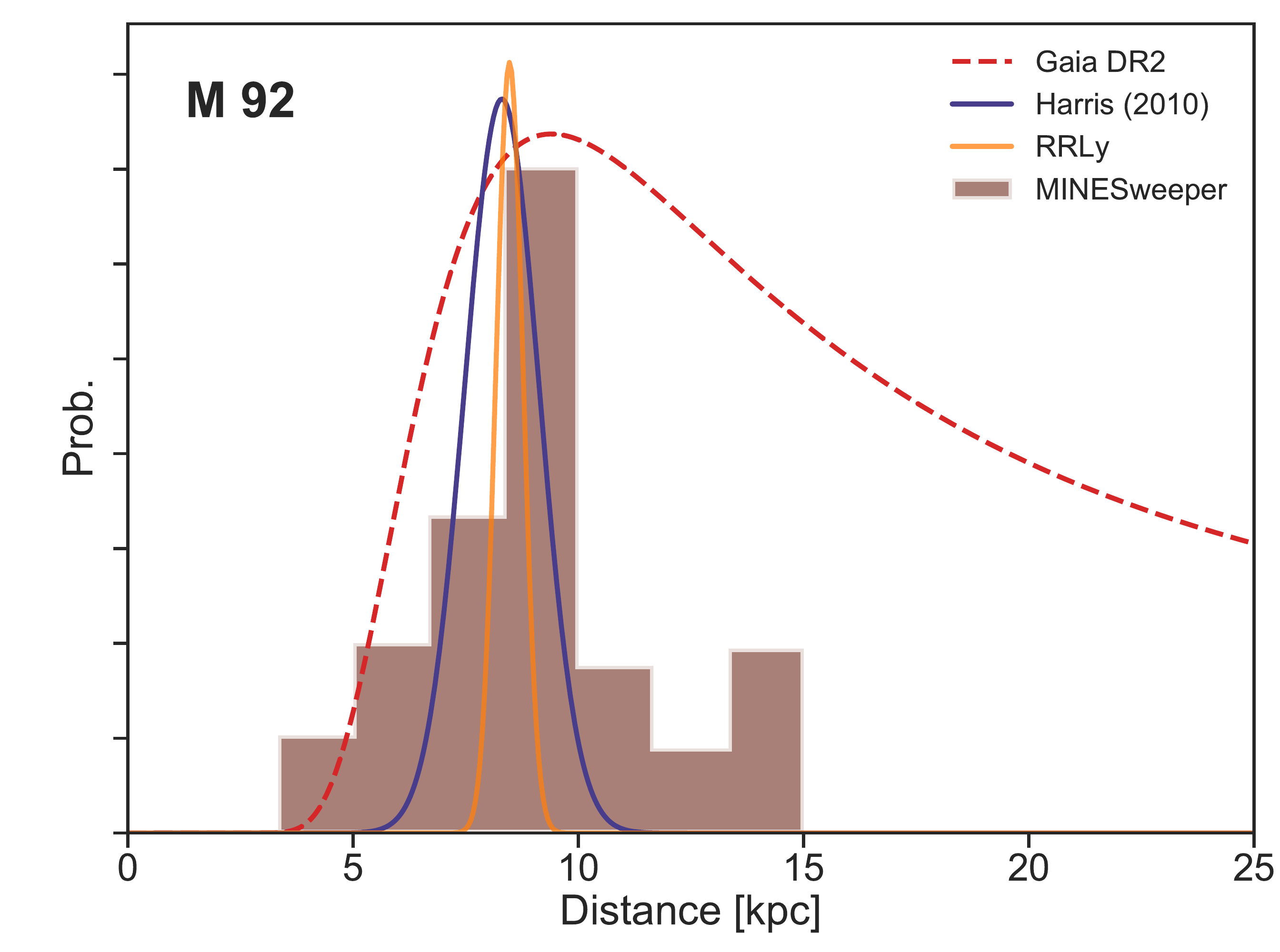}
    \caption{Derived distances for stars in six star clusters. \MS-derived distances include the {\it Gaia} DR2 priors. Individual stellar distances have been weighted by their \MS--derived posterior probabilities. The red dashed line shows the distance constraints based on {\it Gaia} DR2 parallaxes alone \citep[][]{GaiaHRD2018,GaiaGC2018}.  The other lines are for cluster distances measured using alternative techniques (see text for references).  The \MS-derived distances are in good agreement with the existing independent constraints.}
    \label{fig.cluster.dist}
\end{figure*}

\begin{figure*}[ht!]
    \centering
    \includegraphics[width=0.45\linewidth,height=0.25\textheight]{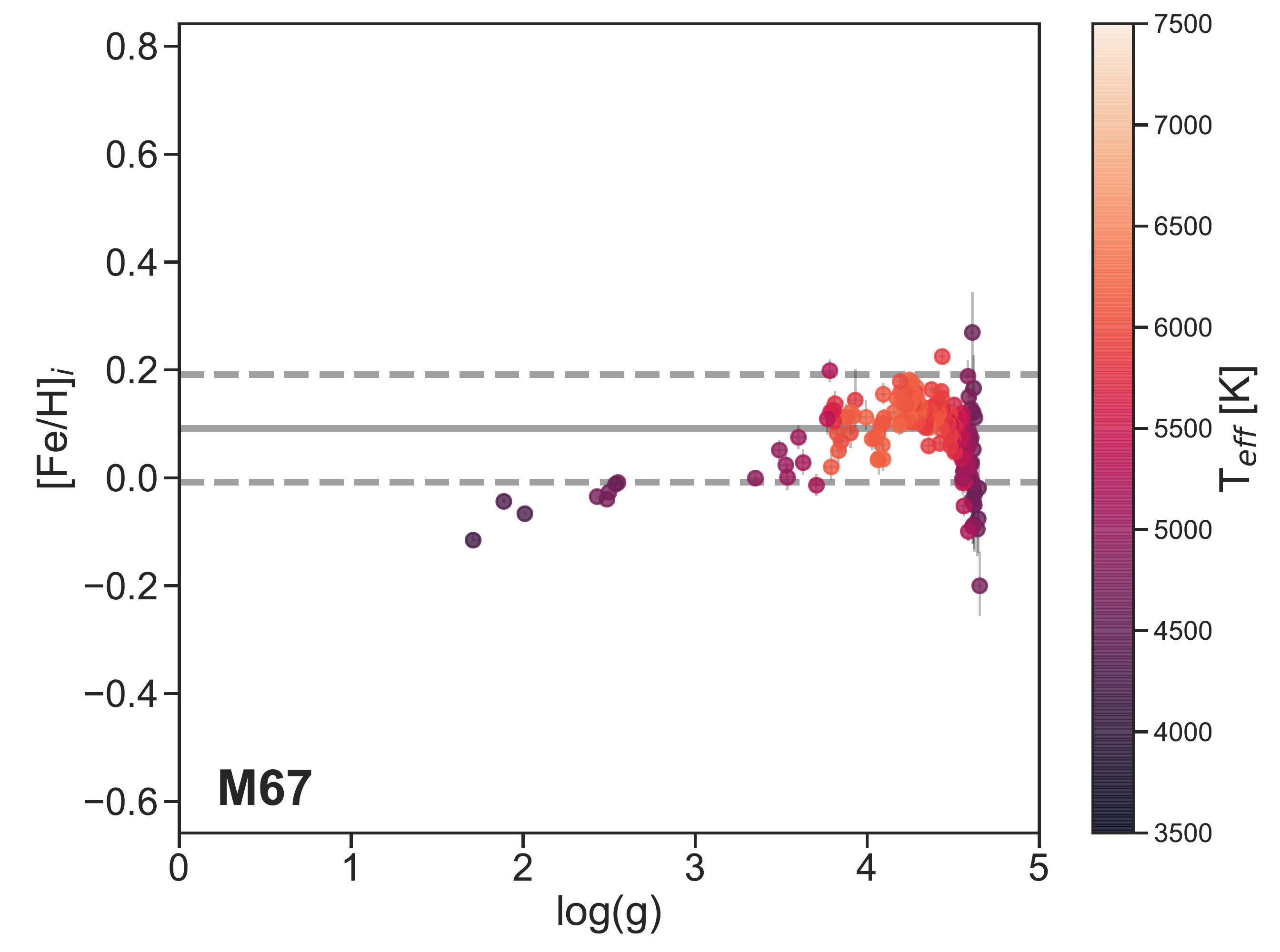}
    \includegraphics[width=0.45\linewidth,height=0.25\textheight]{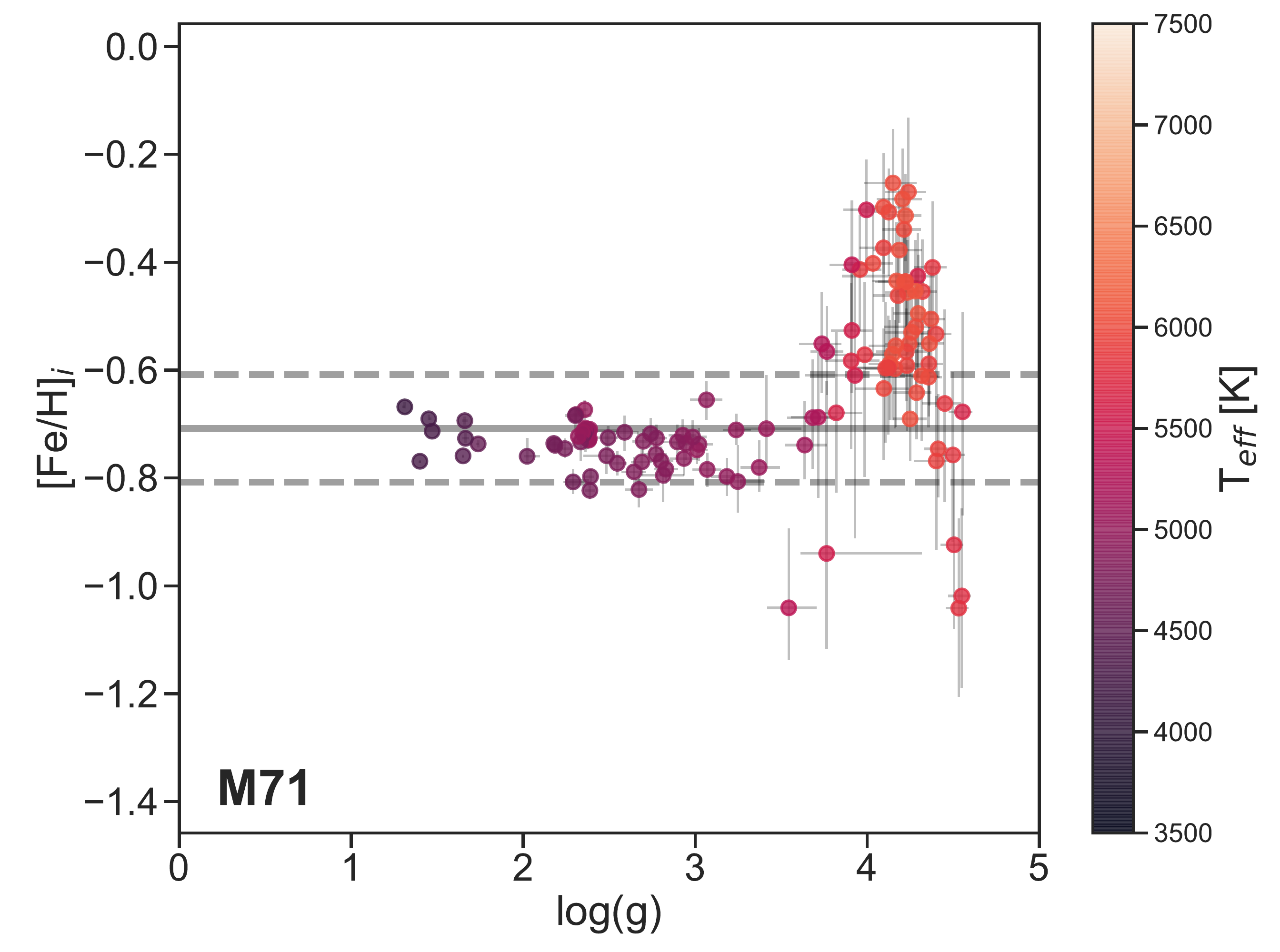}
    \includegraphics[width=0.45\linewidth,height=0.25\textheight]{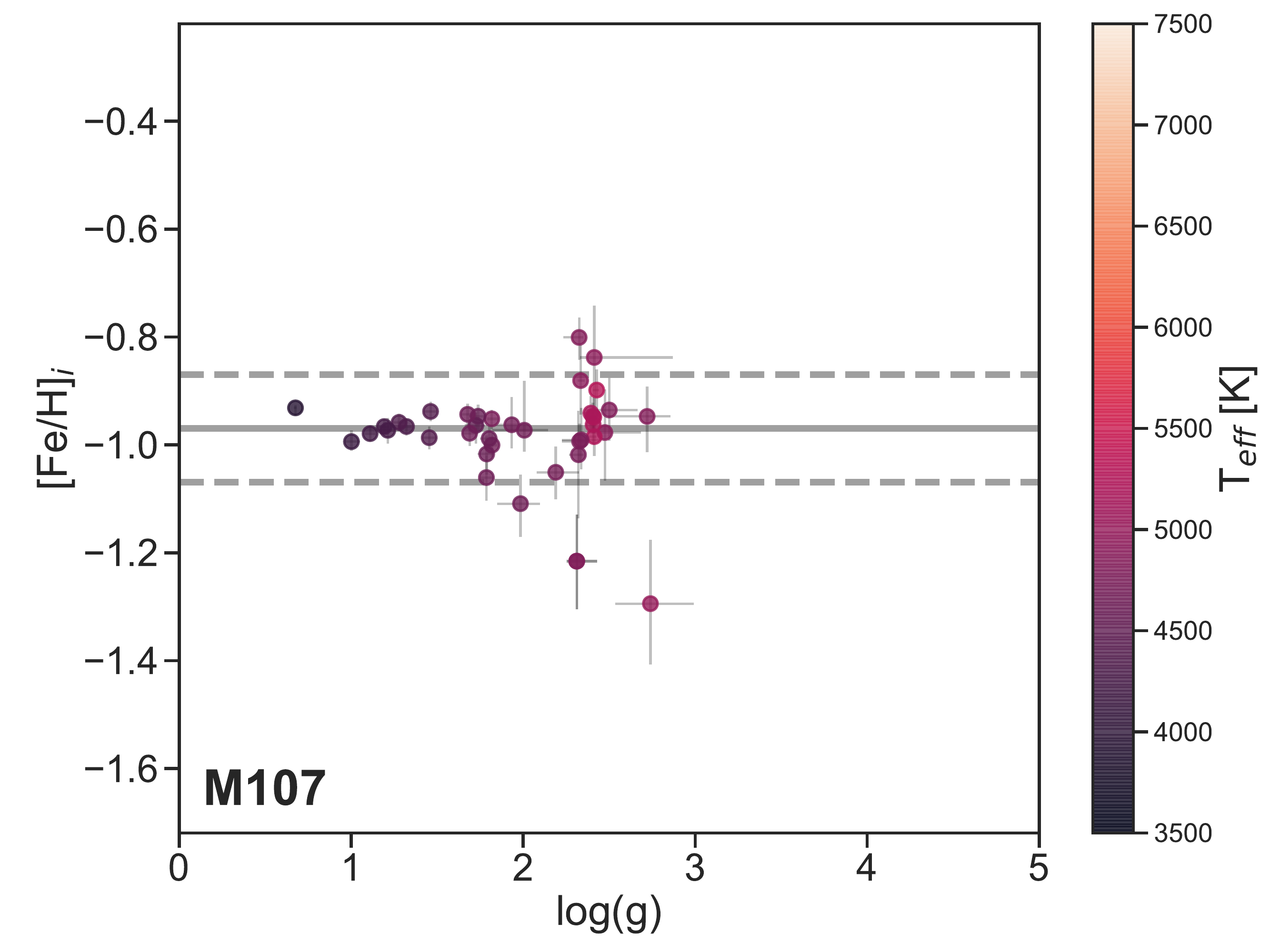}
    \includegraphics[width=0.45\linewidth,height=0.25\textheight]{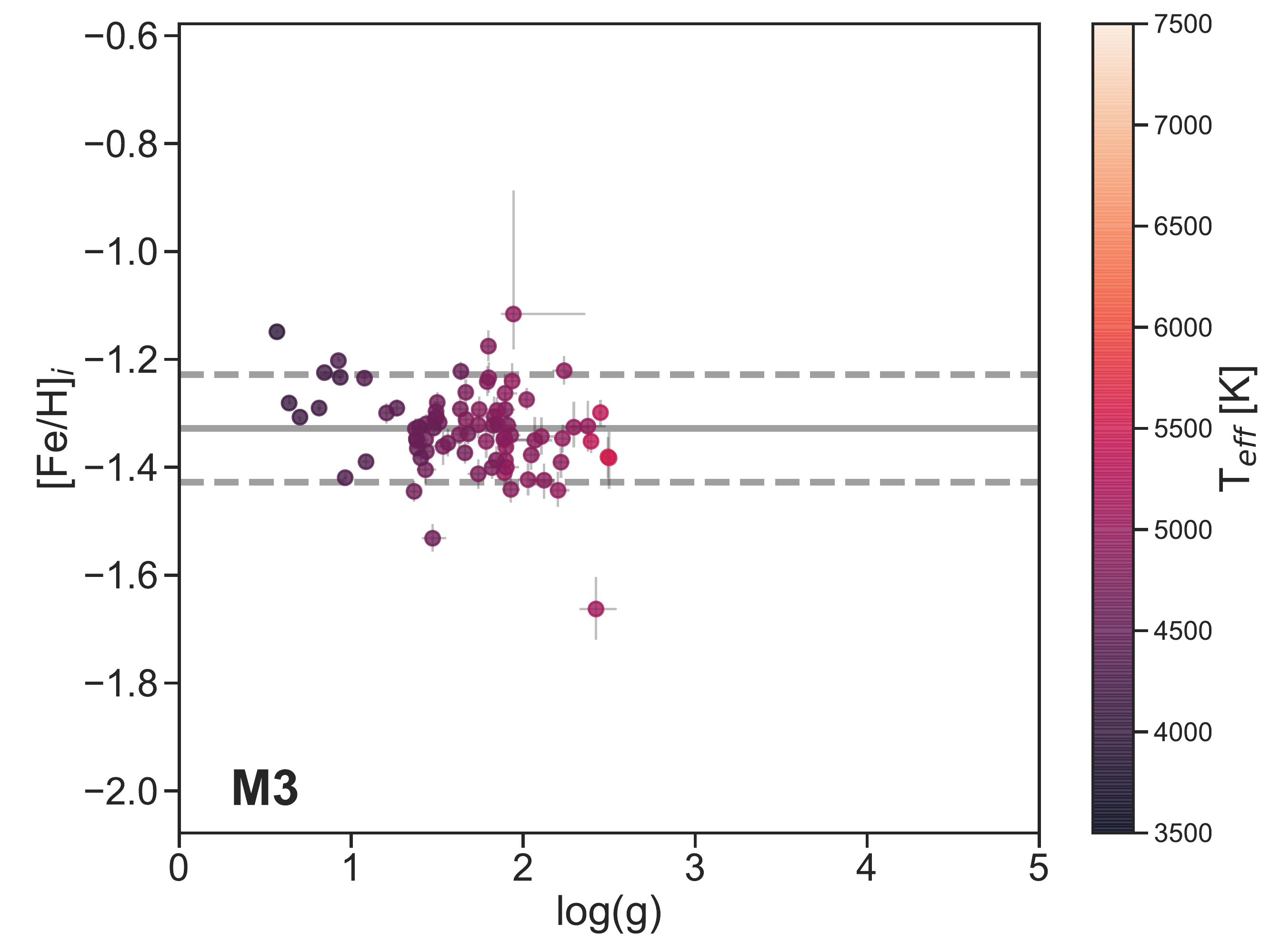}
    \includegraphics[width=0.45\linewidth,height=0.25\textheight]{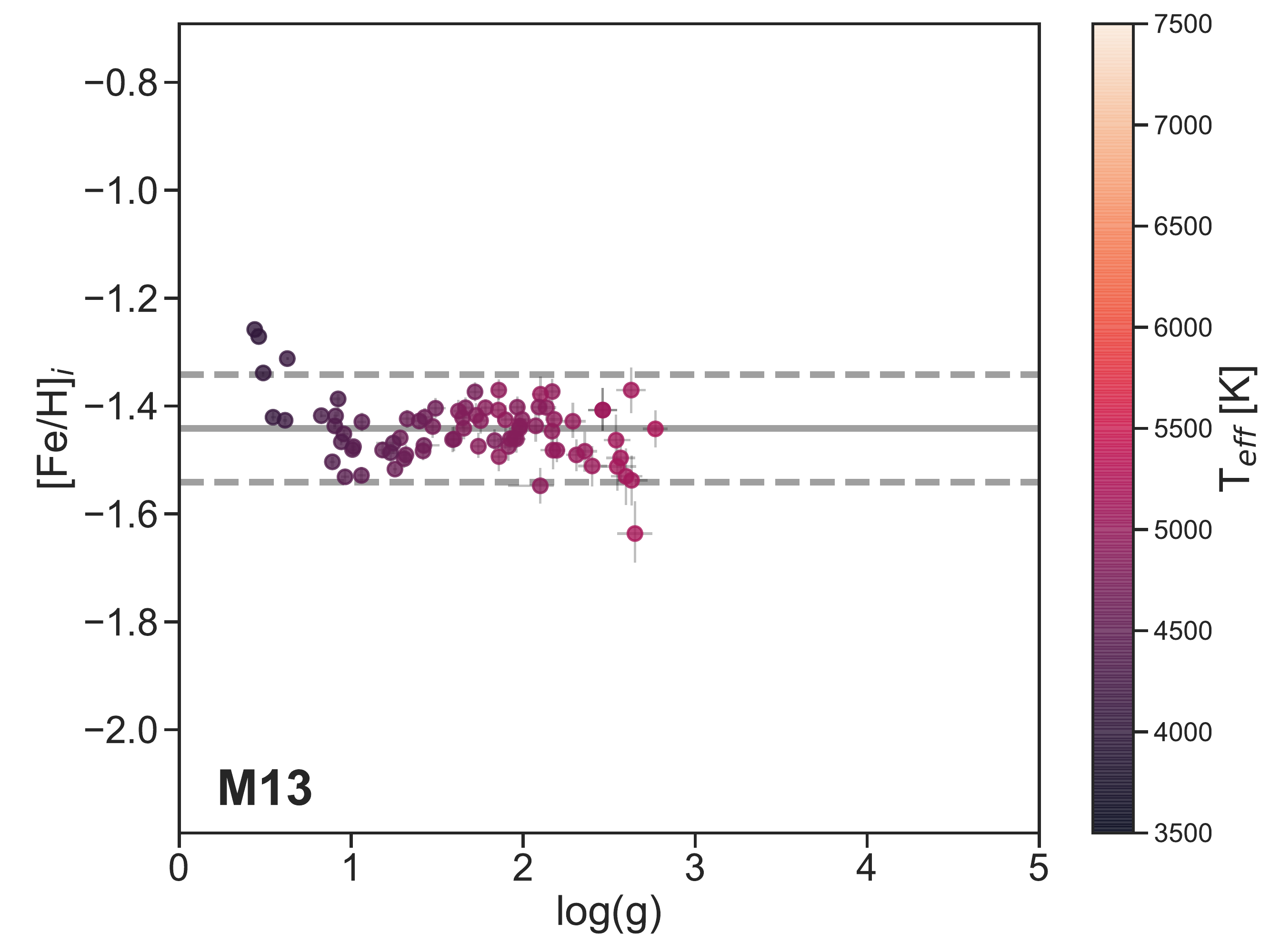}
    \includegraphics[width=0.45\linewidth,height=0.25\textheight]{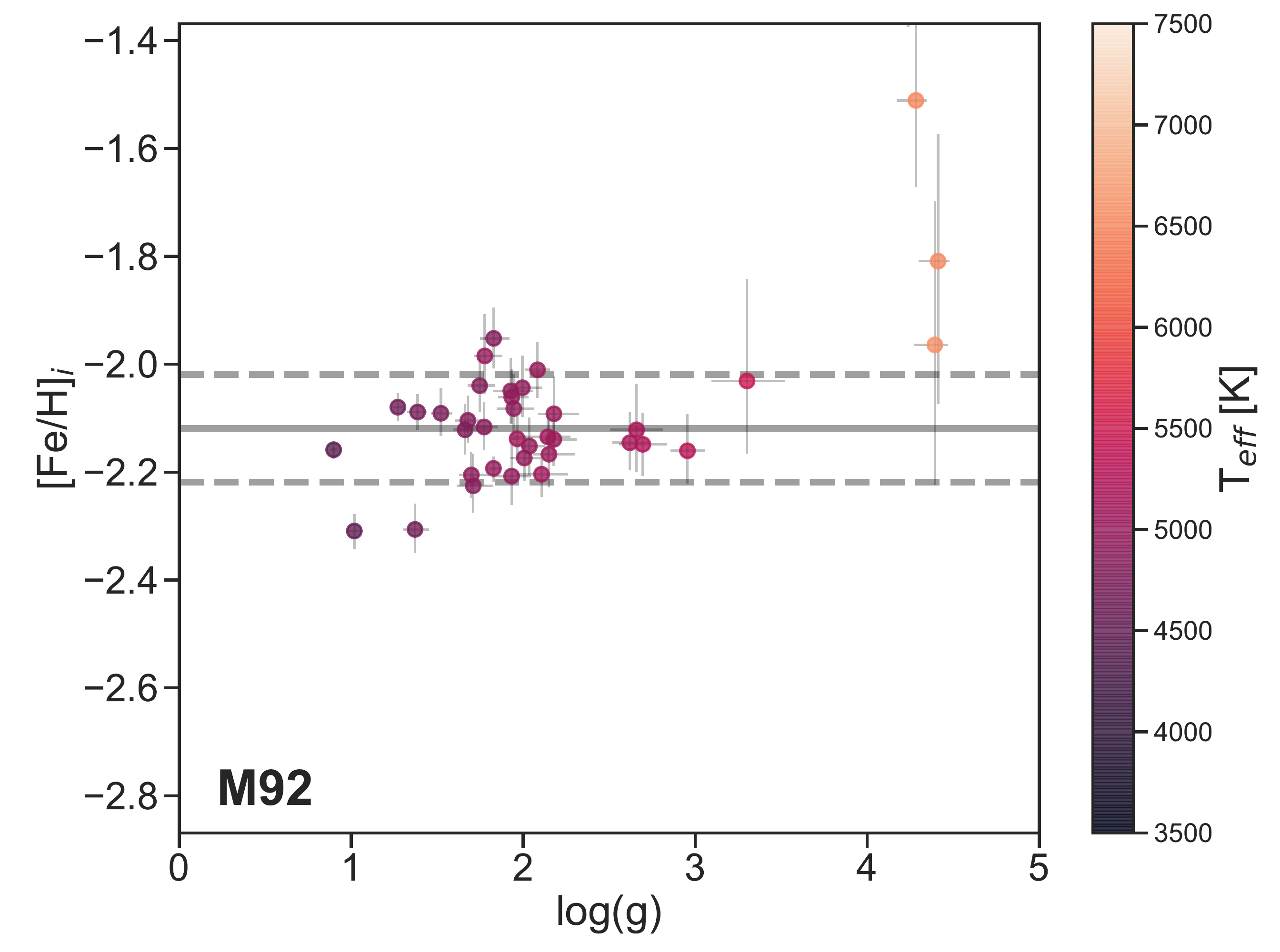}
    \caption{Derived metallicities as a function of \logg\ for six star clusters.  Plotted on the y-axis is the initial stellar \feh\, as opposed to the more commonly reported surface abundance. The color-scale of the points indicates the \teff\, of the stars. The dashed black lines indicate the weighted median \feh$_i$ for each cluster and the grey dashed lines mark $\pm0.1$ around the median value and is intended to guide the eye.}
    \label{fig.cluster.feh}
\end{figure*}

\begin{figure*}[ht!]
    \centering
    \includegraphics[width=0.45\linewidth,height=0.25\textheight]{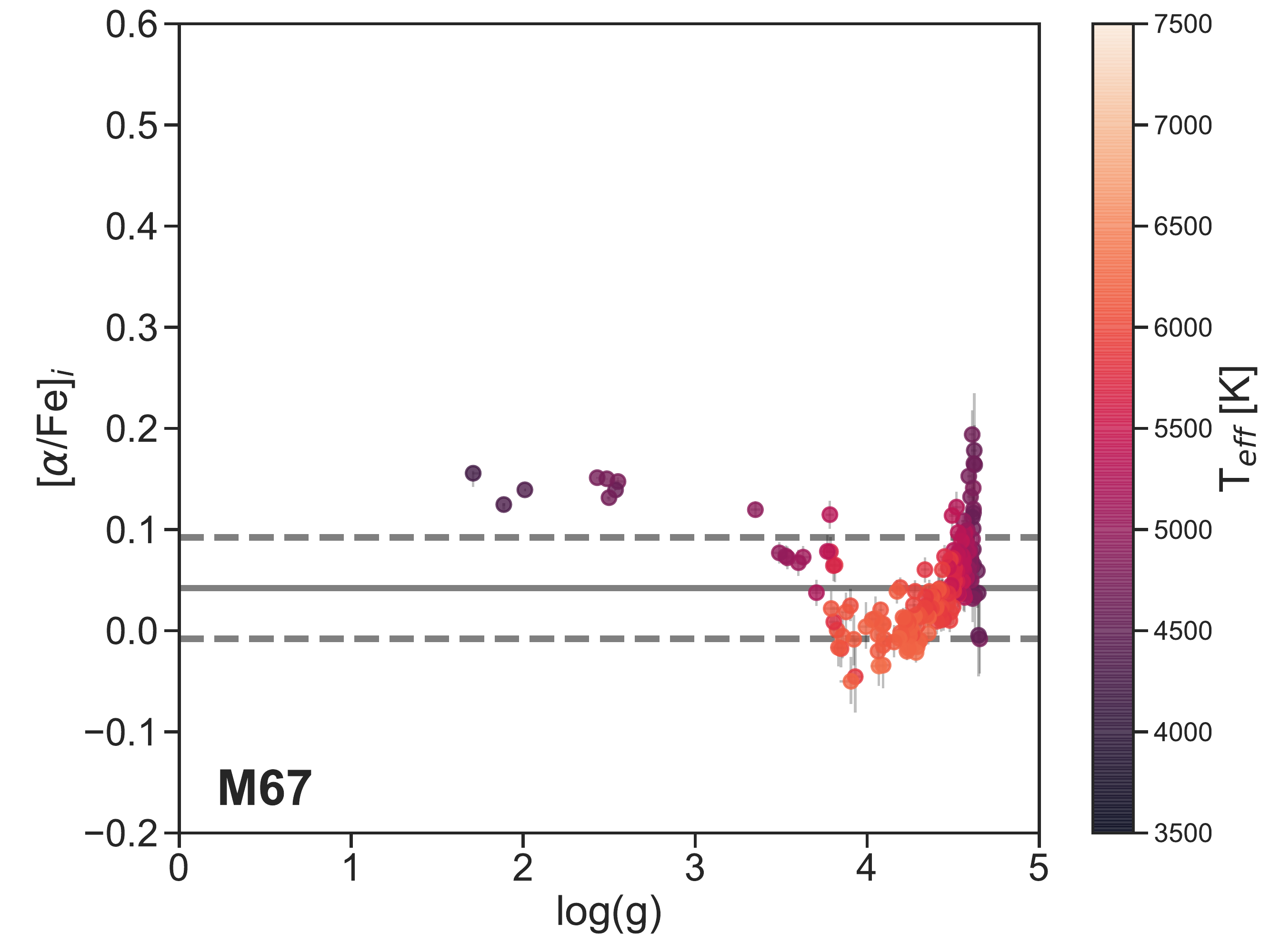}
    \includegraphics[width=0.45\linewidth,height=0.25\textheight]{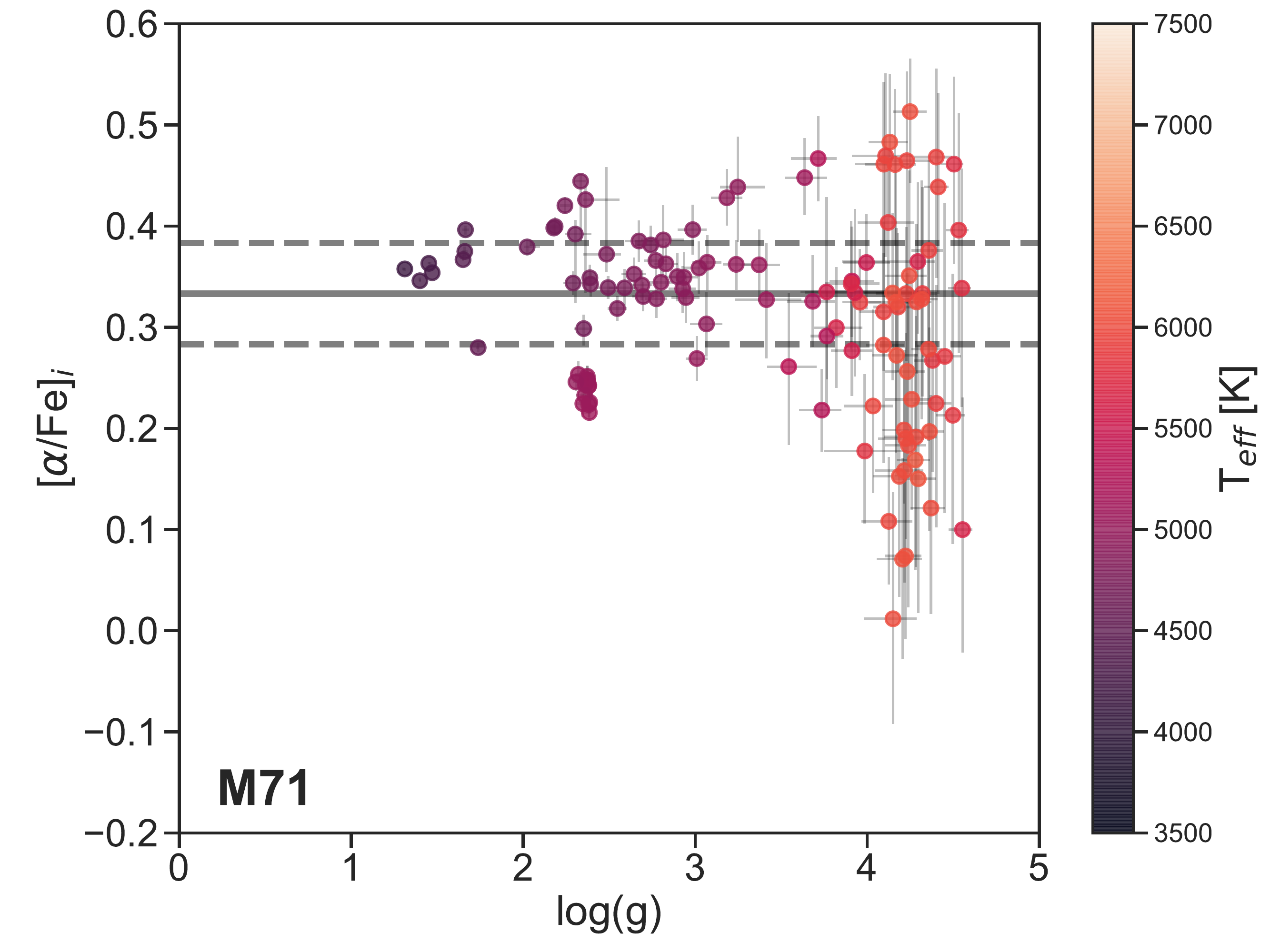}
    \includegraphics[width=0.45\linewidth,height=0.25\textheight]{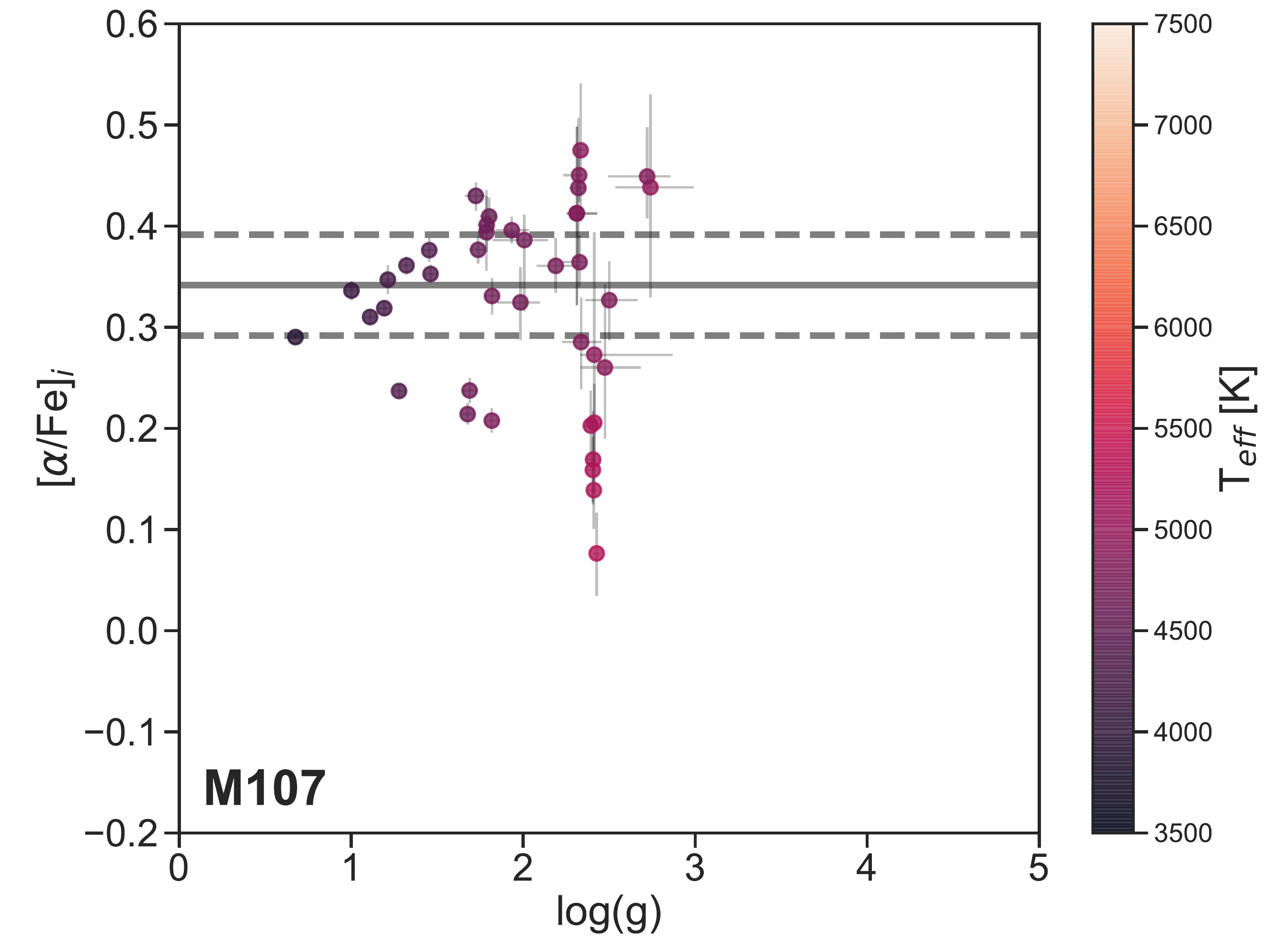}
    \includegraphics[width=0.45\linewidth,height=0.25\textheight]{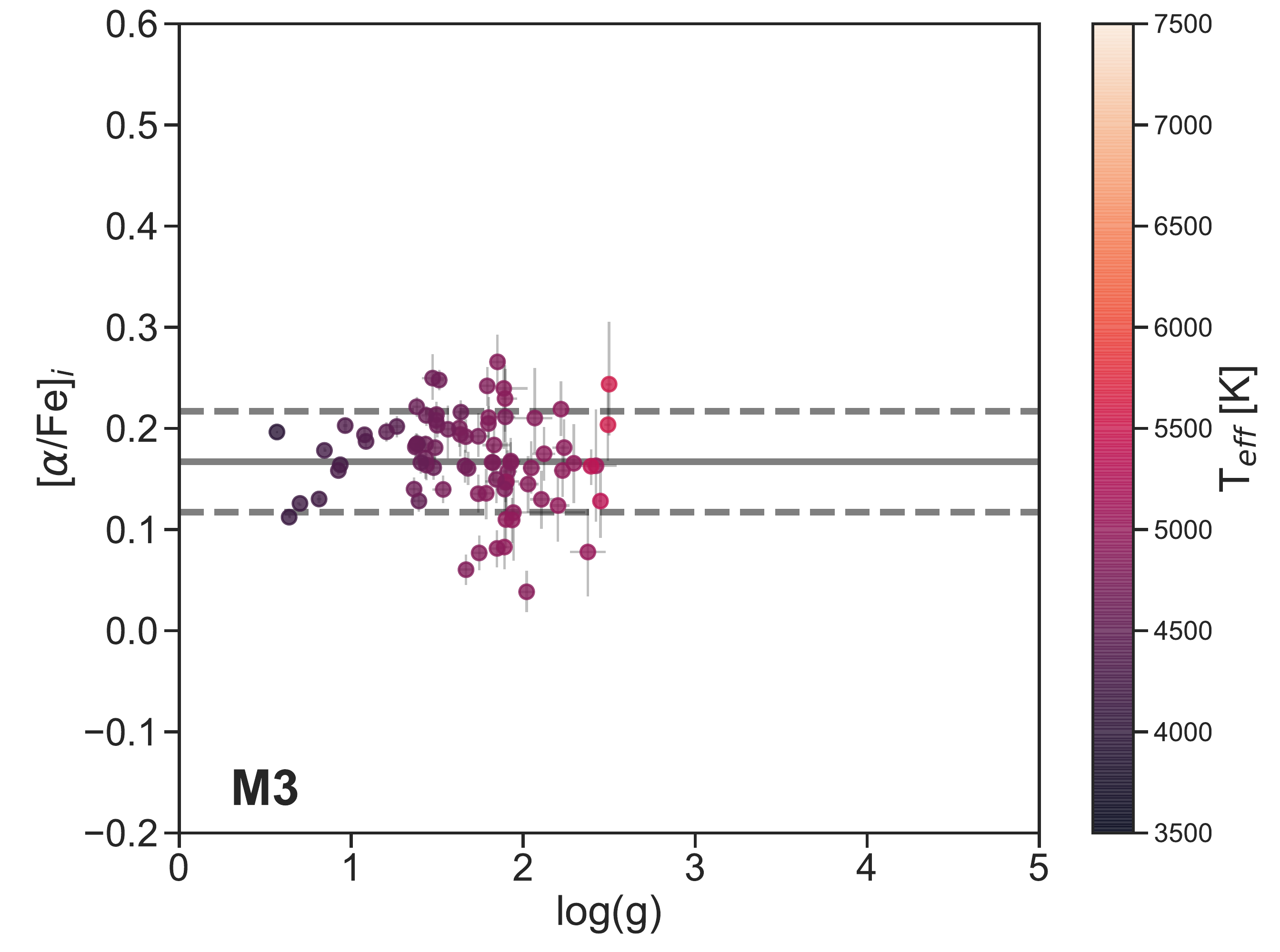}
    \includegraphics[width=0.45\linewidth,height=0.25\textheight]{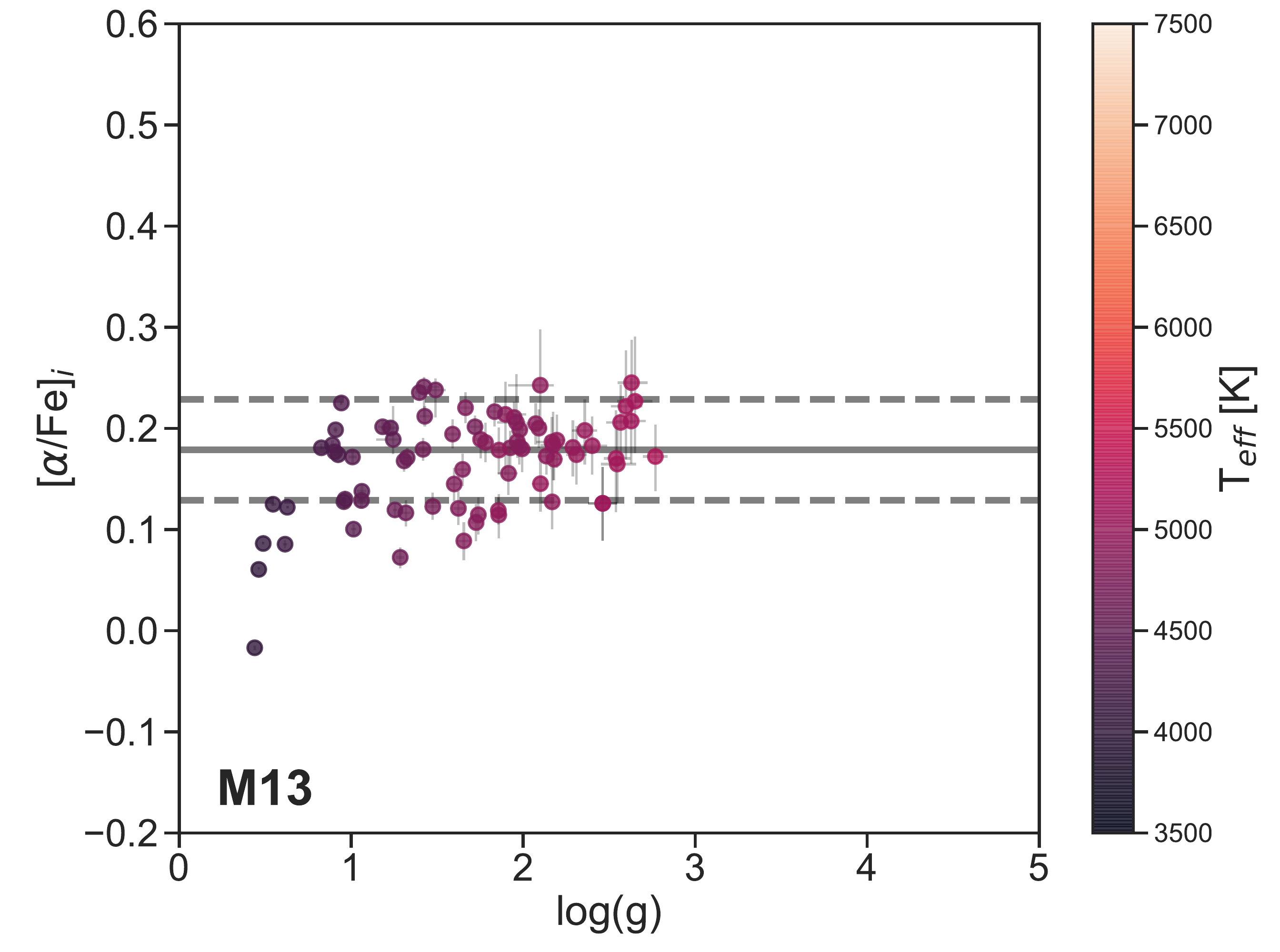}
    \includegraphics[width=0.45\linewidth,height=0.25\textheight]{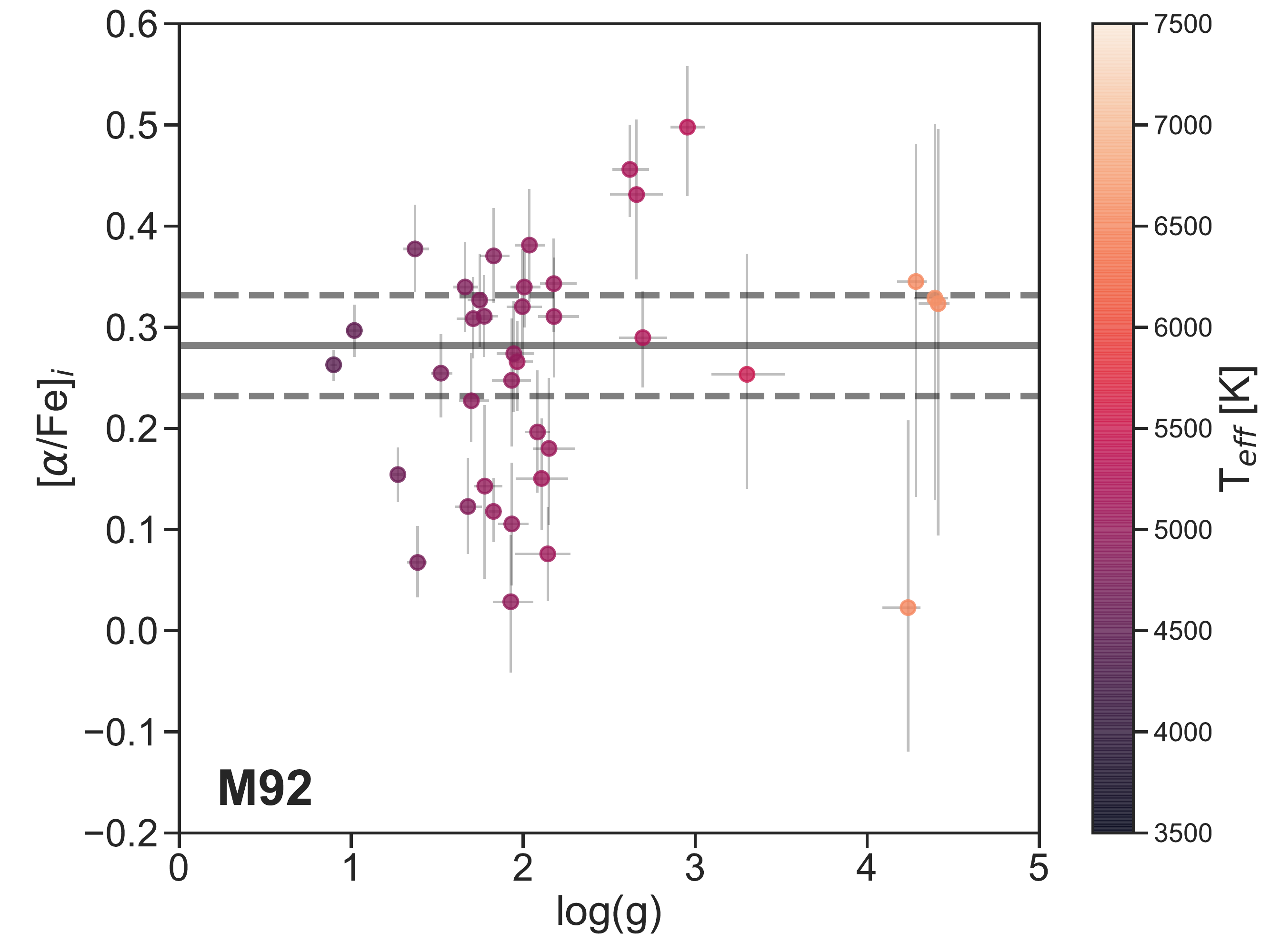}
    \caption{As in Figure \ref{fig.cluster.feh}, but now for \afe\, as a function of \logg.  Grey dashed lines mark $\pm0.05$ around the median value for each the cluster and is intended to guide the eye.}
    \label{fig.cluster.afe}
\end{figure*}

\begin{figure*}[htb!]
    \centering
    \includegraphics[width=0.9\linewidth]{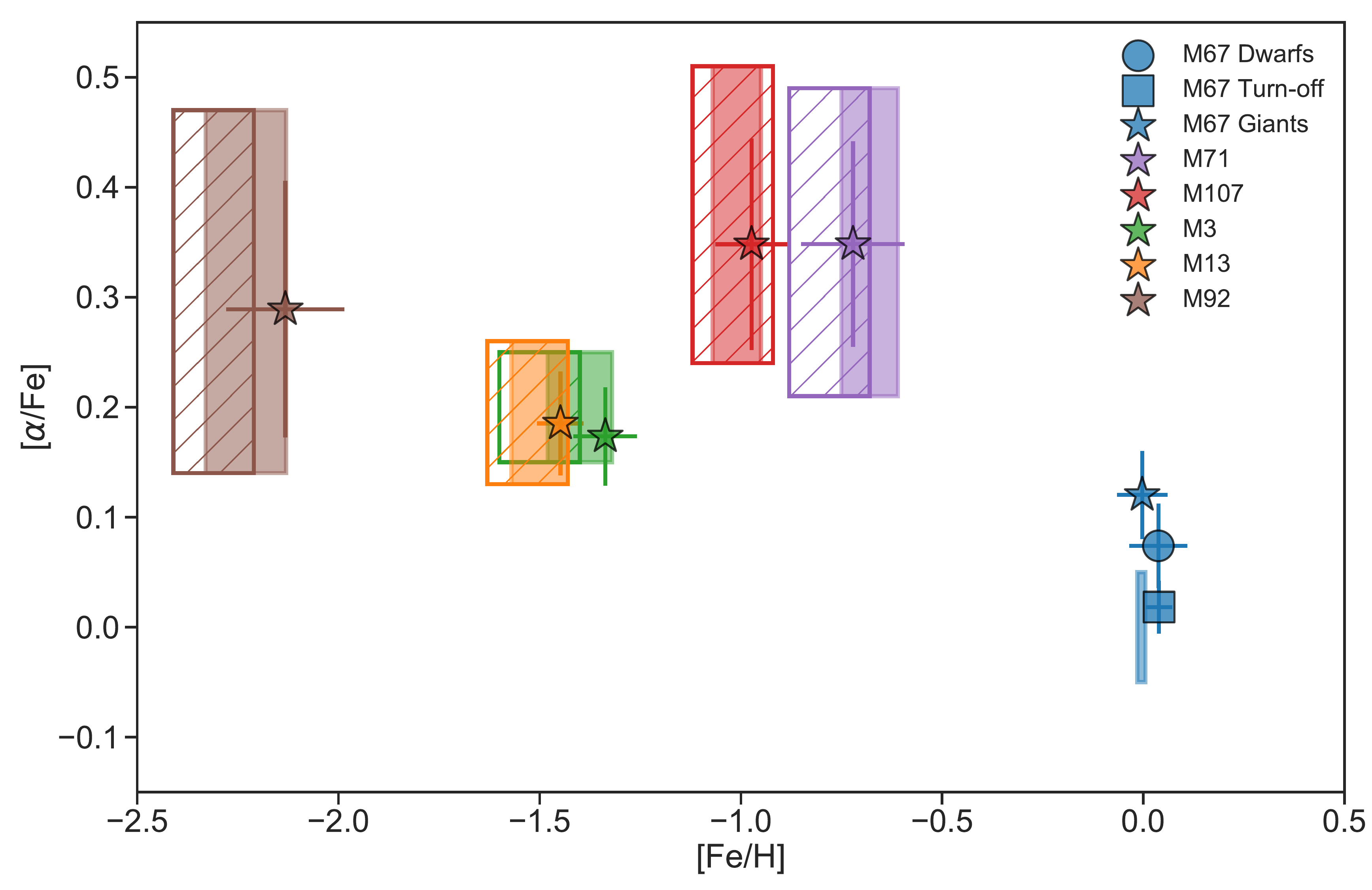}
    \caption{
    Distributions of surface metallicity and $\alpha$-abundance (\afe) for the open cluster M67 (blue) and globular clusters M71 (purple), M107 (red), M3 (green), M13 (orange), and M92 (brown). The star points and associated error bars are the weighted median and $1\sigma$ values from the \MS\, analysis. The range of literature values for each cluster are given by the shaded and hatched regions \citep[for the metallicity scales of][respectively]{Meszaros2015,Carretta2009}. The M67 results are presented separately for dwarfs (circle), turn-off stars (square), and giants (star).  Our \feh\ values are in good agreement with the literature, given the $\approx0.1$ dex systematic uncertainty in the overall metallicity scale.  It is difficult to draw meaningful conclusions from the comparison in \afe\ given the large range in literature estimates and the inconsistent definitions of ``$\alpha$".}
    \label{fig.cluster.afefeh}
\end{figure*}

 \begin{figure*}[htb!]
    \centering
    \includegraphics[width=0.33\linewidth]{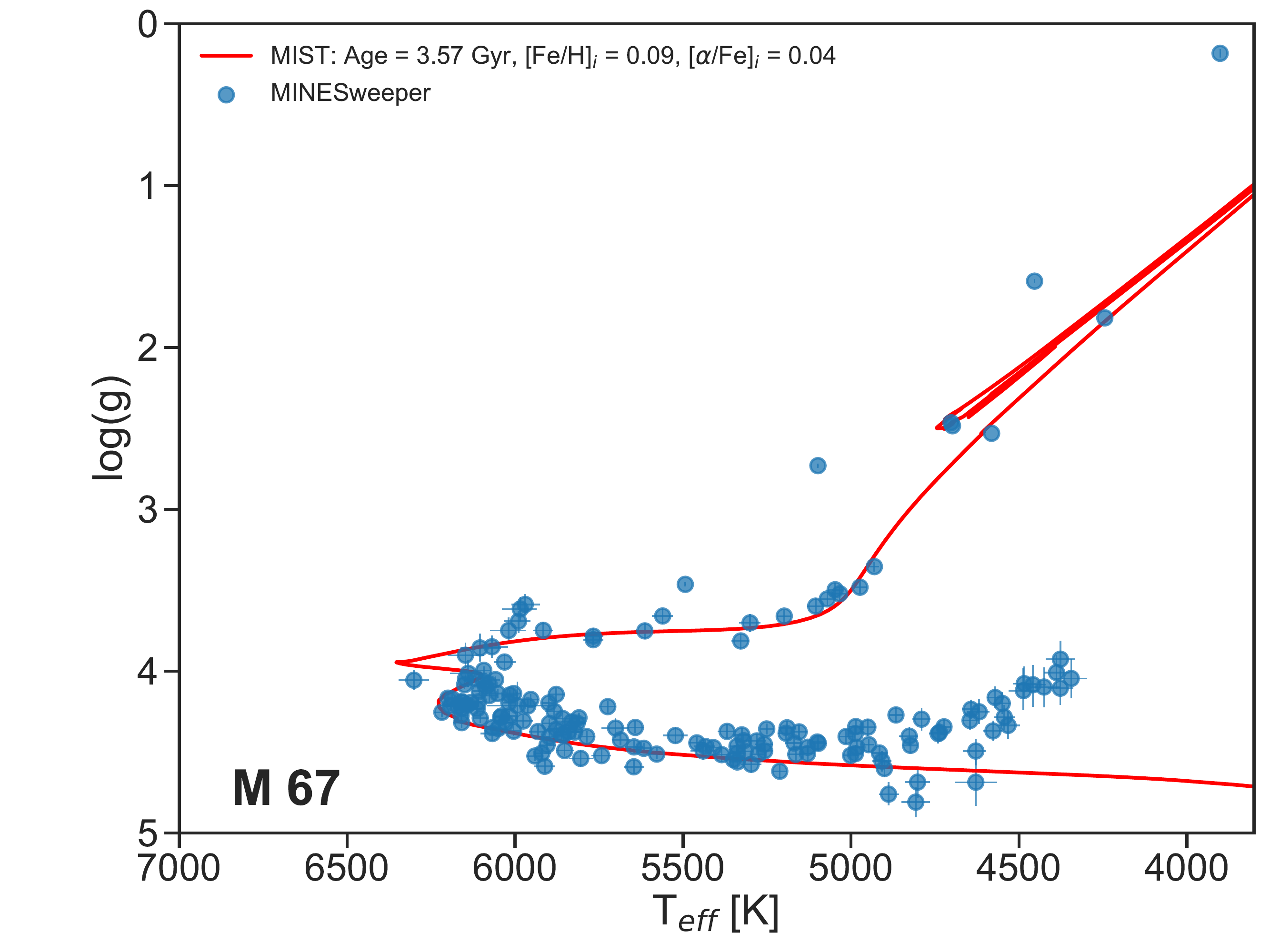}
    \includegraphics[width=0.33\linewidth]{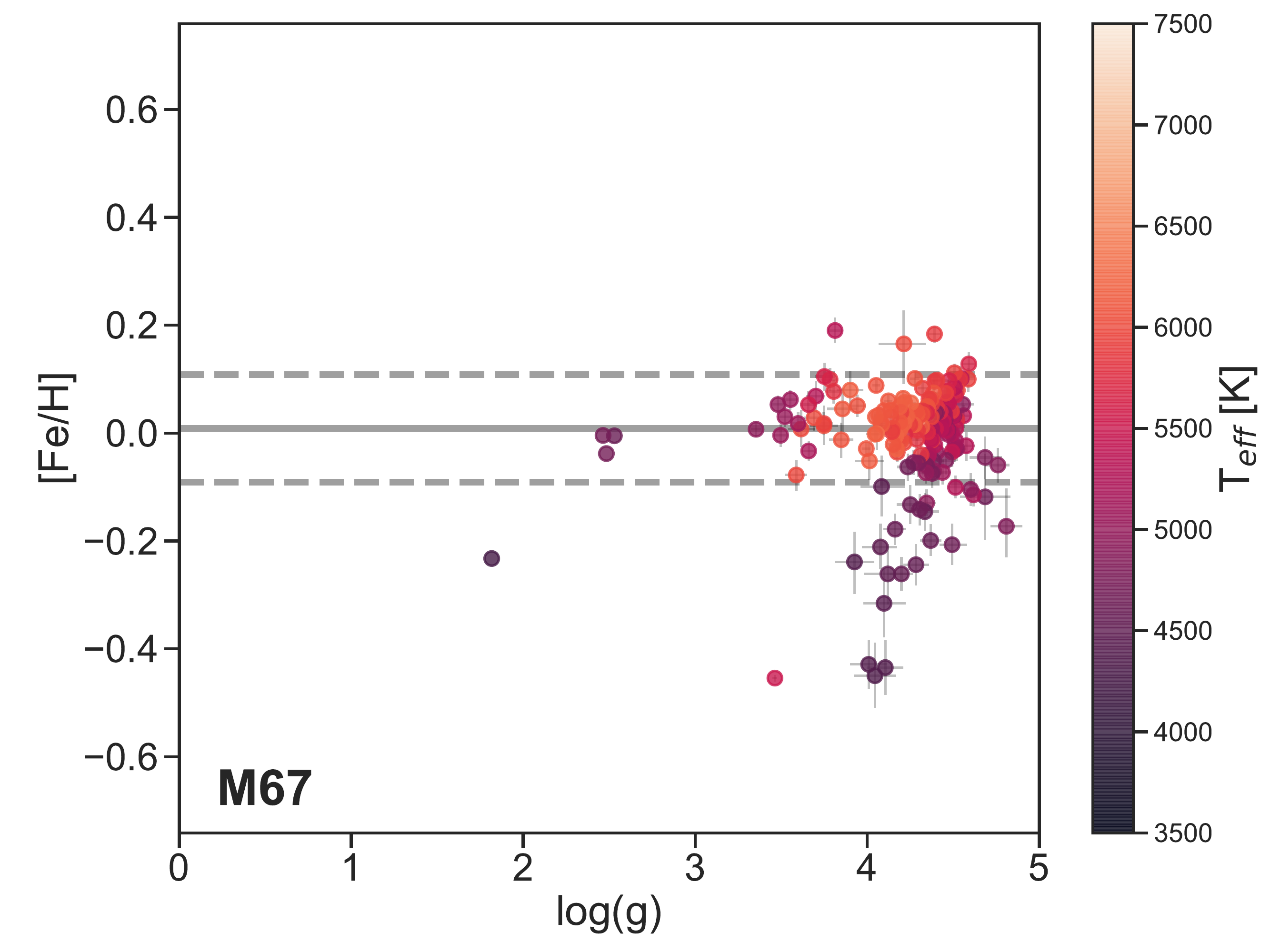}
    \includegraphics[width=0.33\linewidth]{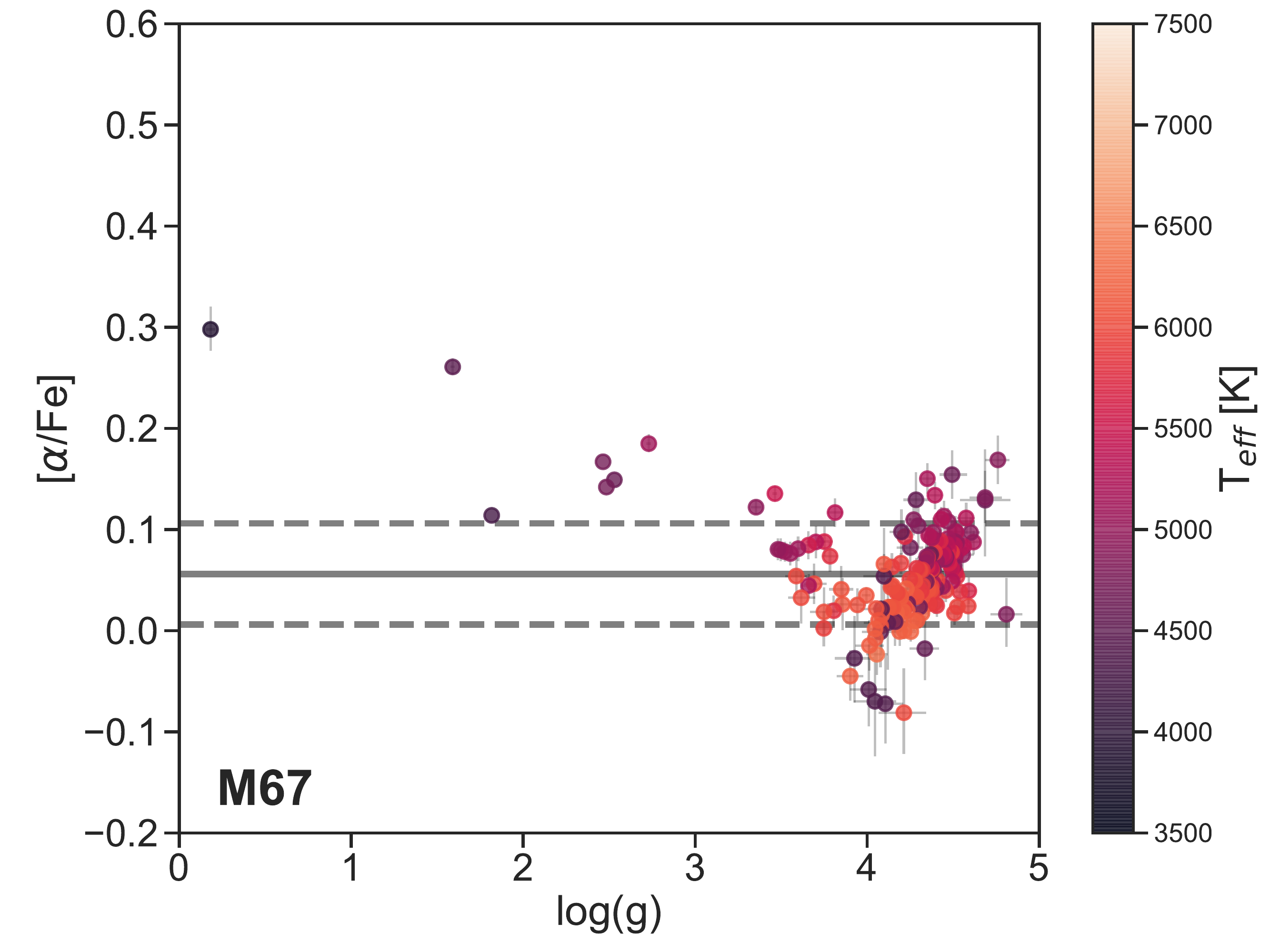}
    \caption{Stellar parameters for M67 cluster members derived without isochrone constraints.  The Kiel diagram (left panel) shows reasonable agreement with the expected isochrone locus (red line), though there are several mis-placed giants, and an upturn in \logg\ amongst the coolest dwarfs (compare with Figure \ref{fig.cluster.Kiel}).  Surface \feh\ (center panel), and surface \afe\ (right panel) are shown as a function of \logg\ (compare with Figures \ref{fig.cluster.feh} and \ref{fig.cluster.afe}).  The star with \logg $<1$ has \feh $<-1$, which is below the bottom range of the middle panel.  It is noteworthy that the trend between \afe\ and \logg\ is very similar to the case with isochrone constraints (Figure \ref{fig.cluster.afe}), suggesting that that this behavior is due to a limitation in our spectral models.
    }
    \label{fig.cluster.m67tp}
\end{figure*}

\subsection{Star Clusters}
\label{sec.modtest.clusters}

Another important validation test for \MS\, is to fit stars belonging to well studied star clusters.  These populations offer important tests for several reasons: first, a subset of open and globular clusters have been extensively studied for decades and so their spectroscopic parameters and distances are relatively well-known.  Second, stars within a cluster are all at approximately the same distance.  And third, all stars within a cluster were born with the same abundances, so inspection of \feh$_i$ and \afe$_i$ as a function of \teff\, and \logg\, offer a very strong test of the precision of the reported abundances.  This third point is complicated by two effects: first, element diffusion and mixing of the surface layers of a star will result in differences between photospheric abundances and the initial abundances of a star \citep[see e.g.,][]{Dotter2017}.  \texttt{MIST} includes both of these effects, so our reported initial abundances should be independent of evolutionary phase.  Second, it is now well-known that globular clusters harbor substantial light element abundance variations {\it within} individual clusters \citep[e.g.,][]{Gratton2004}.  This includes variation in helium and oxygen, which can significantly alter a star's location in the HR diagram.  These caveats must be kept in mind when using globular clusters as calibrators.

\subsubsection{Data}

Here we fit data collected for members of clusters spanning a wide range of metallicities: globular clusters M92, M13, M3, M107, and M71, as well as the solar-metallicity open cluster M67. Literature parameters for each individual cluster are given in Table \ref{tab.clusters}. For the globular clusters, the \feh\, values are taken from \citet[][]{Meszaros2015} by default, with a nominal uncertainty of 0.05 dex.  We also quote \feh\, values from \citet{Carretta2009} in some cases to highlight the literature uncertainty associated with the metallicity scale.  Ages for the globular clusters are adopted from \citet[][]{Dotter2010}. For the open cluster M67, literature \feh\, and \afe\, are adopted from \citet[][]{Liu2016}, while a literature age is adopted from \citet[][]{Stello2016} who estimate the age from astroseismology of M67 giants.  

For \afe, we adopt a range for each cluster, determined by the literature compilation of \citet{Pritzl2005} and the results presented in \citet{Carretta2009} and \citet{Meszaros2015}.  The latter two authors measure [Mg/Fe], which we adopt as a proxy for \afe.  We note that it is difficult to report a single value of ``\afe" for a cluster for two reasons.  First, the $\alpha$-elements do not vary in lock-step and so formally ``\afe" should be some appropriate mass-weighted average.  Second, the light $\alpha$ elements (e.g., O, Mg) show star-to-star variation within clusters, in some cases with $\approx1$ dex variation \citep[e.g.,][]{Carretta2009b}.  So a large sample of stars must be studied in order to report a reliable average for a cluster.   

Medium-resolution spectra were collected for stars in these clusters as part of the calibration phase of the H3 Survey using the Hectochelle instrument on the MMT (see Section \ref{sec.modtest.h3}).  Owing to the varying distances, in some clusters spectra were obtained for stars over a wide range of \logg, while in others spectra were collected only for the giants.  Targets were identified by a spatial selection and the requirement that they reside near an isochrone appropriate for the cluster parameters.

We fit both the spectroscopy and, where available, broadband photometry with \MS.  We include the {\it Gaia} DR2 parallax as a Gaussian prior.  The DR2 parallaxes have known systematic zero-point issues.  Following \citet{Lindegren2018}, \citet{Schonrich2019}, and \citet{Leung2019}, we therefore add an offset of $+0.05$ to the {\it Gaia} parallaxes and also add this value in quadrature to the reported uncertainties.

For each cluster, likely members were identified through a combination of coordinates, radial velocities, and {\it Gaia} proper motions and parallaxes. Stars with SNR$_{\rm spec}<3$ are removed from further analysis.  The median SNR of the cluster members are 9, 15, 20, 25, 31, and 35 for M71, M107, M92, M67, M3, and M13, respectively.  We tabulate the results of our \MS\, modeling for likely members in Table \ref{tab.clusters}, where we provide the weighted median values for the cluster members and the reported errors are the 68th percentile range for each parameter.  

\subsubsection{Results}


Figure \ref{fig.cluster.Kiel} shows the resulting Kiel diagram for each cluster based on this analysis, along with \texttt{MIST} isochrones at the derived median cluster metallicity and age.  In the case of the clusters with no or only a few TO stars (M107, M3, M13, and M92) we use the literature age for the displayed isochrone.  Most points within a cluster lie along a single isochrone, which is a non-trivial success of the method because, while an individual star must reside on an isochrone, there was no guarantee that all stars would reside on the {\it same} isochrone.   M71 presents an interesting case, as the core He burning sequence is unusually extended in \teff\ for such a metal-rich cluster.  This extension might reflect the observed spread in light elements for this cluster \citep{Cordero2015}.  This effect is not included in our isochrone tables, so the tension there is not surprising.


In Figure \ref{fig.cluster.dist} we compare the \MS\ distances for cluster members to several literature sources.  We show the {\it Gaia} DR2 cluster parallax estimates from \citet{GaiaGC2018}, in which we include the $+0.05$ zero-point correction and also add that as an additional term in the error budget.  For the nearby clusters (in particular M67) the {\it Gaia} parallaxes offer a useful constraint, while for the most distant clusters they provide very weak constraints.  We also include the \citet{Harris2010} tabulation of distances for globular clusters (with a 10\% distance uncertainty for reference), and RR Lyrae standard candles where available \citep{Benedict2011, Deras2018, Hernitschek2019, Deras2019}.  Finally, for M67 we include distances measured from eclipsing binaries \citep{Yakut2009, Sandquist2018} and from astroseismology of giants in the cluster \citep{Stello2016}.  Overall the \MS-derived distances are in good agreement with the literature values.


Figures \ref{fig.cluster.feh} and \ref{fig.cluster.afe} show measured \feh$_i$ and \afe$_i$ values as a function of \logg\, for each cluster.  These plots are valuable for diagnosing systematic trends in the derived abundances, especially since the reported abundances are the {\it initial}, rather than the surface values (the former should not vary with evolutionary state of the star).  In each panel the points are color-coded by \teff.  The weighted medians are shows as black dashed lines, and grey lines indicate $\pm0.1$ for \feh$_i$ and $\pm0.05$ for \afe$_i$ and are meant to guide the eye.

We begin by discussing M67.  This cluster has the most members in our sample at 181 spanning the widest range in \logg.  In addition, the {\it Gaia} parallaxes are high SNR for this cluster ($\langle$SNR$_\pi\rangle=28$), and this cluster has the highest metallicity in the sample, so the data place very demanding constraints on the model.  There is a clear systematic trend in \feh$_i$ with \logg.  In fact, \feh$_i$ is monotonically related to the derived \teff, with a mean offset between the cool and warm stars of $\approx0.15$ dex.   Similar systematic behavior is seen in \afe$_i$.  There are several possible reasons for these offsets including mis-matches between the \texttt{MIST} isochrones, limitations in the atomic line lists, non-LTE effects, or our choice to fix the microturbulence parameter (see Section \ref{sec.m67} below for additional discussion).   Trends of this magnitude as a function of \logg\ or \teff\ are common in the literature \citep[e.g.,][]{Valenti2005, Adibekyan2012, Holtzman2015}, and in particular for studies focusing on M67 \citep{Gao2018, Souto2019}.

M71 shows broadly similar behavior to M67 for \feh$_i$ in which the hotter stars prefer higher metallicities compared to the cooler giants.  Amongst the giants, there is no noticeable systematic trend between \feh$_i$ and \logg.  Moreover, Figure \ref{fig.cluster.afe} shows no trend between \afe$_i$ and either \logg\, or \teff.  However, the uncertainties are larger when compared to the other clusters, as the SNR of the M71 spectra are the lowest amongst our sample (M71 also has high reddening of $A_{\rm V}\approx0.7$).

The remaining four clusters show broadly similar behavior.  Aside from a small number of TO stars in M92, the cluster members are confined to $0<$\logg$<3$. Amongst the giants there is no mean trend between \feh$_i$ and \logg\, in any of M107, M3, M13 and M92.  Moreover, the scatter amongst the measured \feh$_i$ is $<0.1$ dex for all four clusters.  Similar conclusions hold for \afe$_i$, although the scatter is slightly larger for M107 and M92.  


Figure \ref{fig.cluster.afefeh} shows our measured \feh\, and \afe\ values for the six clusters along with 68th percentile ranges.  Here we show the average surface abundances in order to compare to literature parameters.  For M67 we divided the sample into dwarfs (\logg\, $\ge$ 4.4), turn-off stars ($4.0<$\logg\,$<4.4$), and giants (\logg\ $\le4.0$).  For M67 the literature abundances are adopted from \citet[][]{Liu2016}, while for other clusters we show the results from both \citet{Meszaros2015} and \citet{Carretta2009}.  These two sources have a $\approx0.1$ dex offset in their overall metallicity scale.

Based on these star cluster tests we conclude that \MS\ is able to recover the known abundances of \feh\ and \afe\ for clusters that span a wide range in metallicity.  We do not detect any systematic behavior between measured abundances and other stellar parameters (\logg\ and \teff) amongst the giant populations of these clusters (\logg$<3$).  In the two clusters with dwarf stars (M67 and M71) we identify systematics at the $0.1-0.2$ dex level.  Resolving this issue will be the subject of future work.

\begin{figure*}[t!]
\center
    \includegraphics[width=0.95\linewidth]{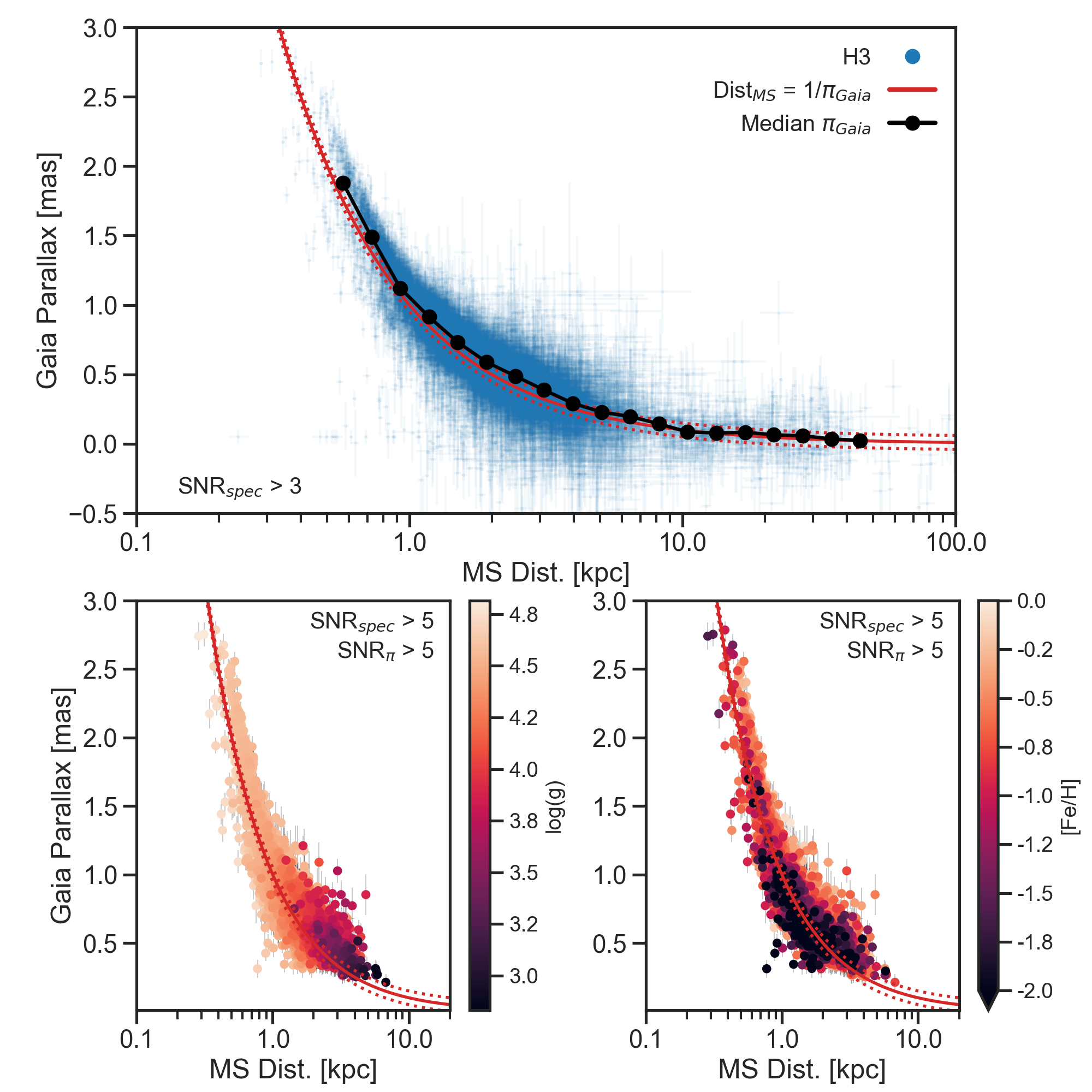}
    \caption{
    Comparison of {\it Gaia} parallaxes to measured \MS\ distances for a set of stars observed as part of the H3 Survey. Top panel shows the comparison for the full sample. The red line shows the $1-1$ line and dotted red lines indicate the {\it Gaia} DR2 zeropoint uncertainty of 0.05 mas.  Black symbols show the binned median Gaia parallax in \MS\ distance bins. The errors on the binned medians are smaller than the datapoints. Bottom panels show the same comparison for a subset of the data with high quality parallaxes and H3 spectra. The color of the points show the individual stellar surface gravity (left) and metallicity (right).  \MS\ distances in this figure were computed {\it without} including {\it Gaia} parallax priors.}
    \label{fig.h3.paradist}
\end{figure*}

\subsubsection{Testing the Effect of the Isochrone Prior with M67}
\label{sec.m67}

Figures \ref{fig.cluster.feh} and \ref{fig.cluster.afe} showed systematic offsets in the derived abundances as a function of \logg\ within individual clusters where both giant and dwarf stars were available.  In this section we test if those offsets can be attributed to the requirement that stars lie on model isochrones.  In particular, we have refit all of the M67 data using \MS\ in a mode where the star is not required to lie on a model isochrone.  In practice, this means that only the surface parameters (\teff, \logg, \feh\, \afe, \Vrad, \Vstar, along with \Av) were fit to the data.  


Figure \ref{fig.cluster.m67tp} shows the result of this test.  The Kiel diagram (left panel) shows overall good agreement with the expected isochrone locus.  A few giants are displaced to substantially higher \logg\ than allowed by the isochrones.  More significantly, the cool dwarfs (\teff$<5000$ K) display an upturn in the Kiel diagram that is in strong disagreement with the expected isochrone behavior.  This behavior has been seen in many previous analyses \citep[e.g.,][]{Adibekyan2012, Bensby2014, Holtzman2015}, although the physical origin of the trend is uncertain.

The middle and right panels of Figure \ref{fig.cluster.m67tp} show the surface \feh\ and \afe\ abundances vs. \logg\ for the fits where stars are not required to lie on a model isochrone.  Compared with the default fits in Figure \ref{fig.cluster.feh}, the isochrone-free results show increased scatter for the dwarfs, and a few giants at very low metallicity (\feh$<-1.0$, which is off the plot in Figure \ref{fig.cluster.m67tp}).  Perhaps more interesting is the behavior of \afe\ with \logg.  Here the result is very similar to the default fits shown in Figure \ref{fig.cluster.afe}, although with slightly larger scatter.  The fact that the U-shaped behavior is clear in both fitting modes strongly suggests that the cause of the behavior is due to the underlying spectral models.  Possible reasons include microturbulence and non-LTE effects.

\subsection{Application to the H3 Survey}
\label{sec.modtest.h3}

In this section we present a final test of the \MS-derived distances by comparing the spectrophotometric distances to {\it Gaia} parallaxes.  For this test we utilize data collected through the H3 Survey \citep{Conroy2019a}.  The survey is obtaining spectra of 200,000 stars in high latitude fields ($|b|>30^\circ$) in order to map the stellar halo of the Galaxy.  H3 employs the Hectochelle instrument on the MMT, which is delivering $R\approx32,000$ spectra over the wavelength range $5150-5300$\AA.  The targets are relatively bright ($15<r<18$) and hence have good SED coverage from Pan-STARRS, {\it Gaia}, 2MASS, and WISE (and SDSS where available).  The main sample selection function is simple, consisting only of magnitude and {\it Gaia} parallax cuts.

\MS\ is being used to derive stellar parameters for the H3 Survey.  In its default mode, the {\it Gaia} parallaxes are used as a prior in the fitting.  However, in order to provide an independent test of the \MS-based distances, we have re-fit a subset of the H3 data excluding the {\it Gaia} parallaxes.

The result of this test is shown in Figure \ref{fig.h3.paradist}.  In the top panel we show all stars that have SNR$_{\rm spec}>3$, H3 quality flag $=0$, and a stellar broadening \Vstar $<5$ \kms.  The latter cut was necessary as there is a failure mode where \MS\ occasionally attempts to fit a giant solution with a large broadening, when the star is in fact a dwarf (using the {\it Gaia} parallax as a prior tends to alleviate this failure mode).  The one-to-one line is shown as a red line, and binned medians and error on the medians are shown as black circles and error bars (the latter are generally too small to be seen in the plot).  We plot the {\it Gaia} parallaxes with a zero-point offset of 0.05 mas \citep{Leung2019, Schonrich2019}; in addition, red dotted lines show $\pm0.05$ mas systematic uncertainties in the {\it Gaia} parallaxes \citep{Lindegren2018, Schonrich2019}.  The lower panels show results for a subset of data with higher spectroscopic SNR and with a {\it Gaia} parallax SNR$_{\pi}>5$.  In the lower panels the points are color-coded by \logg\, (left panel) and \feh\, (right panel).

Overall the agreement is good, especially at $d>1$ kpc, which provides an independent check that the \MS-based distances are reliable. At $d<1$ kpc the \MS-based distances tend to be over-estimates compared to {\it Gaia} parallaxes.  From inspection of the lower panels it is clear that the discrepancy can be traced to metal-rich dwarf stars.  Such stars are challenging to model not only because their cool temperatures result in strong atomic and molecular absorption feature but also because stellar isochrones have difficulty reproducing observations in this regime \citep[e.g.,][]{Feiden12, Choi2016}.  

We provide one additional test of the \MS\ pipeline.  The uniform H3 field placement resulted in serendipitous observations of the Sextans dwarf spheroidal galaxy.  These stars were visually identified as a large overdensity of distant stars in one field.  Closer inspection revealed 26 Sextans stars with S/N$>2$ at the radial velocity associated with Sextans \citep[224 \kms;][]{Walker2009b}.  The mean S/N of these stars is on the low end of the survey (3.3), and so they offer a good test of our ability to recover distances and metallicities at low S/N.  Furthermore, being so distant, these stars all have low \logg\, ($0.7<$\logg$<1.4$).

For the 26 Sextans members we find a mean heliocentric distance of 83 kpc with a $1\sigma$ scatter of 8 kpc.  The mean reported $1\sigma$ errors for this sample is 5 kpc, implying that the reported errors are under-estimated by 60\%.  \citet{Lee2009b} report a distance to Sextans of 86 kpc based on CMD fitting that included the horizontal branch and main sequence turnoff.  We therefore conclude that our reported distances are accurate.  Furthermore, the scatter amongst the individual measurements implies a distance precision of 10\% for upper giant branch stars even at low SNR.

We turn now to the metallicities for Sextans.  We find a mean metallicity of $\langle$[Fe/H]$\rangle=-1.92$ for our 26 Sextans members.  \citet{Kirby2011} report a mean metallicity of $-1.93$ for Sextans \citep[see also][]{Walker2009b}.  From this comparison we conclude that our metallicities are accurate, even for low-\logg\, giants at low S/N.

\section{Summary \& Future Directions}
\label{sec.future}

In this paper we have presented \MS, a new program to estimate stellar parameters and distances by combining broadband photometry, spectra, and parallaxes.  We presented a suite of mock tests to gain intuition for the measurement precision as a function of stellar evolutionary phase, \feh$_i$, SNR$_{\rm spec}$, and SNR$_{\pi}$.  A key result is that at low parallax SNR ($<1$), distances can be determined to better than 20\% even with low SNR spectra, and as precise as 10\% for RGB and 5\% for main sequence stars.  

We then presented several tests of the method using data.  Moderate-resolution, high SNR spectra of seven benchmark stars including Arcturus, Procyon, and the Sun were fit and the resulting parameters agree well with literature values in most cases.  Moderate resolution spectra and broadband photometry of star clusters spanning a wide range of metallicities were also fit.  The derived metallicities, \afe, distances, and ages also agree well with literature estimates.  We identified a modest level of systematic uncertainties in the \feh$_i$ and \afe$_i$ values as a function of \logg\, and \teff\, within each cluster, at the $\approx0.1$ dex level.  The systematic behavior is mostly confined to differences between dwarfs and giants (or low and high \teff).   The magnitude of the systematic behavior is similar to what has been reported in previous work \citep[e.g.,][]{Cohen2005, Meszaros2015, Holtzman2015}.  Systematic behavior is not apparent within the giant branch.  Finally, we fit 22,000 stellar spectra from the H3 Survey (Conroy et al. 2019).  In these fits the {\it Gaia} parallaxes were not used, and the resulting measured distances were then compared to the independent {\it Gaia} parallaxes.  The comparison showed good agreement as a function of distance, \logg, and \feh.  We also showed that with H3 data \MS\, recovers accurate distances and metallicities for the Sextans dwarf galaxy located at 86 kpc and with a mean metallicity of $-1.9$.

There are several directions for future improvement.  On the spectral modeling side, we currently have fixed the microturbulence parameter.  However, this parameter is known to vary with spectral type and evolutionary phase \citep[e.g.,][]{Ramirez2013, Jofre2014}, and so should be included as an additional parameter in the fit.  Stacked residuals have revealed several locations where the line list needs improvement.

On the fitting side, stellar binarity is not currently included in \MS. However, it should be possible to identify binaries in cases where the {\it Gaia} parallax SNR is large and the star is on the main sequence.  In such cases the ``star" will sit above the main sequence owing to the combined light of the two stars \citep[e.g.,][]{Widmark2018}. A way forward is to model the stellar spectrum and SED with a combined model for two stars parameterized by a ratio of secondary to primary mass \citep[for a similar binary modeling approach, see ][]{ElBadry2018}.  Finally, it would be straightforward to include prior information beyond parallax, e.g., interferometric and/or asteroseismic constraints \citep[e.g.,][]{Mints2017}.

\MS\, was developed specifically to measure stellar parameters for the H3 Survey, but it is a general tool that can be applied to spectrophotometric data of any kind.  It is the ideal tool for extracting the maximum amount of information from ongoing and future stellar spectroscopic surveys in the {\it Gaia} era and beyond.


\acknowledgments

We thank Joseph Rodriguez for helping us with the name \MS, and Robert Kurucz for his extensive and important work managing and updating his archive of atomic and molecular transition data and the \texttt{ATLAS}/\texttt{SYNTHE} suite of programs for the general astronomical community. YST is grateful to be supported by the NASA Hubble Fellowship grant HST-HF2-51425.001 awarded by the Space Telescope Science Institute.

Observations reported here were obtained at the MMT Observatory, a joint facility of the Smithsonian Institution and the University of Arizona. We thank the Hectochelle operators Chun Ly, ShiAnne Kattner, Perry Berlind, and Mike Calkins, as well as the support staff at the CfA and U. of Arizona. The computations in this paper were run on the Odyssey/Cannon cpu cluster supported by the FAS Division of Science, Research Computing Group at Harvard University. We thank the staff at Harvard FAS research computing, in particular Scott Yockel and Paul Edmon for their continued support and troubleshooting. 

This work has made use of data from the European Space Agency (ESA) mission {\it Gaia} (\url{https://www.cosmos.esa.int/gaia}), processed by the {\it Gaia} Data Processing and Analysis Consortium (DPAC, \url{https://www.cosmos.esa.int/web/gaia/dpac/consortium}). Funding for the DPAC has been provided by national institutions, in particular the institutions participating in the {\it Gaia} Multilateral Agreement.



\end{CJK*}


\begin{thebibliography}{87}
\expandafter\ifx\csname natexlab\endcsname\relax\def\natexlab#1{#1}\fi

\bibitem[{{Adibekyan} {et~al.}(2012){Adibekyan}, {Sousa}, {Santos}, {Delgado
  Mena}, {Gonz{\'a}lez Hern{\'a}ndez}, {Israelian}, {Mayor}, \&
  {Khachatryan}}]{Adibekyan2012}
{Adibekyan}, V.~Z., {Sousa}, S.~G., {Santos}, N.~C., {Delgado Mena}, E.,
  {Gonz{\'a}lez Hern{\'a}ndez}, J.~I., {Israelian}, G., {Mayor}, M., \&
  {Khachatryan}, G. 2012, \aap, 545, A32

\bibitem[{{Anders} {et~al.}(2019){Anders}, {Khalatyan}, {Chiappini},
  {et~al.}}]{Anders2019}
{Anders}, F., {Khalatyan}, A., {Chiappini}, C., {et~al.} 2019, \aap, 628, A94

\bibitem[{{Anderson} {et~al.}(2018){Anderson}, {Hogg}, {Leistedt},
  {Price-Whelan}, \& {Bovy}}]{Anderson2018}
{Anderson}, L., {Hogg}, D.~W., {Leistedt}, B., {Price-Whelan}, A.~M., \&
  {Bovy}, J. 2018, \aj, 156, 145

\bibitem[{{Bailer-Jones} {et~al.}(2018){Bailer-Jones}, {Rybizki}, {Fouesneau},
  {Mantelet}, \& {Andrae}}]{Bailer-Jones2018}
{Bailer-Jones}, C.~A.~L., {Rybizki}, J., {Fouesneau}, M., {Mantelet}, G., \&
  {Andrae}, R. 2018, \aj, 156, 58

\bibitem[{{Benedict} {et~al.}(2011){Benedict}, {McArthur}, {Feast},
  {et~al.}}]{Benedict2011}
{Benedict}, G.~F., {McArthur}, B.~E., {Feast}, M.~W., {et~al.} 2011, \aj, 142,
  187

\bibitem[{{Bensby} {et~al.}(2014){Bensby}, {Feltzing}, \& {Oey}}]{Bensby2014}
{Bensby}, T., {Feltzing}, S., \& {Oey}, M.~S. 2014, \aap, 562, A71

\bibitem[{{Binney} {et~al.}(2014){Binney}, {Burnett}, {Kordopatis},
  {et~al.}}]{Binney2014}
{Binney}, J., {Burnett}, B., {Kordopatis}, G., {et~al.} 2014, \mnras, 437, 351

\bibitem[{{Blanco-Cuaresma} {et~al.}(2014){Blanco-Cuaresma}, {Soubiran},
  {Jofr{\'e}}, \& {Heiter}}]{Blanco-Cuaresma2014}
{Blanco-Cuaresma}, S., {Soubiran}, C., {Jofr{\'e}}, P., \& {Heiter}, U. 2014,
  \aap, 566, A98

\bibitem[{{Bland-Hawthorn} \& {Gerhard}(2016)}]{BlandHawthorn2016}
{Bland-Hawthorn}, J. \& {Gerhard}, O. 2016, \araa, 54, 529

\bibitem[{{Breddels} {et~al.}(2010){Breddels}, {Smith}, {Helmi},
  {et~al.}}]{Breddels2010}
{Breddels}, M.~A., {Smith}, M.~C., {Helmi}, A., {et~al.} 2010, \aap, 511, A90

\bibitem[{{Burnett} \& {Binney}(2010)}]{Burnett2010}
{Burnett}, B. \& {Binney}, J. 2010, \mnras, 407, 339

\bibitem[{{Cardelli} {et~al.}(1989){Cardelli}, {Clayton}, \&
  {Mathis}}]{Cardelli1989}
{Cardelli}, J.~A., {Clayton}, G.~C., \& {Mathis}, J.~S. 1989, \apj, 345, 245

\bibitem[{{Carretta} {et~al.}(2009{\natexlab{a}}){Carretta}, {Bragaglia},
  {Gratton}, {D'Orazi}, \& {Lucatello}}]{Carretta2009}
{Carretta}, E., {Bragaglia}, A., {Gratton}, R., {D'Orazi}, V., \& {Lucatello},
  S. 2009{\natexlab{a}}, \aap, 508, 695

\bibitem[{{Carretta} {et~al.}(2009{\natexlab{b}})}]{Carretta2009b}
{Carretta}, E. {et~al.} 2009{\natexlab{b}}, \aap, 505, 117

\bibitem[{{Choi} {et~al.}(2016){Choi}, {Dotter}, {Conroy}, {Cantiello},
  {Paxton}, \& {Johnson}}]{Choi2016}
{Choi}, J., {Dotter}, A., {Conroy}, C., {Cantiello}, M., {Paxton}, B., \&
  {Johnson}, B.~D. 2016, \apj, 823, 102

\bibitem[{{Cohen} \& {Mel{\'e}ndez}(2005)}]{Cohen2005}
{Cohen}, J.~G. \& {Mel{\'e}ndez}, J. 2005, \aj, 129, 303

\bibitem[{{Connelly} {et~al.}(2012){Connelly}, {Bizzarro}, {Krot}, {Nordlund},
  {Wielandt}, \& {Ivanova}}]{Connelly2012}
{Connelly}, J.~N., {Bizzarro}, M., {Krot}, A.~N., {Nordlund}, {\r{A}}.,
  {Wielandt}, D., \& {Ivanova}, M.~A. 2012, Science, 338, 651

\bibitem[{{Conroy} {et~al.}(2019){Conroy}, {Bonaca}, {Cargile}, {Johnson},
  {Caldwell}, {Naidu}, {Zaritsky}, {Fabricant}, {Moran}, {Rhee},
  {Szentgyorgyi}, {Berlind}, {Calkins}, {Kattner}, \& {Ly}}]{Conroy2019a}
{Conroy}, C., {Bonaca}, A., {Cargile}, P., {Johnson}, B.~D., {Caldwell}, N.,
  {Naidu}, R.~P., {Zaritsky}, D., {Fabricant}, D., {Moran}, S., {Rhee}, J.,
  {Szentgyorgyi}, A., {Berlind}, P., {Calkins}, M.~L., {Kattner}, S., \& {Ly},
  C. 2019, \apj, 883, 107

\bibitem[{{Cordero} {et~al.}(2015){Cordero}, {Pilachowski}, {Johnson}, \&
  {Vesperini}}]{Cordero2015}
{Cordero}, M.~J., {Pilachowski}, C.~A., {Johnson}, C.~I., \& {Vesperini}, E.
  2015, \apj, 800, 3

\bibitem[{{Creevey} {et~al.}(2019){Creevey}, {Grundahl}, {Th{\'e}venin},
  {Corsaro}, {Pall{\'e}}, {Salabert}, {Pichon}, {Collet}, {Bigot}, {Antoci}, \&
  {Andersen}}]{Creevey2019}
{Creevey}, O., {Grundahl}, F., {Th{\'e}venin}, F., {Corsaro}, E., {Pall{\'e}},
  P.~L., {Salabert}, D., {Pichon}, B., {Collet}, R., {Bigot}, L., {Antoci}, V.,
  \& {Andersen}, M.~F. 2019, \aap, 625, A33

\bibitem[{{Creevey} {et~al.}(2012){Creevey}, {Th{\'e}venin}, {Boyajian},
  {et~al.}}]{Creevey2012}
{Creevey}, O.~L., {Th{\'e}venin}, F., {Boyajian}, T.~S., {et~al.} 2012, \aap,
  545, A17

\bibitem[{{Deras} {et~al.}(2019){Deras}, {Arellano Ferro}, {L{\'a}zaro},
  {Bustos Fierro}, {Calder{\'o}n}, {Muneer}, \& {Giridhar}}]{Deras2019}
{Deras}, D., {Arellano Ferro}, A., {L{\'a}zaro}, C., {Bustos Fierro}, I.~H.,
  {Calder{\'o}n}, J.~H., {Muneer}, S., \& {Giridhar}, S. 2019, \mnras, 486,
  2791

\bibitem[{{Deras} {et~al.}(2018){Deras}, {Arellano Ferro}, {Muneer},
  {Giridhar}, \& {Michel}}]{Deras2018}
{Deras}, D., {Arellano Ferro}, A., {Muneer}, S., {Giridhar}, S., \& {Michel},
  R. 2018, Astronomische Nachrichten, 339, 603

\bibitem[{{Dotter}(2016)}]{Dotter2016}
{Dotter}, A. 2016, \apjs, 222, 8

\bibitem[{{Dotter} {et~al.}(2017){Dotter}, {Conroy}, {Cargile}, \&
  {Asplund}}]{Dotter2017}
{Dotter}, A., {Conroy}, C., {Cargile}, P., \& {Asplund}, M. 2017, \apj, 840, 99

\bibitem[{{Dotter} {et~al.}(2010){Dotter}, {Sarajedini}, {Anderson},
  {Aparicio}, {Bedin}, {Chaboyer}, {Majewski}, {Mar{\'\i}n-Franch}, {Milone},
  \& {Paust}}]{Dotter2010}
{Dotter}, A., {Sarajedini}, A., {Anderson}, J., {Aparicio}, A., {Bedin}, L.~R.,
  {Chaboyer}, B., {Majewski}, S., {Mar{\'\i}n-Franch}, A., {Milone}, A., \&
  {Paust}, N. 2010, \apj, 708, 698

\bibitem[{{El-Badry} {et~al.}(2018){El-Badry}, {Rix}, {Ting}, {Weisz},
  {Bergemann}, {Cargile}, {Conroy}, \& {Eilers}}]{ElBadry2018}
{El-Badry}, K., {Rix}, H.-W., {Ting}, Y.-S., {Weisz}, D.~R., {Bergemann}, M.,
  {Cargile}, P., {Conroy}, C., \& {Eilers}, A.-C. 2018, \mnras, 473, 5043

\bibitem[{{Evans} {et~al.}(2018){Evans}, {Riello}, {De Angeli}, {Carrasco},
  {Montegriffo}, {Fabricius}, {Jordi}, {Palaversa}, {Diener}, \&
  {Busso}}]{Evans2018}
{Evans}, D.~W., {Riello}, M., {De Angeli}, F., {Carrasco}, J.~M.,
  {Montegriffo}, P., {Fabricius}, C., {Jordi}, C., {Palaversa}, L., {Diener},
  C., \& {Busso}, G. 2018, \aap, 616, A4

\bibitem[{{Feiden} \& {Chaboyer}(2012)}]{Feiden12}
{Feiden}, G.~A. \& {Chaboyer}, B. 2012, \apj, 757, 42

\bibitem[{{Gaia Collaboration} {et~al.}(2018{\natexlab{a}}){Gaia
  Collaboration}, {Babusiaux}, {van Leeuwen}, {Barstow},
  {et~al.}}]{GaiaHRD2018}
{Gaia Collaboration}, {Babusiaux}, C., {van Leeuwen}, F., {Barstow}, M.~A.,
  {et~al.} 2018{\natexlab{a}}, \aap, 616, A10

\bibitem[{{Gaia Collaboration} {et~al.}(2018{\natexlab{b}}){Gaia
  Collaboration}, {Helmi}, {van Leeuwen}, {McMillan}, {Massari}, {Antoja},
  {Robin}, {Lindegren}, {Bastian}, \& {Arenou}}]{GaiaGC2018}
{Gaia Collaboration}, {Helmi}, A., {van Leeuwen}, F., {McMillan}, P.~J.,
  {Massari}, D., {Antoja}, T., {Robin}, A.~C., {Lindegren}, L., {Bastian}, U.,
  \& {Arenou}, F. 2018{\natexlab{b}}, \aap, 616, A12

\bibitem[{{Gaia Collaboration} {et~al.}(2016){Gaia Collaboration}, {Prusti},
  {de Bruijne}, {Brown}, {Vallenari}, {Babusiaux}, {Bailer-Jones}, {Bastian},
  {Biermann}, \& {Evans}}]{Gaia2016}
{Gaia Collaboration}, {Prusti}, T., {de Bruijne}, J.~H.~J., {Brown}, A.~G.~A.,
  {Vallenari}, A., {Babusiaux}, C., {Bailer-Jones}, C.~A.~L., {Bastian}, U.,
  {Biermann}, M., \& {Evans}, D.~W. 2016, \aap, 595, A1

\bibitem[{{Gao} {et~al.}(2018){Gao}, {Lind}, {Amarsi}, {Buder}, {Dotter},
  {Nordlander}, {Asplund}, {Bland-Hawthorn}, {de Silva}, {D'Orazi}, {Freeman},
  {Kos}, {Lewis}, {Lin}, {Martell}, {Schlesinger}, {Sharma}, {Simpson},
  {Zucker}, {Zwitter}, {da Costa}, {Anguiano}, {Horner}, {Hyde}, {Kafle},
  {Nataf}, {Reid}, {Stello}, {Ting}, \& {Galah Collaboration}}]{Gao2018}
{Gao}, X., {Lind}, K., {Amarsi}, A.~M., {Buder}, S., {Dotter}, A.,
  {Nordlander}, T., {Asplund}, M., {Bland-Hawthorn}, J., {de Silva}, G. h.~M.,
  {D'Orazi}, V., {Freeman}, K.~C., {Kos}, J., {Lewis}, G.~F., {Lin}, J.,
  {Martell}, S.~L., {Schlesinger}, K.~J., {Sharma}, S., {Simpson}, J.~D.,
  {Zucker}, D.~B., {Zwitter}, T., {da Costa}, G., {Anguiano}, B., {Horner}, J.,
  {Hyde}, E.~A., {Kafle}, P.~R., {Nataf}, D.~M., {Reid}, W., {Stello}, D.,
  {Ting}, Y.-S., \& {Galah Collaboration}. 2018, \mnras, 481, 2666

\bibitem[{{Gratton} {et~al.}(2004){Gratton}, {Sneden}, \&
  {Carretta}}]{Gratton2004}
{Gratton}, R., {Sneden}, C., \& {Carretta}, E. 2004, \araa, 42, 385

\bibitem[{{Harris}(2010)}]{Harris2010}
{Harris}, W.~E. 2010, arXiv e-prints, arXiv:1012.3224

\bibitem[{{Heiter} {et~al.}(2015){Heiter}, {Jofr{\'e}}, {Gustafsson}, {Korn},
  {Soubiran}, \& {Th{\'e}venin}}]{Heiter2015}
{Heiter}, U., {Jofr{\'e}}, P., {Gustafsson}, B., {Korn}, A.~J., {Soubiran}, C.,
  \& {Th{\'e}venin}, F. 2015, \aap, 582, A49

\bibitem[{{Hernitschek} {et~al.}(2019){Hernitschek}, {Cohen}, {Rix}, {Magnier},
  {Metcalfe}, {Wainscoat}, {Waters}, {Kudritzki}, \&
  {Burgett}}]{Hernitschek2019}
{Hernitschek}, N., {Cohen}, J.~G., {Rix}, H.-W., {Magnier}, E., {Metcalfe}, N.,
  {Wainscoat}, R., {Waters}, C., {Kudritzki}, R.-P., \& {Burgett}, W. 2019,
  \apj, 871, 49

\bibitem[{{Hogg} {et~al.}(2010){Hogg}, {Bovy}, \& {Lang}}]{Hogg2010}
{Hogg}, D.~W., {Bovy}, J., \& {Lang}, D. 2010, arXiv e-prints, arXiv:1008.4686

\bibitem[{{Holtzman} {et~al.}(2015){Holtzman}, {Shetrone}, {Johnson},
  {et~al.}}]{Holtzman2015}
{Holtzman}, J.~A., {Shetrone}, M., {Johnson}, J.~A., {et~al.} 2015, \aj, 150,
  148

\bibitem[{{Jofr{\'e}} {et~al.}(2019){Jofr{\'e}}, {Heiter}, \&
  {Soubiran}}]{Jofre2019}
{Jofr{\'e}}, P., {Heiter}, U., \& {Soubiran}, C. 2019, \araa, 57, 571

\bibitem[{{Jofr{\'e}} {et~al.}(2014){Jofr{\'e}}, {Heiter}, {Soubiran},
  {et~al.}}]{Jofre2014}
{Jofr{\'e}}, P., {Heiter}, U., {Soubiran}, C., {et~al.} 2014, \aap, 564, A133

\bibitem[{{Jofr{\'e}} {et~al.}(2015){Jofr{\'e}}, {Heiter}, {Soubiran},
  {et~al.}}]{Jofre2015}
---. 2015, \aap, 582, A81

\bibitem[{{J{\o}rgensen} \& {Lindegren}(2005)}]{Jorgensen2005}
{J{\o}rgensen}, B.~R. \& {Lindegren}, L. 2005, \aap, 436, 127

\bibitem[{{Juri{\'c}} {et~al.}(2008){Juri{\'c}}, {Ivezi{\'c}}, {Brooks},
  {et~al.}}]{Juric2008}
{Juri{\'c}}, M., {Ivezi{\'c}}, {\v{Z}}., {Brooks}, A., {et~al.} 2008, \apj,
  673, 864

\bibitem[{{Kirby} {et~al.}(2011){Kirby}, {Lanfranchi}, {Simon}, {Cohen}, \&
  {Guhathakurta}}]{Kirby2011}
{Kirby}, E.~N., {Lanfranchi}, G.~A., {Simon}, J.~D., {Cohen}, J.~G., \&
  {Guhathakurta}, P. 2011, \apj, 727, 78

\bibitem[{{Kroupa}(2001)}]{Kroupa2001}
{Kroupa}, P. 2001, \mnras, 322, 231

\bibitem[{{Kurucz}(1970)}]{Kurucz1970}
{Kurucz}, R.~L. 1970, SAO Special Report, 309

\bibitem[{{Kurucz}(1993)}]{Kurucz1993}
---. 1993, {SYNTHE spectrum synthesis programs and line data}

\bibitem[{{Kurucz} \& {Avrett}(1981)}]{Kurucz1981}
{Kurucz}, R.~L. \& {Avrett}, E.~H. 1981, SAO Special Report, 391

\bibitem[{{Lee} {et~al.}(2009){Lee}, {Yuk}, {Park}, {Harris}, \&
  {Zaritsky}}]{Lee2009b}
{Lee}, M.~G., {Yuk}, I.-S., {Park}, H.~S., {Harris}, J., \& {Zaritsky}, D.
  2009, \apj, 703, 692

\bibitem[{{Leistedt} \& {Hogg}(2017)}]{Leistedt2017}
{Leistedt}, B. \& {Hogg}, D.~W. 2017, \aj, 154, 222

\bibitem[{{Leung} \& {Bovy}(2019)}]{Leung2019}
{Leung}, H.~W. \& {Bovy}, J. 2019, \mnras, 489, 2079

\bibitem[{{Lindegren} {et~al.}(2018){Lindegren}, {Hern{\'a}ndez}, {Bombrun},
  {et~al.}}]{Lindegren2018}
{Lindegren}, L., {Hern{\'a}ndez}, J., {Bombrun}, A., {et~al.} 2018, \aap, 616,
  A2

\bibitem[{{Liu} {et~al.}(2016){Liu}, {Asplund}, {Yong}, {Mel{\'e}ndez},
  {Ram{\'\i}rez}, {Karakas}, {Carlos}, \& {Marino}}]{Liu2016}
{Liu}, F., {Asplund}, M., {Yong}, D., {Mel{\'e}ndez}, J., {Ram{\'\i}rez}, I.,
  {Karakas}, A.~I., {Carlos}, M., \& {Marino}, A.~F. 2016, \mnras, 463, 696

\bibitem[{{Luri} {et~al.}(2018){Luri}, {Brown}, {Sarro}, {Arenou},
  {Bailer-Jones}, {Castro-Ginard}, {de Bruijne}, {Prusti}, {Babusiaux}, \&
  {Delgado}}]{Luri18}
{Luri}, X., {Brown}, A.~G.~A., {Sarro}, L.~M., {Arenou}, F., {Bailer-Jones},
  C.~A.~L., {Castro-Ginard}, A., {de Bruijne}, J., {Prusti}, T., {Babusiaux},
  C., \& {Delgado}, H.~E. 2018, \aap, 616, A9

\bibitem[{{Ma{\'\i}z Apell{\'a}niz} \& {Weiler}(2018)}]{MaizApellaniz2018}
{Ma{\'\i}z Apell{\'a}niz}, J. \& {Weiler}, M. 2018, \aap, 619, A180

\bibitem[{{M{\'e}sz{\'a}ros} {et~al.}(2015){M{\'e}sz{\'a}ros}, {Martell},
  {Shetrone}, {Lucatello}, {Troup}, {Bovy}, {Cunha},
  {Garc{\'\i}a-Hern{\'a}ndez}, {Overbeek}, \& {Allende Prieto}}]{Meszaros2015}
{M{\'e}sz{\'a}ros}, S., {Martell}, S.~L., {Shetrone}, M., {Lucatello}, S.,
  {Troup}, N.~W., {Bovy}, J., {Cunha}, K., {Garc{\'\i}a-Hern{\'a}ndez}, D.~A.,
  {Overbeek}, J.~C., \& {Allende Prieto}, C. 2015, \aj, 149, 153

\bibitem[{{Mints} \& {Hekker}(2017)}]{Mints2017}
{Mints}, A. \& {Hekker}, S. 2017, \aap, 604, A108

\bibitem[{{Mints} \& {Hekker}(2018)}]{Mints2018}
---. 2018, \aap, 618, A54

\bibitem[{Paszke {et~al.}(2017)Paszke, Gross, Chintala, Chanan, Yang, DeVito,
  Lin, Desmaison, Antiga, \& Lerer}]{Paszke2017}
Paszke, A., Gross, S., Chintala, S., Chanan, G., Yang, E., DeVito, Z., Lin, Z.,
  Desmaison, A., Antiga, L., \& Lerer, A. 2017, in NIPS-W

\bibitem[{{Paxton} {et~al.}(2011){Paxton}, {Bildsten}, {Dotter}, {Herwig},
  {Lesaffre}, \& {Timmes}}]{Paxton2011}
{Paxton}, B., {Bildsten}, L., {Dotter}, A., {Herwig}, F., {Lesaffre}, P., \&
  {Timmes}, F. 2011, \apjs, 192, 3

\bibitem[{{Pont} \& {Eyer}(2004)}]{Pont2004}
{Pont}, F. \& {Eyer}, L. 2004, \mnras, 351, 487

\bibitem[{{Portillo} {et~al.}(2019){Portillo}, {Speagle}, \&
  {Finkbeiner}}]{Portillo2019}
{Portillo}, S. K.~N., {Speagle}, J.~S., \& {Finkbeiner}, D.~P. 2019, arXiv
  e-prints, arXiv:1902.02374

\bibitem[{{Pritzl} {et~al.}(2005){Pritzl}, {Venn}, \& {Irwin}}]{Pritzl2005}
{Pritzl}, B.~J., {Venn}, K.~A., \& {Irwin}, M. 2005, \aj, 130, 2140

\bibitem[{{Ram{\'{\i}}rez} {et~al.}(2013){Ram{\'{\i}}rez}, {Allende Prieto}, \&
  {Lambert}}]{Ramirez2013}
{Ram{\'{\i}}rez}, I., {Allende Prieto}, C., \& {Lambert}, D.~L. 2013, \apj,
  764, 78

\bibitem[{{Rodriguez} {et~al.}(2017){Rodriguez}, {Zhou}, {Vanderburg},
  {Eastman}, {Kreidberg}, {Cargile}, {Bieryla}, {Latham}, {Irwin}, \&
  {Mayo}}]{Rodriguez2017}
{Rodriguez}, J.~E., {Zhou}, G., {Vanderburg}, A., {Eastman}, J.~D.,
  {Kreidberg}, L., {Cargile}, P.~A., {Bieryla}, A., {Latham}, D.~W., {Irwin},
  J., \& {Mayo}, A.~W. 2017, \aj, 153, 256

\bibitem[{{Sahlholdt} {et~al.}(2019){Sahlholdt}, {Feltzing}, {Lindegren}, \&
  {Church}}]{Sahlholdt2019}
{Sahlholdt}, C.~L., {Feltzing}, S., {Lindegren}, L., \& {Church}, R.~P. 2019,
  \mnras, 482, 895

\bibitem[{{Sanders} \& {Das}(2018)}]{Sanders2018}
{Sanders}, J.~L. \& {Das}, P. 2018, \mnras, 481, 4093

\bibitem[{{Sandquist} {et~al.}(2018){Sandquist}, {Mathieu}, {Quinn}, {Pollack},
  {Latham}, {Brown}, {Esselstein}, {Aigrain}, {Parviainen}, \&
  {Vanderburg}}]{Sandquist2018}
{Sandquist}, E.~L., {Mathieu}, R.~D., {Quinn}, S.~N., {Pollack}, M.~L.,
  {Latham}, D.~W., {Brown}, T.~M., {Esselstein}, R., {Aigrain}, S.,
  {Parviainen}, H., \& {Vanderburg}, A. 2018, \aj, 155, 152

\bibitem[{{Schlafly} \& {Finkbeiner}(2011)}]{Schlafly2011}
{Schlafly}, E.~F. \& {Finkbeiner}, D.~P. 2011, \apj, 737, 103

\bibitem[{{Schlegel} {et~al.}(1998){Schlegel}, {Finkbeiner}, \&
  {Davis}}]{Schlegel1998}
{Schlegel}, D.~J., {Finkbeiner}, D.~P., \& {Davis}, M. 1998, \apj, 500, 525

\bibitem[{{Sch{\"o}nrich} \& {Bergemann}(2014)}]{Schonrich2014}
{Sch{\"o}nrich}, R. \& {Bergemann}, M. 2014, \mnras, 443, 698

\bibitem[{{Sch{\"o}nrich} {et~al.}(2019){Sch{\"o}nrich}, {McMillan}, \&
  {Eyer}}]{Schonrich2019}
{Sch{\"o}nrich}, R., {McMillan}, P., \& {Eyer}, L. 2019, \mnras, 487, 3568

\bibitem[{{Smiljanic} {et~al.}(2014){Smiljanic}, {Korn}, {Bergemann},
  {et~al.}}]{Smiljanic2014}
{Smiljanic}, R., {Korn}, A.~J., {Bergemann}, M., {et~al.} 2014, \aap, 570, A122

\bibitem[{{Souto} {et~al.}(2019){Souto}, {Allende Prieto}, {Cunha},
  {Pinsonneault}, {Smith}, {Garcia-Dias}, {Bovy}, {Garc{\'\i}a-Hern{\'a}ndez},
  {Holtzman}, {Johnson}, {J{\"o}nsson}, {Majewski}, {Shetrone}, {Sobeck},
  {Zamora}, {Pan}, \& {Nitschelm}}]{Souto2019}
{Souto}, D., {Allende Prieto}, C., {Cunha}, K., {Pinsonneault}, M., {Smith},
  V.~V., {Garcia-Dias}, R., {Bovy}, J., {Garc{\'\i}a-Hern{\'a}ndez}, D.~A.,
  {Holtzman}, J., {Johnson}, J.~A., {J{\"o}nsson}, H., {Majewski}, S.~R.,
  {Shetrone}, M., {Sobeck}, J., {Zamora}, O., {Pan}, K., \& {Nitschelm}, C.
  2019, \apj, 874, 97

\bibitem[{{Speagle}(2020)}]{Speagle2020a}
{Speagle}, J.~S. 2020, \mnras, 493, 3132

\bibitem[{{Stello} {et~al.}(2016){Stello}, {Vanderburg}, {Casagrande},
  {Gilliland}, {Silva Aguirre}, {Sandquist}, {Leiner}, {Mathieu}, \&
  {Soderblom}}]{Stello2016}
{Stello}, D., {Vanderburg}, A., {Casagrande}, L., {Gilliland}, R., {Silva
  Aguirre}, V., {Sandquist}, E., {Leiner}, E., {Mathieu}, R., \& {Soderblom},
  D.~R. 2016, \apj, 832, 133

\bibitem[{{Szentgyorgyi} {et~al.}(2011){Szentgyorgyi}, {Furesz}, {Cheimets},
  {et~al.}}]{Szentgyorgyi2011}
{Szentgyorgyi}, A., {Furesz}, G., {Cheimets}, P., {et~al.} 2011, \pasp, 123,
  1188

\bibitem[{{Takeda} {et~al.}(2007){Takeda}, {Ford}, {Sills}, {Rasio}, {Fischer},
  \& {Valenti}}]{Takeda2007}
{Takeda}, G., {Ford}, E.~B., {Sills}, A., {Rasio}, F.~A., {Fischer}, D.~A., \&
  {Valenti}, J.~A. 2007, \apjs, 168, 297

\bibitem[{{Ting} {et~al.}(2019){Ting}, {Conroy}, {Rix}, \&
  {Cargile}}]{Ting2019}
{Ting}, Y.-S., {Conroy}, C., {Rix}, H.-W., \& {Cargile}, P. 2019, \apj, 879, 69

\bibitem[{{Valenti} \& {Fischer}(2005)}]{Valenti2005}
{Valenti}, J.~A. \& {Fischer}, D.~A. 2005, \apjs, 159, 141

\bibitem[{{Walker} {et~al.}(2009){Walker}, {Mateo}, \&
  {Olszewski}}]{Walker2009b}
{Walker}, M.~G., {Mateo}, M., \& {Olszewski}, E.~W. 2009, \aj, 137, 3100

\bibitem[{{Wang} {et~al.}(2016){Wang}, {Shi}, {Zhao}, {et~al.}}]{Wang2016}
{Wang}, J., {Shi}, J., {Zhao}, Y., {et~al.} 2016, \mnras, 456, 672

\bibitem[{{Widmark} {et~al.}(2018){Widmark}, {Leistedt}, \&
  {Hogg}}]{Widmark2018}
{Widmark}, A., {Leistedt}, B., \& {Hogg}, D.~W. 2018, \apj, 857, 114

\bibitem[{{Xue} {et~al.}(2014){Xue}, {Ma}, {Rix}, {et~al.}}]{Xue2014}
{Xue}, X.-X., {Ma}, Z., {Rix}, H.-W., {et~al.} 2014, \apj, 784, 170

\bibitem[{{Xue} {et~al.}(2015){Xue}, {Rix}, {Ma}, {Morrison}, {Bovy}, {Sesar},
  \& {Janesh}}]{Xue2015}
{Xue}, X.-X., {Rix}, H.-W., {Ma}, Z., {Morrison}, H., {Bovy}, J., {Sesar}, B.,
  \& {Janesh}, W. 2015, \apj, 809, 144

\bibitem[{{Yakut} {et~al.}(2009){Yakut}, {Zima}, {Kalomeni}, {van Winckel},
  {Waelkens}, {De Cat}, {Bauwens}, {Vu{\v{c}}kovi{\'c}}, {Saesen}, \& {Le
  Guillou}}]{Yakut2009}
{Yakut}, K., {Zima}, W., {Kalomeni}, B., {van Winckel}, H., {Waelkens}, C., {De
  Cat}, P., {Bauwens}, E., {Vu{\v{c}}kovi{\'c}}, M., {Saesen}, S., \& {Le
  Guillou}, L. 2009, \aap, 503, 165

\end{thebibliography}
\end{document}